\documentclass[a4paper,11pt]{article}
\pdfoutput=1
\usepackage{jheppub}
\usepackage{graphicx}
\usepackage[figuresright]{rotating}
\usepackage{bm,amsmath,amssymb}
\usepackage[mathscr]{eucal}
\usepackage{multirow}
\usepackage[dvipsnames]{xcolor}
\usepackage{mathrsfs}
\usepackage{mathtools}
\usepackage{slashed}
\usepackage{graphics}
\usepackage{graphicx}
\usepackage{subcaption}
\usepackage{dsfont}
\usepackage{longtable}
\usepackage{bbm} 
\usepackage{float}
\usepackage{tcolorbox}
\usepackage{lipsum}
\usepackage{cancel}
\usepackage{enumerate}
\usepackage{braket}
\usepackage{makecell}
\usepackage{soul}

\usepackage{transparent}

\definecolor{lcolor}{rgb}{0.,0.0,0.}
\definecolor{citcolor}{rgb}{0,0.,0.5}

\newcommand{\vect}[1]{\boldsymbol{#1}_{\perp}}
\newcommand{\kt}{\vect{k}}

\newcommand{\pt}{\vect{p}}
\newcommand{\ptb}{\vect{\bar p}}

\newcommand{\qt}{\vect{q}}

\newcommand{\bt}{\vect{b}}

\newcommand{\ktone}{\boldsymbol{k_{1\perp}}}
\newcommand{\kttwo}{\boldsymbol{k_{2\perp}}}

\newcommand{\xt}{\vect{x}}
\newcommand{\rtb}{\vect{\bar r}}
\newcommand{\qtb}{\vect{\bar q}}
\newcommand{\btb}{\vect{\bar b}}

\newcommand{\Qt}{\vect{Q}}
\newcommand{\Qcalt}{\vect{\mathcal Q}}
\newcommand{\Bcalt}{\vect{\mathcal B}}
\newcommand{\Rcalt}{\vect{\mathcal R}}
\newcommand{\Pcalt}{\vect{\mathcal P}}
\newcommand{\Qcaltb}{\vect{\bar \mathcal Q}}
\newcommand{\Bcaltb}{\vect{\bar \mathcal B}}
\newcommand{\Rcaltb}{\vect{\bar \mathcal R}}
\newcommand{\Pcaltb}{\vect{\bar \mathcal P}}

\newcommand{\Ptz}{\boldsymbol{P}_{0\perp}}

\newcommand{\rt}{\vect{r}}

\newcommand{\rtone}{\boldsymbol{r_{1\perp}}}
\newcommand{\rttwo}{\boldsymbol{r_{2\perp}}}

\newcommand{\et}{\vect{\epsilon}}

\newcommand{\der}{\mathrm{d}}

\newcommand{\Tr}{\mathrm{Tr}}

\renewcommand{\arraystretch}{1.2}

\newcommand{\qhat}{\hat{q}}

\title{Open quantum system approach to the transverse momentum broadening of a colour dipole}

\author[a]{François Arleo, }
\author[a]{Pietro Benzoni, }
\author[a]{Paul Caucal, }
\author[a]{Pol Bernard Gossiaux}
 \affiliation[a]{SUBATECH UMR 6457 (IMT Atlantique, Université de Nantes, IN2P3/CNRS), 4 rue Alfred Kastler, 44307 Nantes, France}

\emailAdd{francois.arleo@subatech.in2p3.fr}
\emailAdd{benzoni@subatech.in2p3.fr}
\emailAdd{caucal@subatech.in2p3.fr}
\emailAdd{pol-bernard.gossiaux@subatech.in2p3.fr}

\abstract{
Using the open quantum systems formalism, we study the propagation of a quark-antiquark pair propagating through a dense QCD plasma of size $L$ and transverse momentum broadening transport coefficient $\hat q$, and we derive the Lindblad evolution equation for the density matrix of the system for a transversely homogeneous medium.
We focus on the boosted regime where the opening angle $\theta_{q\bar q}$ of this effective colour dipole satisfies $\theta_{q\bar q}\ll 1$. In the correlation limit where the quark-antiquark relative transverse momentum $p_\perp$ is much larger than the imbalance $q_\perp$ as well as the medium typical transverse momentum scale $Q_s=\sqrt{\hat q L}$, we demonstrate that the resulting Wigner distribution displays quasi factorisation between a hard factor describing the hard splitting producing the $q\bar q$ pair and the transverse momentum imbalance $q_\perp$-distribution encoding the broadening induced by the medium. The factorisation is violated by a ``colour decoherence'' factor that controls the $\theta_{q\bar q}$ dependence of the $q_\perp$-distribution through the ratio $\theta_{q\bar q}/\theta_c$, with $\theta_c \sim (\qhat L^3)^{-1/2}$. The open quantum systems approach enables us to clarify the role of this critical angle $\theta_c$ and its associated critical time $t_c$ in the genuine quantum decoherence of the density matrix in colour and kinematic space: in particular, $t_c$ controls both the suppression of the off-diagonal elements of the density matrix in colour space and the transition between singlet and octet states. We find, however, that colour decoherence sets in earlier than the full decoherence of the density matrix, thereby marking the onset of classical behaviour in the system. Finally, we investigate the corrections beyond the quasi-factorised picture due to the quantum diffusion term in $p_\perp$ of the Lindblad equation and we find that these corrections are mild for Gaussian initial wave-packets.
}
\begin{document}
\maketitle
\newpage 

\section{Introduction}

Jet quenching physics~\cite{Blaizot:2015lma,Connors:2017ptx,Cao:2020wlm,Apolinario:2022vzg,Cao:2024pxc,Mehtar-Tani:2025rty}, broadly referring to the suppression of hard particles and jets in ultra-relativistic heavy-ion collisions compared to proton-proton collisions~\cite{Bjorken:1982tu}, also encompasses a wide range of fascinating emergent phenomena in Quantum Chromodynamics (QCD). These include, for instance, the suppression of the 
medium-induced 
gluon emission spectrum as given by the  Baier, Dokshitzer, Mueller, Peigné, Schiff and Zakharov (BDMPS-Z) spectrum~\cite{Baier:1996kr,Baier:1996sk,Zakharov:1996fv,Zakharov:1997uu,Gyulassy:2000er,Wiedemann:2000za}, also known as the Landau-Pomeranchuk-Migdal effect~\cite{Landau:1953um,Migdal:1956tc} in QCD, the energy loss 
experienced 
by a hard parton through a turbulent gluon cascade~\cite{Blaizot:2013vha,Blaizot:2013hx} leading to the thermalisation of the soft particles~\cite{Baier:2000sb,Schlichting:2020lef,Mehtar-Tani:2022zwf}, or the emergence of anomalous transverse momentum broadening for quarks and gluons propagating through quark-gluon plasma (QGP)~\cite{Liou:2013qya,Blaizot:2014bia,Iancu:2014kga,Caucal:2021lgf}. One of the key objectives of the heavy-ion collision experiments at the LHC and RHIC (in particular with the recent sPHENIX program at the latter~\cite{Belmont:2023fau}) is to reveal these phenomena and to characterise quantitatively jet suppression in order to extract the transport parameters of the quark-gluon plasma~\cite{JET:2013cls,Andres:2016iys,Arleo:2017ntr,Feal:2019xfl,Mehtar-Tani:2021fud,JETSCAPE:2024cqe,Pablos:2025cli}, such as the transverse momentum broadening coefficient $\hat q$, which measures the average transverse momentum squared acquired by a fast parton 
per unit of time via collisions.

In this article, we focus on one of these emergent phenomena, commonly referred to as ``colour (de)coherence'' in the jet quenching literature. Since the earliest studies on this topic~\cite{Mehtar-Tani:2010ebp,Mehtar-Tani:2011hma,Mehtar-Tani:2011vlz,Mehtar-Tani:2012mfa,Casalderrey-Solana:2011ule,Casalderrey-Solana:2012evi}, it has been shown that a colour dipole initially in a colour singlet state can emit large-angle gluons when propagating through a dense medium --- a process that is forbidden in vacuum due to quantum interference effects and which leads to the famous angular ordering property of time-like parton cascades in jet physics~\cite{Dokshitzer:1991wu,Ellis:1996mzs}. Colour decoherence is characterised by a single angular scale, known as the critical angle $\theta_c$, conventionally defined as 
\begin{equation}
    \theta_c=\frac{2}{\sqrt{\qhat L^3}}\,,
\end{equation}
for a dense medium of size $L$. This angle represents the minimal opening angle above which the medium resolves 
the internal colour structure 
of the dipole.
Below this threshold, the medium cannot influence the propagation of the antenna through transverse momentum broadening or radiative processes. Recently, several theoretical advances have been made in order to improve the 
quantitative understanding of colour decoherence, including quark mass corrections~\cite{Armesto:2011ir}, the dilute medium limit~\cite{Mehtar-Tani:2011lic,Vaidya:2026yfa}, the role of the dipole’s finite formation time~\cite{Abreu:2024wka}, the calculation of the antenna radiation pattern beyond the harmonic approximation~\cite{Kuzmin:2025fyu} within the improved opacity expansion~\cite{Mehtar-Tani:2019tvy,Mehtar-Tani:2019ygg}, and the incorporation of anisotropies in QCD matter~\cite{Barata:2024bqp}.

From a phenomenological perspective, colour decoherence leaves its imprint on jet substructure observables~\cite{Marzani:2019hun} and is therefore the subject of intense experimental investigation~\cite{Apolinario:2024equ}. In particular, it has been shown~\cite{Caucal:2018dla,Caucal:2020xad} that it leads to a relative increase of the soft hadron yield within jets as quantified by the jet fragmentation function measured in nucleus-nucleus collisions as compared to proton-proton collisions and that the critical angle $\theta_c$ can be pinned down~\cite{Caucal:2019uvr,Caucal:2021cfb,Pablos:2022mrx,Cunqueiro:2023vxl} by measuring jet substructure observables such as the $\theta_g$ distribution defined after Soft Drop grooming~\cite{Larkoski:2014wba}. More recently, energy-energy correlators (see~\cite{Moult:2025nhu} and references therein for a review) within jets have also emerged as a possible probe of the critical angle $\theta_c$~\cite{Andres:2022ovj,Andres:2023xwr}, due to their strong sensitivity to dynamics occurring at a specific angular scale.

In this work, we aim to approach the phenomenon of colour decoherence from the perspective of open quantum systems (OQS), where the notions of coherence and decoherence are more rigorously defined. In general, the coherence of a quantum state manifests through the presence of off-diagonal elements in the system's density operator, reflecting the fact that the initial state is a superposition of quantum states. When such a system interacts with an environment containing many degrees of freedom, the off-diagonal elements of the subsystem’s density matrix typically decay exponentially (in a given basis), and the density operator becomes diagonal after a timescale of the order of the decoherence time. 
From a physical standpoint, it therefore seems natural to ask whether the phenomenon of colour decoherence, as discussed in the jet quenching literature, can be formalised within this framework. In this respect, our objective differs from that of the recent studies~\cite{Vaidya:2020cyi,Mehtar-Tani:2024smp,Mehtar-Tani:2025xxd} on transverse-momentum broadening and jet energy loss within the open quantum system framework. In those works, the primary goal is to construct a theoretically controlled effective field theory for jet quenching, with open quantum system techniques serving mainly as a convenient tool to factorize the dynamics at the QGP scales from that at the jet scales. By contrast, our emphasis here is on exploiting the formalism to gain deeper insight into the phenomenon of colour decoherence itself. 

Moreover, beyond its conceptual appeal, addressing this problem using the standard tools of quantum mechanics can offer valuable guidance for advancing existing results in the literature, as demonstrated by recent insights from the open quantum systems approach to the dynamics of heavy quarks produced in a quark-gluon plasma. While earlier descriptions of heavy-quark dynamics in QGP largely relied on semiclassical approaches --- such as rate equations for quarkonium dissociation and regeneration or Fokker-Planck/Langevin equations for heavy-quark transport, see e.g.~\cite{Liu:2009nb,Nendzig:2014qka,Du:2015wha} --- the open quantum system framework has indeed emerged as a more rigorous, first-principles description~\cite{Akamatsu:2011se,Blaizot:2017ypk,Brambilla:2016wgg}. By treating heavy quarks and quarkonia as a subsystem coupled to the QGP environment, the open quantum system framework provides a consistent quantum-mechanical description of decoherence, dissipation, medium-induced correlations, and regeneration across different temperature regimes~\cite{Brambilla:2017zei,Yao:2018nmy,Delorme:2024rdo,Daddi-Hammou:2025hdz,Armesto:2026fit}.

In this paper, we investigate the transverse momentum broadening experienced by a colour dipole as it propagates through a medium, and its interplay with colour decoherence. Our work builds upon the recent study~\cite{Barata:2023uoi}, in which the transverse-momentum broadening of a single quark was analysed, uncovering several novel time
scales 
(such as the time scale $t_2$ discussed below) and angular scales governing the transition from quantum to classical dynamics in quark propagation. As in~\cite{Barata:2023uoi}, and in contrast to~\cite{Vaidya:2020cyi}, where a Lindblad equation was also derived for a single massless quark subsystem, we do not restrict our analysis to the single-scattering regime. Instead, we resum multiple soft scatterings to all orders in the description of the dipole propagation. This resummation is crucial~\cite{Mehtar-Tani:2011lic} to account for the exponential suppression of the off-diagonal elements of the subsystem’s density matrix.

With a view towards future phenomenological applications, we consider a kinematic regime in which the dipole is highly boosted 
in the medium rest frame
and has a small opening angle $\theta_{q\bar q}$, smaller or of the order of the jet radius $R$ used in jet reconstruction at RHIC and the LHC (between $R=0.1$ and $R=0.4$). This configuration should be contrasted with that of a dijet pair produced back-to-back in the center-of-mass frame~\cite{Mueller:2016gko,Mueller:2016xoc,Chen:2016cof,Chen:2016vem,Tannenbaum:2017afg,STAR:2017hhs}; rather, it naturally arises, for instance, from the decay of a boosted heavy boson. Moreover, such a boosted configuration in the medium rest frame naturally arises in the propagation of a dipole through cold nuclear matter. As a result, our findings can also be applied in this context, provided that an appropriate modification is made in the description of the environment (cold versus hot).
For non colour-singlet dipoles, it also encompasses hard splitting inside quark or gluon jets which can be experimentally selected using jet substructure techniques (see e.g.~\cite{Banfi:2006hf,Gallicchio:2011xq,Gallicchio:2012ez,Metodiev:2018ftz,Larkoski:2019nwj}). The advantage of considering boosted dipoles is that one can exploit the presence of three widely separated scales: the jet energy $E$, the relative transverse momentum of the pair $p_\perp \sim E \theta_{q\bar q}$, and its total transverse momentum $q_\perp$, such that $E \gg p_\perp \gg q_\perp$. In this regime, the initial condition for the quantum master equation satisfied by the Wigner function can be factorised into two Wigner functions describing, respectively, the center-of-mass and the relative degrees of freedom of the pair. It is a priori not obvious whether such a factorised property will be preserved by the time evolution inside the medium. One of our main results is that factorisation of the Wigner function of the quark-antiquark pair after propagation through the medium is actually violated: the Wigner density for the center-of-mass degrees of freedom (providing the transverse momentum imbalance distribution of the $q\bar q$ pair) still depends on the hard scale $p_\perp$ via the ratio $\theta_{q\bar q}/\theta_c$, or equivalently $L/t_c$ with the critical time $t_c$ of order of $t_c\sim (\qhat \theta_{q\bar q}^2)^{-1/3}$.

This factorisation violation by hot medium effects should be contrasted with the transverse-momentum-dependent (TMD) factorisation~\cite{Collins:2011zzd,Boussarie:2023izj} familiar from the description of~initial state effects in cold nuclear matter~\cite{Arleo:2025oos}, where the cross section for e.g.~dijet production can be written~\cite{delCastillo:2020omr,delCastillo:2021znl} as the product of a hard factor, depending only on the hard scale of the process, and a TMD distribution encoding the dependence on the dijet momentum imbalance (the ``soft'' scale) and the associated cold nuclear matter effects such as gluon saturation in nuclei~\cite{Dominguez:2011wm}. Unlike in the initial-state case, however, transverse-momentum-dependent factorisation is violated in our timelike final state process: the transverse momentum imbalance distribution strongly depends on the ratio $\theta_{q\bar q}/\theta_c$, and thus on the hard scale $p_\perp$ since $\theta_{q\bar q}\sim p_\perp/E$. This behaviour explicitly demonstrates the interplay between transverse momentum broadening and the quantum phenomena related to $\theta_c$ and $t_c$, which we now discuss.

As expected on physical grounds, the transverse momentum imbalance distribution displays the following behaviour. If the $q\bar q$ dipole is initially in a colour-singlet state, the dependence on $\theta_{q\bar q}/\theta_c$ is such that, for wide-angle dipoles with $\theta_{q\bar q}\gg\theta_c$,
the $q_\perp$-distribution coincides with that of two independent quarks, undergoing transverse diffusion with typical mean squared value $\sim Q_s^2\equiv\qhat L$, where $\qhat$ is conventionally defined in the fundamental representation throughout this paper. As a result, its squared second moment scales as 
$\langle q_\perp^2 \rangle = 2\, \qhat L$. By contrast, for small-angle dipoles with
$\theta_{q\bar q}\ll \theta_c$, the $q\bar q$ pair does not acquire any transverse momentum broadening. The physical intuition behind this suppression of transverse diffusion is that the dipole is ``too small to be resolved by the medium". Our approach, however, makes explicit the quantum-mechanical nature of the underlying mechanism encoded in this statement. 

First, we find that the time scale $t_c \sim (\qhat \theta_{q\bar q}^2)^{-1/3}$ governs the suppression of the off-diagonal elements of the density matrix in colour space. From the perspective of the open quantum system formalism, it is therefore appropriate to interpret $t_c$ as the colour decoherence time.
However, while it is indeed true that $t_c$ or $\theta_c$ govern colour decoherence, the fundamental quantum process at work in the $\theta_{q\bar q}/\theta_c$ dependence of the $q_\perp$-distribution is that of a two-level system undergoing decay from one state (the colour-singlet state of the dipole) to another (the colour-octet state) after a time which is also set by $t_c$. After $t_c$, the dipole propagates as a colour octet ``gluon'' state and thus suffers broadening, unlike the singlet state. In addition, we also find that the complete decoherence of the density 
matrix due to interactions between the subsystem and the thermal bath happens much later than $t_c$. For an initial density matrix in the momentum basis whose off-diagonal support has a typical width of order $\Lambda$, the corresponding classicalization time scales as $t_2 \sim (E^2/(\qhat \Lambda^2))^{1/3}$. For high-energy dipoles, this time scale is parametrically larger than $t_c$, and typically larger than $L$ as well, meaning that the dipole, while classical in colour space if $L\ge t_c$, remains a genuine quantum object when escaping the medium.

As the Lindblad quantum master equation we derive does not assume the hierarchy $p_\perp\gg q_\perp\sim Q_s$ from the outset, we can also assess the importance of power corrections in the ratio $Q_s/p_\perp$. These corrections, coming from the quantum $p_\perp$-diffusion operator in the master equation, are investigated through the calculation of the singlet-to-octet transition probability, which is the quantity of greatest phenomenological importance since it enters various other processes beyond transverse-momentum broadening, such as the medium-induced gluon emission rate from a colour dipole. For an initial Gaussian wave-packet, we find that even for relatively low-$p_\perp$ quark-antiquark pairs, the effect of the quantum diffusion term in $p_\perp$ yields only a mild correction to the transition probability obtained when neglecting it. Albeit obtained for a specific functional form of the initial condition, these results provide indications that the formula obtained from the simplified quantum master equation, which neglects power corrections in $Q_s/p_\perp$, should be phenomenologically robust. Indeed, by virtue of the linearity of the quantum master equation, our conclusions can be extended to initial conditions that are not necessarily Gaussian, provided they can be expressed as linear superpositions of Gaussian wave packets with widths within the range where the $p_\perp$-diffusion term has only a mild effect on each component.

Our paper is structured as follows. In the coming section, we specify our notation and properly define the boosted $q\bar q$ dipole configuration on which this study is focused. Section~\ref{sec:lindblad} outlines the derivation of the quantum master equation for the $q\bar q$ pair evolving in the “environment” formed by a weakly coupled quark-gluon plasma, assumed to be transversely homogeneous. In section~\ref{sec:initial}, we briefly discuss the initial conditions we consider when analytically solving the quantum master equation for the Wigner function or the density matrix. The solution to this quantum master equation, for an initial condition factorized in $p_\perp$ and $q_\perp$ and neglecting the $p_\perp$ diffusion term, is obtained and discussed in section~\ref{sec:factorisation-violation}. This section contains our main analytical results, which exhibit the violation of the aforementioned factorisation in the $p_\perp$ and $q_\perp$ dependence of the Wigner function. The next section extends the previous discussion; however, the emphasis is placed on the quantum notion of coherence, which is the main advantage of using the open quantum system formalism to interpret the results physically. Finally, section~\ref{sec:pt-diffusion} is devoted to the corrections induced by the quantum $p_\perp$ diffusion term in the quantum master equation.

The paper is supplemented by three appendices. The first presents the calculation of the cross section for the transverse-momentum broadening of a $q\bar q$ dipole using standard Feynman diagram techniques. It provides a cross-check of the results obtained within the open quantum system formalism. The second appendix gives additional details on the derivation of the Lindblad equation describing a colour dipole propagating through a QGP in thermal equilibrium. Finally, the third appendix collects lengthy analytic expressions required to derive the Gaussian solution for the colour-singlet component of the Wigner distribution when the quantum $\pt$-diffusion term is included.

\section{Boosted regime and back-to-back limit}
\label{sec:kinematics}

\begin{figure}
    \centering
    \includegraphics[width=0.6\linewidth]{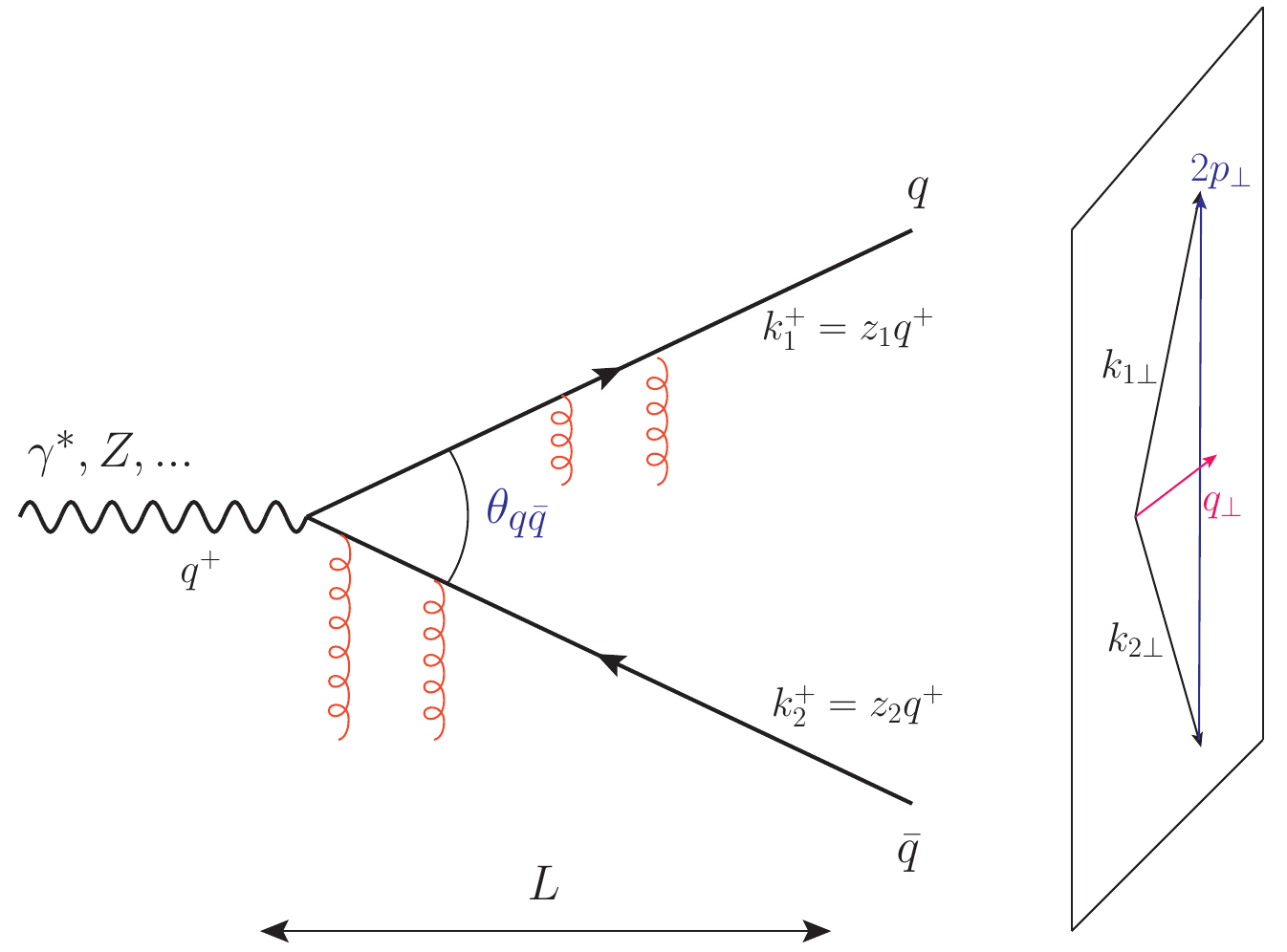}
    \caption{Geometrical configuration of a boosted $q\bar q$ colour dipole propagating through a dense QCD medium of longitudinal size $L$. In the transverse plane projection, we use $z_1=z_2=1/2$ in the definition of $p_\perp$ for the sake of simplicity.}
    \label{fig:qqbar-geometry}
\end{figure}

We start by introducing our notation and specifying the kinematic regime we are primarily interested in throughout this paper, for which the cross-section simplifies considerably, even in the presence of the medium, while still featuring the decoherence effects we aim to study. We note $\pt$ and $\qt$ the relative and total transverse momentum of the $q\bar q$ pair, defined as
\begin{align}
    \pt&=z_2\ktone-z_1\kttwo\,,
    \label{eq:mom-def:rel}
    \\
    \qt&= \ktone+\kttwo\,,
    \label{eq:mom-def:tot}
\end{align}
where $k_i^\mu$ are the four-vectors of the partons and $z_i=k_i^+/(k_1^++k_2^+)$ their light-cone longitudinal momentum fraction w.r.t.~the parent particle\footnote{Throughout this paper, we use the light-cone coordinates 
    $x^\pm=(x^0\pm x^3)/\sqrt{2}$ 
    and $\xt=(x^1,x^2)$ for any four vector $x^\mu=(x^0,x^1,x^2,x^3)$.}. It is useful to define in an analogous way also the relative and center-of mass transverse position of the $q\bar q$ pair as $\rt=\rtone-\rttwo$ and $\bt=z_1\rtone+z_2\rttwo$.
    
    We also introduce the reduced and total plus light-cone component of the $q\bar q$ pair
\begin{align}
    \frac{1}{E}&=\frac{1}{k_1^+}+\frac{1}{k_2^+}\,,\quad q^+=k_1^+ + k_2^+\,,
    \label{eq:momplus:def}
\end{align}
where $E$ plays the roles of the total mass in analogy with the heavy-quark case. Note that for symmetric splitting with $z_{1},z_2=O(1)$, which is the typical situation for a quark antiquark pair produced from the decay of an electroweak boson, one has $E\sim q^+$ and the dijet invariant mass $M_{q\bar q}^2=(k_1+k_2)^2=p_\perp^2/(z_1z_2)\sim p_\perp^2$. The opening angle of the $q\bar q$ pair is related to these kinematic variables as
\begin{align}
    1-\cos(\theta_{q\bar q})&=\frac{p_\perp^2}{E^2}\,.
\end{align}
The highly boosted regime corresponds to $E\gg p_\perp$ such that $\theta_{q\bar q}\ll 1$ and more precisely, expanding the cosine function, $\theta_{q\bar q}^2=2p_\perp^2/E^2$.  In practice, we have in mind $\theta_{q\bar q}\lesssim 0.1\div 0.4$, such that $\theta_{q\bar q}$ is small, but can still be comparable to typical jet reconstruction radii used in jet analysis in heavy-ion collisions.

At leading order in the strong coupling $\alpha_s$ (no radiative corrections) and in the vacuum, the quark-antiquark pair is produced exactly back-to-back in the transverse plane (transverse with respect to the parent particle direction), meaning that the cross-section is proportional to $\delta(\qt)$. Beyond LO, the broadening of the imbalance distribution by QCD radiation can be described using Sudakov resummation techniques. Here, our main objective is to investigate, using the open quantum system formalism, what happens in the presence of a medium. We nevertheless keep this back-to-back configuration as our baseline and therefore systematically assume that the transverse-momentum imbalance is small compared to the relative transverse momentum, $p_\perp\gg q_\perp$. In addition, there exists an intrinsic transverse-momentum scale associated with the medium, given by the total transverse-momentum broadening acquired by a quark propagating over a distance $L$, namely $Q_s^2\equiv \qhat L$.  We also assume that the medium “saturation” scale $Q_s$ is much smaller than $p_\perp$, i.e.~$p_\perp \gg Q_s$; however, the two scales $q_\perp$ and $Q_s$ can be of the same order, yielding a non-trivial imprint of the medium on the transverse-momentum distribution of the quark-antiquark pair. To summarise, we focus on the kinematic regime
\begin{align}
E \gg p_\perp \gg q_\perp ,Q_s\,,
\label{eq:kinematics:hierarchy}
\end{align}
except in the last section~\ref{sec:pt-diffusion}, where we address corrections of order $\sim Q_s/p_\perp$ in the Wigner phase-space distribution. We shall refer to the hierarchy $p_\perp\gg q_\perp,Q_s$ as the back-to-back limit, while the ordering $E\gg p_\perp$ defines the boosted regime. We note that this regime is physically relevant for the phenomenology of high energy jets in heavy-ion collisions such as those produced at the LHC. For $\qhat\sim 1.5$ GeV$^2$/fm and $L\sim 4$ fm, we have $\theta_c\sim 0.05$ and $Q_s\sim 2.45$ GeV, such that, for $E\sim 100\div 1000$ GeV, $p_\perp \gtrsim 5\div 50$ GeV when $\theta_{q\bar q}$ is of the order of the critical angle $\theta_c$.

In the analysis presented in the following sections, we assume that the quark-antiquark pair already exists and do not explicitly discuss its production mechanism (except partially when introducing the initial condition of the system’s density matrix). Nevertheless, it is important to keep in mind that, for the process to be physically relevant, the $q\bar q$ dipole must be produced inside the medium, with a formation time $t_f$ (much) shorter than $L$ . The formation time can be estimated as $t_f \sim E/p_\perp^2 \sim 1/(\theta_{q\bar q}^2 E)$, which implies $\theta_{q\bar q}^2 \gtrsim 1/(E L)$. While this constraint is not very restrictive at sufficiently high energy, one should remember that $\theta_{q\bar q}$ cannot be arbitrarily small if the dipole is to be produced inside the medium.

\section{Lindblad evolution of the quark-antiquark dipole density matrix}
\label{sec:lindblad}

In this section, we use the open quantum system framework to derive the Lindblad master equation for the evolution of the $q\bar q$ dipole, whose kinematic regime has been described in the previous section, inside a dense QCD medium, assumed to be transversely homogeneous over transverse distances much larger than the transverse size of the dipole. In particular, we first rely on the high-energy limit, in which the dynamics of, say, the quark closely resemble those of a particle with an effective mass given by its plus light-cone momentum $k_1^+$ moving in the two-dimensional transverse plane to its direction of motion. The dynamics of the quark-antiquark pair can then be described using standard non-relativistic quantum mechanics~\cite{Blaizot:2015lma}. It follows that the light-cone hamiltonian for the dipole inside the quark-gluon plasma takes the simple form
\begin{equation}
H
=
H_0\otimes\mathbbm{1}_E
+
H_{\rm int} + \mathbbm{1}_S\otimes H_\text{pl},
\label{eq:ham:dipole}
\end{equation}
where $H_0$ contains the usual free kinetic terms for the two independent particles with effective masses $k_1^+$ and $k_2^+$:
\begin{equation}
H_0=\frac{\hat p_{\perp,q}^2}{2k_1^+}\otimes\mathbbm{1}_{\bar q}+\mathbbm{1}_{q}\otimes\frac{\hat p_{\perp,\bar{q}}^2}{2k_2^+},
\label{eq:ham:kinetic}
\end{equation}
while the weak-coupling interaction term between the dipole and the plasma reads
\begin{equation}
H_\text{int} = -g\int\der^2\xt \ \hat n^a(\xt)
\otimes A^-_a(t,\xt)\,,
\label{eq:ham:interaction}
\end{equation}
with
\begin{equation}
\hat{n}^a(\xt)
=
\delta(\xt-\hat{x}_{\perp,q})t^a\otimes\mathbbm{1}_{\bar{q}}-\mathbbm{1}_q\otimes\delta(\xt-\hat{x}_{\perp,\bar{q}})\Tilde{t}^a\,.
\label{eq:numberdensity:dipole}
\end{equation}
The identity operators $\mathbbm{1}_E$, $\mathbbm{1}_S$, and $\mathbbm{1}_{q/\bar q}$ respectively act on the Hilbert space of the environment, the subsystem made of the $q\bar q$ pair, and the quark/antiquark themselves. In Eq.\,\eqref{eq:ham:kinetic} and Eq.\,\eqref{eq:numberdensity:dipole}, $\hat x_{\perp,q/\bar q}$ and $\hat p_{\perp,q/\bar q}$ are the quantum position and momentum operators in the two-dimensional transverse space. $H_\text{int}$ describes the interaction between the colour field $A^-$ and the colour charges of the pair. (We work in the light-cone gauge $A^+=0$.) Note that, because the quark and antiquark are moving along the positive light-cone direction, the coupling between the colour charges and the gauge field $A_a^\mu$ only involves the minus component of the latter: the coupling with the transverse component is sub-eikonal, i.e.~suppressed by powers of the large $k_1^+$ or $k_2^+$ components. Finally, in Eq.\,\eqref{eq:ham:dipole}, the hamiltonian $H_\text{pl}$ generates the evolution of the environment degrees of freedom, which we are not interested in describing, as the main aim of the open quantum system approach is to trace them out in order to focus on the subsystem and its interactions with the environment. We note that the form of $H$ in Eq.\,\eqref{eq:ham:interaction} can also be obtained from the QCD Hamiltonian in light-front quantization~\cite{Dirac:1949cp,Kogut:1969xa,Bjorken:1970ah} projected on the Fock subspace with a quark-antiquark pair only.

Another physical situation in which this system can be described through a non-relativistic evolution is the heavy-mass limit in quarkonia physics. The same form of Hamiltonian for a quark-antiquark pair as the one in Eq.\,\eqref{eq:ham:dipole} emerges from the Non-Relativistic QCD effective field theory~\cite{Akamatsu:2020ypb}, with the main difference being that it is formulated in a (3+1) --- instead of a (2+1) --- dimensional spacetime. This close analogy permits us to follow the steps of the derivation of the heavy quark-antiquark master equation \cite{Blaizot:2017ypk}, tailoring it to the new scales and approximations of the light-quark evolution at high energy.

\subsection{Quantum master equation}

In the Schr\"{o}dinger picture the density matrix evolves according to the Von Neumann equation
\begin{equation}
\frac{\der\rho_{S+\rm pl}(t)}{\der t} = -i[H,\rho_{S+\rm pl}(t)]\,,
\label{eq:dens_mat:eqmot:schr-pic}
\end{equation}
where we use the notation $t$ for the light-cone ``time'' $x^+$.
As mentioned above, we are interested only in the evolution of the subsystem. Therefore, we trace out all the environment degrees of freedom by applying the partial trace $\rho(t) \equiv \Tr_\text{pl}\left(\rho_{S+\rm pl}(t)\right)$, assuming that at the initial time $t_0$ the density matrix is factorized as $\rho_{S+\rm pl}(t_0)=\rho_S(t_0)\otimes \rho_{\text{pl}}(t_0)$. The details of the intermediate steps are similar to those performed in~\cite{Blaizot:2017ypk} and have been reported in appendix~\ref{app:master-equation} for completeness. Let us highlight the feature of the calculation of most physical interest:
\begin{itemize}
    \item 
        we assume a weak coupling between the subsystem and the environment, allowing us to use 
        a factorized form for the density matrix at all times $\rho_{S+\rm pl}(t)=\rho_S(t)\otimes\rho_{\rm pl}$,  
        and providing time locality in the evolution equation,
    \item 
        Markovianity is then obtained owing to the fact that the correlation time of the environment $\sim 1/(gT)$ is smaller than the characteristic system ``relaxation'' time of the order of the mean-free path $\sim 1/(g^2 T)$ for a QGP in thermal equilibrium at temperature $T$,
    \item
        the kinematic hierarchy of scales $E\gg p_\perp$ enables one to rule out friction contributions giving longitudinal energy loss from the master equation, which are subdominant for jets propagating in the QGP,
    \item while, \emph{a priori}, $A_a^-$ represents the total colour field --- including both that of the medium and that mediating the interaction between the quark and the antiquark --- causality ensures that the binding potential does not contribute, and only medium-induced correlations remain. These are encoded in the imaginary part of the two-point function of the medium gauge field. In contrast to the quarkonium case~\cite{Blaizot:2017ypk}, the real part, which would account for the screening of the binding potential, vanishes, again as a consequence of causality. Further details are provided in appendix~\ref{app:master-equation}.
\end{itemize}
These steps permit to arrive to the evolution equation
\begin{align}    
\frac{\der\rho(t)}{\der t}+i[H_0,\rho(t)]=
    \frac{g^2}{2}\int\der^2\xt&\int\der^2\xt' \ W(\xt-\xt')\nonumber\\
    &\times
            \left(\{\hat n^a(\xt)\hat n^a(\xt'),\rho(t)\}-2\hat n^a(\xt')\rho(t)\hat n^a(\xt)\right)\,,
    \label{eq:dm:EoM}    
\end{align}
where the potential $W(\rt)\equiv -\mathfrak{Im}(\Delta(\omega=0,\rt))$ is defined as the imaginary value of the zero frequency part of the time-ordered gauge field two-point function in the QGP state~$\rho_{\rm pl}$
\begin{align}
    \left\langle T \left(A_a^-(t,\rt)A_b^-
    (0
    ,\boldsymbol{0}_\perp)\right)\right\rangle_{\rm pl}\equiv -i\delta_{ab}\Delta(t,\rt)\,,\quad\Delta(\omega,\rt)=\int_{-\infty}^\infty \der t e^{i\omega t}\Delta(t,\rt)\,,\label{eq:Wdef}
\end{align}
with $\langle \mathcal{O} \rangle_{\rm pl}\equiv \textrm{Tr}(\rho_{\rm pl}\mathcal{O})$ is the quantum and statistical average value in the state of the quark-gluon plasma described by the density matrix $\rho_{\rm pl}$. In the quarkonia literature, it has already been shown that the operator on the r.h.s.~of the master equation can be expressed in the Lindblad form and that it is responsible for describing the collisions between the propagating partons and the QGP \cite{Blaizot:2017ypk, Delorme:2024rdo}. The functional form of $W(\rt)$ in the case of a thermal state for $\rho_{\rm pl}$ will be shown in the next subsection~\ref{sub:ho-approx}. At this stage, we have only assumed that the two-point function depends on the transverse-coordinate differences, which amounts to assuming that the medium is transversely homogeneous, at least over transverse scales larger than the transverse size of the quark-antiquark dipole.

At this stage of the derivation, we have not yet used the definition of the number density in Eq.\,\eqref{eq:numberdensity:dipole}. This means that the expression in Eq.\,\eqref{eq:dm:EoM} stands for any number of quarks and antiquarks.
In particular, if one restricts ourselves to the single quark case by choosing the number density
$
    \hat{n}^a_q(\xt)
    =
    \delta(\xt-\hat{x}_{\perp,q})t^a
$ and, by imposing the colour structure
$
    \rho= \rho_0\mathbbm{1}+\rho_8^bt^b
$ for the density matrix, we recover the master equation presented in \cite{Barata:2023uoi}.
Then, in order to adapt Eq.\,\eqref{eq:dm:EoM} specifically to the quark-antiquark antenna case, we now introduce the number density in Eq.\,\eqref{eq:numberdensity:dipole}. 

Before computing the matrix elements of the density operator in position space, we consider its colour structure. In colour space, the density operator is a $9\times 9$ matrix belonging to the representation $(\bar3\otimes3)\otimes(3\otimes\bar3)$ of (real) dimension 81, which is indeed the dimension of the vector space of colour density matrices (recall that $\rho$ must be hermitian). This representation can be decomposed as
\begin{equation}
    (\bar3\otimes3)\otimes(3\otimes\bar3)
    = 1\oplus \underbrace{(8\otimes1)}_8 \oplus \underbrace{(1\otimes8)}_8 \oplus \underbrace{(8\otimes8)}_{1\oplus\ldots},
    \label{eq:dm:qqbar:reps}
\end{equation}
meaning that the corresponding structure for colour operator reads
\begin{equation}
    \rho
    =
    A(\mathbbm{1}\otimes\mathbbm{1})+
    B_b(t^b\otimes\mathbbm{1})+
    C_c(\mathbbm{1}\otimes\tilde{t}^c)+
    D_{bc}(t^b\otimes\tilde{t}^c).
    \label{eq:dm:qqbar:tensor:general}
\end{equation}
This decomposition should be understood as the most general decomposition of any 9 by 9 density matrix describing the quantum colour state of the system. 

However, there are additional physical constraints on the structure of the density matrix which are specific to the colour degrees of freedom. Indeed, the total density matrix of the subsystem and the environment, $\rho_{S+\rm pl}(t)$, is invariant under any global $SU(3)$ colour transformation, provided that the initial density matrices $\rho_S(t_0)$ and $\rho_{\rm pl}(t_0)$ are individually invariant under global colour rotations and that the full Hamiltonian commutes with the corresponding global colour generators. In particular, $\rho_{\rm pl}(t_0)$ is invariant for a colour-neutral thermal state, while $\rho_s(t_0)$ is invariant for the production mechanisms considered here: $\gamma^*\to q\bar q$ produces a colour-singlet state, whereas $g\to q\bar q$ produces an octet state which becomes colour invariant after averaging over the initial gluon colour. Since the global $SU(3)$ transformation factorizes between the subsystem and the plasma and the partial trace is covariant under such transformations, this invariance is inherited by the reduced density matrix $\rho(t)$. Therefore, $\rho$ must commute with any $SU(3)$ colour transformation on the $3\otimes\bar 3$ vector space, which implies, by the Schur lemma that $\rho$ is proportional to the identity on each irreducible representation of $3\otimes \bar 3=1\oplus 8$. Denoting as $\ket{s}$ the colour singlet state and $\ket{o^c}$ the eight colour octet states, we have then
\begin{equation}
    \rho
    =
    \rho_s\ket{s}\bra{s} + \rho_o\sum_c\ket{o^c}\bra{o^c},
    \label{eq:dm:qqbar:singlet-octet}
\end{equation}
where the singlet $\ket{s}$ and octet $\ket{o^c}$ vectors are normalized to unity, $\braket{s|s}=1$, $\braket{o^a|o^b}=\delta^{ab}$, and are defined by their inner products with eigenstates in the $q\bar{q}$ colour-anticolour space
\begin{equation}
    \braket{\alpha\bar{\alpha}|s}=\delta_{\alpha\bar{\alpha}}\frac{1}{\sqrt{N_c}}=\braket{s|\alpha\bar{\alpha}},
    \qquad
    \braket{\alpha\bar{\alpha}|o^c}=t^c_{\alpha\bar{\alpha}}\sqrt{2},
    \qquad
    \braket{o^c|\alpha\bar{\alpha}}=\tilde{t}^c_{\alpha\bar{\alpha}}\sqrt{2}.
    \label{eq:app:sing-oct-states:def}
\end{equation}
Expressed in this form, it becomes clear that, although $\rho$ is formally a nine-by-nine matrix, it actually resides in a two-dimensional vector space spanned by the matrices $\ket{s}\bra{s}$ and $\sum_c \ket{o^c}\bra{o^c}$. From a quantum-mechanical perspective, this expression shows that coherences of the form $\ket{s}\bra{o^c}\in(1\otimes 8)$ or $\ket{o^c}\bra{o^{c'}}$ (with $c\neq c'$) are not physically allowed. It is therefore not obvious \emph{a priori} that the commonly used term ``colour decoherence,'' associated with the critical angle $\theta_c$, is appropriate in this context. We will examine in section~\ref{sub:colour-decoherence} to what extent this terminology can nevertheless be justified.

To that aim, as well as for the derivation of the quantum master equation, it is useful to also consider the decomposition of $\rho$ in the $\mathbbm{1}\otimes\mathbbm{1}$, $t^b\otimes\tilde{t}^b$ (summed over $b$) basis, obtained by projecting Eq.\,\eqref{eq:dm:qqbar:tensor:general} onto the two-dimensional $1\oplus1$ irreducible representation:
\begin{equation}
    \rho
    =
    \rho_0\mathbbm{1}\otimes\mathbbm{1}+
    \rho_1t^b\otimes\tilde{t}^b.
    \label{eq:dm:qqbar:0-1}
\end{equation}
It can be shown that the two basis with respective coefficients $(\rho_s,\rho_o)$, $(\rho_0,\rho_1)$ of the effective two dimensional space of the colour density matrix are connected by the relations~\cite{Blaizot:2017ypk}
\begin{equation}
    \rho_0 = \frac{1}{N_c^2}\left[\rho_s+(N_c^2-1)\rho_o)\right],
    \qquad
    \rho_1 = \frac{2}{N_c}(\rho_s-\rho_o).
    \label{eq:dm:basis:01-to-so}
\end{equation}

Now that we have an explicit expression in colour space both for the number density $\hat n^a(\xt)$ and the density matrix $\rho$, Eq.\,\eqref{eq:dm:qqbar:0-1}, we can plug them into the master equation Eq.\,\eqref{eq:dm:EoM} and choose to represent it in coordinate space, where $\ket{\rtone\rttwo}$ is the quantum state such that the quark is located at transverse coordinate $\rtone$ and the antiquark is located at transverse coordinate $\rttwo$. The matrix element of the r.h.s.~of Eq.\,\eqref{eq:dm:EoM} becomes
\begin{align}
    &\frac{1}{2}\int\der^2\xt\int\der^2\xt' \ W(\xt-\xt')\nonumber\\
    &\times \bra{\rtone\rttwo}\left(\{\hat n^a(\xt)\hat n^a(\xt'),\rho(t)\}-2\hat n^a(\xt')\rho(t)\hat n^a(\xt)\right)\ket{\rtone'\rttwo'}=\nonumber\\
    &=\left[
        C_F(2W(\boldsymbol{0}_\perp)-W_a)\rho_0+
        \frac{C_F}{2N_c}\left(W_b-W_c\right)\rho_1
    \right](\mathbbm{1}_q\otimes\mathbbm{1}_{\bar{q}})+\nonumber\\
    &+\left[
        (W_b-W_c)\rho_0+
        C_F(2W(\boldsymbol{0}_\perp)-W_c)\rho_1+
        \frac{1}{2N_c}(W_a-2W_b+W_c)\rho_1
    \right](t^a\otimes\tilde{t}^a),
    \label{eq:qqbar:EoM:L2:0-1}
\end{align}
with the slight abuse of notation $\rho_i\equiv\braket{\rtone\rttwo|\rho_i|\rtone'\rttwo'}$ and where we have introduced
\begin{equation}
    W_a\equiv W_{11'}+W_{22'},\qquad
    W_b\equiv W_{21'}+W_{12'},\qquad
    W_c\equiv W_{12}+W_{1'2'}
    \label{eq:wawbwc}
\end{equation}
and $W_{ij}=W(\boldsymbol{r}_{i\perp}-\boldsymbol{r}_{j\perp})$, $W_{ij'}=W(\boldsymbol{r}_{i\perp}-\boldsymbol{r}_{j\perp}')$, $W_{i'j'}=W(\boldsymbol{r}_{i\perp}'-\boldsymbol{r}_{j\perp}')$.
The result in Eq.\,\eqref{eq:qqbar:EoM:L2:0-1} closely resembles the expression found in Eq.\,(F.3) in \cite{Blaizot:2017ypk}. 
The matrix element of the (colour-transparent) kinetic term \eqref{eq:ham:kinetic} reads
\begin{multline}
    \braket{\rtone\rttwo|[H_0,\rho]|\rtone'\rttwo'}
    =\\=
    \left[
    \left(
            \frac{\nabla^2_{\rtone}}{2k_1^+}+\frac{\nabla^2_{\rttwo}}{2k_2^+}
        \right)
        -
        \left(
            \frac{\nabla^2_{\rtone'}}{2k_1^+}+\frac{\nabla^2_{\rttwo'}}{2k_2^+}
        \right)\right]
        \left(\rho_0\mathbbm{1}_q\otimes\mathbbm{1}_{\bar q}+
    \rho_1t^a\otimes\tilde{t}^a\right).
    \label{eq:qqbar:kinetic:me:0-1}
\end{multline}
The dynamical terms of the master equation Eq.\,\eqref{eq:qqbar:EoM:L2:0-1} and Eq.\,\eqref{eq:qqbar:kinetic:me:0-1} derived here show that the colour structure imposed in the initial state is preserved (either in the 0-1 or $s$-$o$ basis), as no other colour factor rather than $\mathbbm{1}_q\otimes\mathbbm{1}_{\bar q}$ and $t^a\otimes\tilde{t}^a$ appears, meaning that singlet and octet components of the density matrix will behave as two coupled species, with a conserved total population, and whose probabilities are determined without any interferences between the two of them.
This property comes from the fact that under this constraint on the colour structure of the initial state, the initial state density matrix commutes with the interaction term of the Hamiltonian Eq.\,\eqref{eq:ham:interaction}.
This is also the reason why the equation ends up with having a unique potential $W(\xt)$. Indeed, if we had taken a density matrix in a non-singlet representation, that means not projecting onto the colour transformation invariant representation $1\oplus1$, we would have had additional potentials, whose role would have been to make the relative density matrix component decay. This is indeed the case of the density matrix of a single quark, which has two different potentials associated to the singlet and the octet, with the latter exponentially decaying \cite{Barata:2023uoi}.

\begin{figure}[t]
    \centering
    \includegraphics[width=0.4\linewidth]{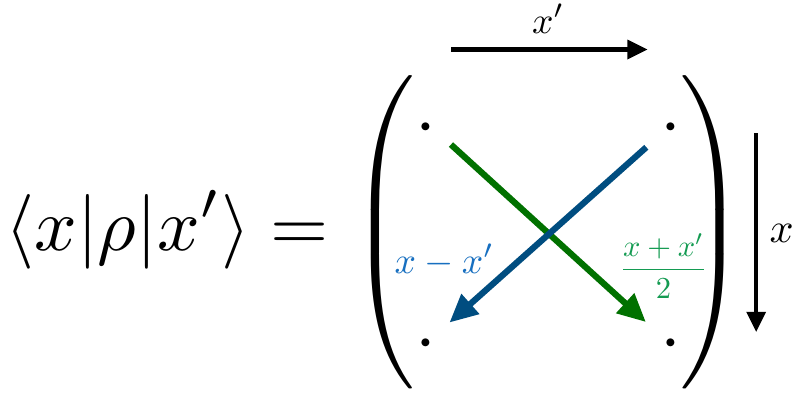}
    \caption{Illustration of the change of variables from $\{x,x'\} \to \{(x+x')/2,x-x'\}$: the average variable $(x+x')/2$ lies along the diagonal of the density matrix, which contains the ``classical'' probability densities, while the difference variable $(x-x')$ spans the off-diagonal elements, thereby quantifying the quantum coherences of the density matrix in the chosen basis.}
    \label{fig:dm:me:cl-vs-qm}
\end{figure}

At this stage, most of the previous developments have focused on adapting the derivation presented in~\cite{Blaizot:2017ypk} to the problem at hand, exploiting the formal similarity between the evolution of a quarkonium state in three-dimensional space and that of a high-energy $q\bar q$ pair in two dimensions. We now proceed to perform a series of manipulations and approximations of the quantum master equation that are tailored to the specific physical process under consideration, whose kinematics were outlined in the previous section. First, before writing the full equation, we perform a change of variables from the four coordinates $\{\rtone, \rttwo, \rtone', \rttwo'\}$ to a different set which can be organized both in terms of \textit{relative} $\{\rt, \Rcalt\}$ and \textit{center-of-mass} $\{\bt, \Bcalt\}$ coordinates which are more convenient as they are Fourier conjugate of the relative $\pt$ and total $\qt$ transverse momenta of the $q\bar q$ pair:
\begin{align}
    \rt&=\frac{(\rtone-\rttwo)+(\rtone'-\rttwo')}{2}\label{eq:c-o-v}\,,\\
    \Rcalt&= (\rtone-\rttwo)-(\rtone'-\rttwo')\,,\\
\bt&=\frac{(z_1\rtone+z_2\rttwo)+(z_1\rtone'+z_2\rttwo')}{2}\,,\\
    \Bcalt&=(z_1\rtone+z_2\rttwo)-(z_1\rtone'+z_2\rttwo')\,.\label{eq:c-o-v-4}
\end{align}
In a given set, we distinguish the ``classical'' $\{\rt,\bt\}$ and ``quantum'' $\{\Rcalt,\Bcalt\}$ coordinates, highlighting the parametrization respectively of the diagonal and off-diagonal (continuously dense) matrix elements, as depicted in Figure~\ref{fig:dm:me:cl-vs-qm}. 
The definition of the variables and their classification is summarized in Table~\ref{tab:variable-table}, where there are also the Fourier counterparts, the momenta, which will be intensively used along the paper. In particular, we define the Wigner distribution as the operator $\rho$ expressed in terms of the phase-space variables $\pt$, $\rt$, $\qt$, and $\bt$. In this representation, these variables play the role of classical phase-space coordinates, which allows the Wigner distribution to be interpreted as a quasi-classical distribution in phase space.
\begin{table}[t]
    \centering
    \renewcommand{\arraystretch}{2.}
    \setlength{\tabcolsep}{4pt}
\begin{tabular}{|c|c|c|c|} \hline & \makecell{\vspace{0.08cm}\textbf{Wigner ``classical'' variable}} & \makecell{\vspace{0.08cm}\textbf{Fourier conjugate}\\ \textbf{``quantum'' variable}} \\ \hline \multirow{2}{*}{relative} & \makecell{\vspace{0.1cm}$\pt=\dfrac{(z_2\ktone-z_1\kttwo)+(...)'}{2}$} & \makecell{\vspace{0.1cm}$\Rcalt=(\rtone-\rttwo)-(...)'$} \\ 
& \makecell{\vspace{0.1cm}$\rt=\dfrac{(\rtone-\rttwo)+(...)'}{2}$} & \makecell{\vspace{0.1cm}$\Pcalt=(z_2\ktone-z_1\kttwo)-(...)'$} \\ \hline \multirow{2}{*}{center-of-mass} & \makecell{\vspace{0.1cm}$\qt=\dfrac{(\ktone+\kttwo)+(...)'}{2}$} & \makecell{\vspace{0.1cm}$\Bcalt=(z_1\rtone+z_2\rttwo)-(...)'$} \\ 
& \makecell{\vspace{0.1cm}$\bt=\dfrac{(z_1\rtone+z_2\rttwo)+(...)'}{2}$} & \makecell{\vspace{0.1cm}$\Qcalt=(\ktone+\kttwo)-(...)'$} \\ \hline \end{tabular}

    \caption{Summary of the notations for the Wigner function arguments and density matrix elements used in this paper. The notation $(...)'$ refers to the same vector as the one in the other parenthesis, but with prime indices, as in Eqs.\,\eqref{eq:c-o-v}-\eqref{eq:c-o-v-4}.}
    \label{tab:variable-table}
\end{table}

Finally, we use the relations in Eq.\,\eqref{eq:dm:basis:01-to-so} to obtain the master equation for the singlet and octet components of the density matrix in position space
\begin{multline}
    \frac{\partial}{\partial t}\begin{pmatrix}\rho_s\\\rho_o\end{pmatrix}-
    i\left(
        \frac{1}{E}
        \nabla_{\rt}\cdot\nabla_{\Rcalt}
        +
        \frac{1}{q^+}
        \nabla_{\bt}\cdot\nabla_{\Bcalt}
    \right)\begin{pmatrix}\rho_s\\\rho_o\end{pmatrix}
    =\\
    =g^2\begin{pmatrix}
        -C_F\Gamma_c & C_F\left(\Gamma_b-\Gamma_a\right)\\
        \frac{\Gamma_b-\Gamma_a}{2N_c} & -C_F \Gamma_a +\frac{\Gamma_a +\Gamma_c-2\Gamma_b}{2N_c}
    \end{pmatrix}
    \begin{pmatrix}\rho_s\\\rho_o\end{pmatrix},
    \label{eq:dm:qqbar:EoM:s-o}
\end{multline}
where $E$ and $q^+$ follow from the same definition as the one in Eq.\,\eqref{eq:momplus:def}, and the imaginary potential has been shifted through the function $\Gamma_i\equiv W_i-2W(\boldsymbol{0}_\perp)$ with $W_a$, $W_b$, $W_c$ defined by Eqs.\,\eqref{eq:wawbwc} followed by the change of variables Eq.\,\eqref{eq:c-o-v}-\eqref{eq:c-o-v-4}, e.g.~$W_{11'}=W(\rtone-\rtone')=W(\Bcalt+z_2\Rcalt)$, $W_{12}=W(\rtone-\rttwo)=W(\Rcalt/2+\rt)$, etc. Equation~\eqref{eq:dm:qqbar:EoM:s-o} constitutes the main result of this section. In the following subsections, we introduce a number of simplifying assumptions that allow us to derive analytic solutions, which will in turn help clarifying the underlying physical mechanisms. 

\subsection{Harmonic potential approximation}
\label{sub:ho-approx}

So far, we have not specified the dependence of the potential $W(\xt)$ --- or equivalently, $\Gamma(\xt)=W(\boldsymbol{x}_\perp)-W(\boldsymbol{0}_\perp)$ --- in the quantum master equation defined by Eq.\,\eqref{eq:Wdef}. If the QGP state described by the density matrix $\rho_{\rm pl}$ is a thermal equilibrium state at temperature $T$, the $W$ potential is obtained from thermal field theory supplemented with hard thermal loop resummation~\cite{Braaten:1989mz}, such that, to leading order in $g$~\cite{Aurenche:2002pd,Arnold:2008vd},
\begin{align}
    \Gamma(\xt)&=\int\frac{\der^2\kt}{(2\pi)^2}\left(1-e^{i\kt\cdot\xt}\right)\frac{m_D^2T}{\kt^2(\kt^2+m_D^2)}\,,\label{eq:WHTL-lo}
\end{align}
with the Debye mass squared $m_D^2=(1+n_F/6)g^2T^2$ and $n_F$ the number of active light flavours. Combining Eq.\,\eqref{eq:dm:qqbar:EoM:s-o} and Eq.\,\eqref{eq:WHTL-lo}, we have all the ingredients necessary to solve the quantum master equation for the density matrix of the quark-antiquark pair, provided some initial condition is chosen.

However, since our aim is to investigate analytically the solution to the quantum master equation Eq.\,\eqref{eq:dm:qqbar:EoM:s-o}, we will perform a further standard approximation consisting in simplifying the function $\Gamma$ for values of $|\xt|$ smaller than $\sim 1/m_D$~\cite{Barata:2020rdn}:
\begin{align}
    g^2C_F\Gamma(\xt)&=\frac{\qhat_0}{4}\xt^2\ln\left(\frac{4e^{2-2\gamma_E}}{\xt^2m_D^2}\right)+O(|\xt|^4m_D^4)\,.
\end{align}
As mentioned in the introduction, our convention is to define the bare quenching parameter $\qhat_0=\alpha_sC_Fm_D^2T$ in the fundamental representation. Here, $\gamma_E\approx 0.577$ is the Euler-Mascheroni constant. 
In addition, in the so-called harmonic approximation, we further neglect the mild logarithmic dependence on $\xt^2$ by redefining $\qhat\equiv\qhat_0\ln(4e^{2-2\gamma_E}Q^2/m_D^2)$ at the scale $Q$ in terms of $\qhat_0$, such that 
\begin{align}
    g^2C_F\Gamma(\xt)&=\frac{\qhat}{4}\xt^2+\frac{\qhat_0}{4}\xt^2\ln\left(\frac{1}{\xt^2Q^2}\right)+O(|\xt|^4m_D^4)\,.\label{eq:harmonic-approx-def}
\end{align}
If the scale $Q$ is chosen close to the typical transverse scale $1/\xt^2$ probed by the process, it is a good approximation to neglect the logarithmic dependence on $1/\xt^2$~\cite{Barata:2020rdn,Barata:2020sav,Barata:2021wuf}. We point out that one can also keep this term and perform the improved opacity expansion~\cite{Mehtar-Tani:2019tvy} around the harmonic potential solution. This would be worth investigating further in the future, in order to relate our calculation to the recent study of the radiation from a QCD antenna within the improved opacity expansion~\cite{Kuzmin:2025fyu}.

In the remainder of this paper, we will systematically employ the harmonic approximation $g^2C_F\Gamma(\xt)=\qhat \xt^2/4$ in the quantum master equation satisfied by the density matrix. Two important caveats should, however, be kept in mind, leaving room for future improvements. First, since the $\Gamma$ functions are evaluated at different transverse coordinate scales in Eq.\,\eqref{eq:dm:qqbar:EoM:s-o}, one should in principle introduce several effective quenching parameters $\qhat$, as the relevant momentum scale $Q$ may differ from term to term. Instead, we will assume a common value for $\qhat$, corresponding to the choice $Q=Q_s=\sqrt{\qhat L}$~\cite{Barata:2020rdn}, which allows us to combine the various contributions on the right-hand side of Eq.\,\eqref{eq:dm:qqbar:EoM:s-o}.

Second, the expansion of the potential to leading order in $\xt^2 m_D^2$ is justified only if all transverse coordinate scales at which $\Gamma$ is evaluated are much smaller than $1/m_D$. The relevant scales are $\rt$, $\Rcalt$, and $\Bcalt$, since by translational invariance the right-hand side of Eq.\,\eqref{eq:dm:qqbar:EoM:s-o} does not depend on the impact parameter $\bt$. By Fourier conjugation, $|\Rcalt|\sim 1/|\pt|$, which is already assumed to be much smaller than $1/Q_s \ll 1/m_D$. Indeed, the ratio $Q_s^2/m_D^2$ scales like $L/\lambda\gg 1$ where $\lambda$ is the mean free path. For the transverse size of the dipole, $|\rt|$, one expects (at least classically) $|\rt|\sim \theta_{q\bar q} t$, since the geometry of the $q\bar q$ dipole implies that the transverse separation between the two particles grows linearly with time. Therefore, one must ensure that at the end of the evolution, i.e.\ for $t=L$, $|r_\perp|\ll 1/m_D$. This condition is satisfied provided $m_D \ll (L\theta_{q\bar q})^{-1}$. Finally, concerning $\Bcalt$, one should first note that, as will be shown in section \ref{subsect5:2}, its typical value decreases with time down to a scale of order $1/q_\perp \sim 1/Q_s \ll 1/m_D$. However, at early times $\Bcalt$ is controlled by the initial condition of the density matrix. One must therefore ensure that the typical initial value of $|\Bcalt|$ is smaller than $1/m_D$ in order to be consistent with the harmonic approximation.

In light of these considerations, we simplify the right-hand side of Eq.\,\eqref{eq:dm:qqbar:EoM:s-o} by adopting the harmonic approximation,
\begin{align}
		g^2\Gamma_a&\simeq
		\frac{\hat q}{4C_F}
        \left[
            2\Bcalt^2
    		+\left(z_1^2+z_2^2\right)\Rcalt^2
    		+2\left(z_2-z_1\right)\Rcalt\cdot\Bcalt
        \right],
		\label{eq:gamma:a:SC}\\
        g^2\Gamma_b&\simeq \frac{\hat q}{4C_F}
        \left[
    		2\Bcalt^2
    		+\frac{1}{2}\left(z_2-z_1\right)^2\Rcalt^2+
    		2\left(z_2-z_1\right)\Rcalt\cdot\Bcalt+2\rt^2
        \right],
		\label{eq:gamma:b:SC}\\
        g^2\Gamma_c&\simeq \frac{\hat q}{4C_F}
    \left[
    	2\rt^2+
    	\frac{1}{2}\Rcalt^2
    \right].
	\label{eq:gamma:c:SC}
	\end{align}
Performing then the Wigner transform $\Rcalt\to\pt$, $\Bcalt\to\qt$, we obtain\footnote{Throughout the paper, we will slightly abuse notation by using $\rho$ to denote both the density matrix $\rho(\pt,\Pcalt;\qt,\Qcalt;t)$ in the momentum basis (or $\rho(\Rcalt,\rt;\Bcalt,\bt;t)$ in the position basis) and the Wigner function $\rho(\pt,\rt;\qt,\bt;t)$, although they depend on different sets of variables. The distinction should be clear from the arguments appearing in each equation.}
\begin{align}
   & \frac{\partial}{\partial t}\begin{pmatrix}\rho_s\\\rho_o\end{pmatrix}
    +\left(
        \frac{1}{E}
        \nabla_{\rt}\cdot\pt
        +
        \frac{1}{q^+}
        \nabla_{\bt}\cdot\qt
    \right)\begin{pmatrix}\rho_s\\\rho_o\end{pmatrix}
    \\
    &=
    \frac{\hat q}{4C_F}
    \begin{array}{c}
    \begin{pmatrix}
        \begin{array}{c}
            -2C_F\rt^2+\frac{C_F}{2}\nabla_{\pt}^2
        \end{array}&
        \begin{array}{c}  
            2C_F\rt^2+\frac{C_F}{2}\nabla_{\pt}^2
        \end{array}\\ 
        \begin{array}{c} 
            \frac{\rt^2}{N_c}
            +\frac{1}{4N_c}\nabla_{\pt}^2
        \end{array}
        & 
        \begin{array}{c}
            -\frac{\rt^2}{N_c}
            +N_c\nabla_{\qt}^2
            +N_c(z_2-z_1)\nabla_{\pt}\cdot\nabla_{\qt}\\
            +\left[\frac{C_F}{2}2(z_1^2+z_2^2)-\frac{1}{2N_C}\left(2z_1z_2+\frac{1}{2}\right)\right]\nabla_{\pt}^2
        \end{array}
    \end{pmatrix}
    \end{array}
    \begin{pmatrix}\rho_s\\
    \rho_o\end{pmatrix}.\label{eq:rho-z1z2-wigner}
\end{align}

Finally, keeping in mind, for instance, a $q\bar q$ produced in the decay of a colour-singlet electroweak boson, which typically favours symmetric longitudinal momentum sharing, we shall throughout this paper restrict ourselves to the symmetric configuration $z_1=z_2=1/2$. In this case, the evolution equation for the singlet–octet density matrix takes a simpler form within the harmonic approximation:
\begin{align}
    &\frac{\partial }{\partial t}\begin{pmatrix}
        \rho_s\\
        \rho_o
    \end{pmatrix}+\left(\frac{1}{E}\pt\cdot\nabla_{\rt}+\frac{1}{q^+}\qt\cdot\nabla_{\bt}\right)\begin{pmatrix}
        \rho_s\\
        \rho_o
    \end{pmatrix}\nonumber\\
    &=\frac{\qhat}{4C_F}\left\{\begin{pmatrix}
        0 & 0 \\
        0 & N_c
    \end{pmatrix}\nabla^2_{\qt}+\begin{pmatrix}
        -2C_F & 2C_F\\
        \frac{1}{N_c} & -\frac{1}{N_c}
    \end{pmatrix}\rt^2+\begin{pmatrix}
        \frac{C_F}{2} &  \frac{C_F}{2}\\
        \frac{1}{4N_c} &  \frac{C_F}{2}-\frac{1}{2N_c}
    \end{pmatrix}\nabla^2_{\pt}\right\}\begin{pmatrix}
        \rho_s\\
        \rho_o
    \end{pmatrix}.
    \label{eq:Wigner-ho-final}
\end{align}
This assumption is not expected to qualitatively affect the conclusions reached in the present study. It is nevertheless interesting to note that asymmetric splittings with $z_1\neq z_2$ generate an additional operator in momentum space, $\nabla_{\pt}\cdot\nabla_{\qt}$, in the Lindblad equation.

We can now give a physical interpretation of the operators in the r.h.s.~of the quantum master equation, anticipating the results presented in the next sections which will describe more quantitatively their properties.
The first and the third terms are the diffusion term for the momentum imbalance $\qt$ and the relative momentum $\pt$ respectively. The colour structure of the matrix in front of the $\nabla^2_{\qt}$ operator has a particularly simple form. It physically describes ``unresolved" scattering of the dipole without colour transition, such that only the diagonal terms are a priori non-vanishing and proportional to the total colour charge of the state: it is thus either zero in the singlet state or $N_c$ in the octet state.

In the Wigner space $(\rt,\bt,\pt,\qt)$ the action of the $\nabla^2_{\qt}$ and $\nabla^2_{\pt}$ operators is responsible for increasing the width of the momentum distribution. In the conjugate variables $\Bcalt$ and $\Rcalt$, the effect is to induce quantum decoherence of the spatial d.o.f., evolving towards a peaked distribution around the diagonal of the density matrix. Both diffusion and decoherence are expected phenomena from the collisions of the system with the environment. 

The second term, proportional to the $\rt^2$ operator, constitutes the main new feature of our study. It drives colour transitions that become increasingly favoured as the relative
separation between the quark and antiquark grows. The off-diagonal components represent loss terms for the colour-singlet and colour-octet states, as they drive transitions between the two colour configurations, while the diagonal components then ensure conservation of the norm of the density matrix. The possibility of having scatterings that resolve the antenna without inducing a colour transition (diagonal component of the matrix in front of the $\rt^2$ operator) is reminiscent of diffractive processes in the dipole picture of Deep Inelastic Scattering, where the interaction with the target is mediated by the Pomeron, which is a colour-singlet exchange and therefore does not alter the colour representation of the dipole. This term introduces an important difficulty, closely related to the issue of factorization of the Wigner function discussed in the next section: as can be seen from Eq.\,\eqref{eq:Wigner-ho-final}, it is straightforward to integrate out the center-of-mass degrees of freedom and derive an evolution equation solely for the relative degrees of freedom. The converse, however, is not possible: integrating over the relative degrees of freedom generates higher moments in the $\rt$ distribution, thereby preventing a complete factorization between the center-of-mass and relative degrees of freedom of the $q\bar q$ dipole.

Finally, it is important to keep in mind that the relative-momentum diffusion operator, proportional to $\nabla^2_{\pt}$, also has a non-diagonal colour structure and therefore induces colour transitions. These transitions become more favourable as $\Rcalt$, the Fourier-conjugate variable to $\pt$ that characterises the "quantumness" of the $q\bar q$ dipole size, increases. We shall see in the next subsection that this colour-transition mechanism is subleading in the regime $p_\perp \gg Q_s$, which is the regime of primary interest here, compared with the colour transitions induced by the $\rt^2$ operator.

It would be highly desirable to obtain an analogue of Eq.\,\eqref{eq:Wigner-ho-final} for other dipole configurations, such as $qg$ and $gg$ (see also~\cite{Zakharov:2018hfz}). In these cases, the colour space is enlarged, with three possible representations: $ 3, \bar{6}, 15$ for the $qg$ dipole and $1, 8, 27$ for the $gg$ dipole. One may anticipate that the $\nabla^2_{\qt}$ operator remains diagonal in this basis, with matrix elements $M_{ij} = C_i \delta_{ij}$ --- where $C_i$ is the Casimir of the representation $i$ ---, although the structure of the remaining matrices is less clear and could provide valuable insight into their physical interpretation.

From a phenomenological perspective, such generalizations could have interesting implications. For instance, a two-prong gluon jet arising from $g \to gg$ may undergo a colour transition into the 27-dimensional representation~\cite{Cougoulic:2025wpl}, leading to significantly enhanced transverse momentum broadening, given that $C_{27} = 8$, compared to $C_8 = 3$ for a single gluon jet or for a $g \to q\bar{q}$ splitting. The reverse transition to a singlet state is expected to be suppressed by powers of $N_c$, and the properties of intermediate transitions, such as $27 \to 8$, also warrant further investigation. Similarly, the $q \to qg$ channel provides another interesting case, as the corresponding Casimir $C_{15} = 16/3$ is much larger than $C_F$.

\subsection{Dimensionless equation in the boosted and back-to-back limit}
\label{sub:scale-analysis}

The quantum master equations derived in the previous subsection, even in their simplest form given by Eq.\,\eqref{eq:Wigner-ho-final} after applying the harmonic approximation, take the form of a multi-species Fokker-Planck equation, with the added complexity that transitions between internal (colour) states are allowed with a rate depending on the relative distance between the two particles. As such, it is notoriously difficult to solve analytically, and to our knowledge, there is no general method to find the solutions of such equations.

However, the situation improves if one imposes additional simplifications motivated by the kinematic regime discussed in Section~\ref{sec:kinematics} where, as we shall see, the term associated with colour transitions dominates over the $\pt$-diffusion term in Eq.\,\eqref{eq:Wigner-ho-final}. Indeed, as discussed in the beginning of this paper, we are primarily interested in the regime $E\gg p_\perp \gg q_\perp \sim Q_s$. We wish to find an approximation for the Wigner function in this kinematic domain. The first inequality implies that $\theta_{q\bar q}\ll 1$, but note that $\theta_{q\bar q}$ can still be comparable to $\theta_c= 2/\sqrt{\qhat L^3}$ since typically, $\theta_c\ll 1$ for realistic values of $\qhat$ and $L$. From the semi-hard scale $Q_s=\sqrt{\qhat L}$ and $E$, one can also define the angular scale $\theta_s$ defined as
\begin{align}
    \theta_s\equiv\frac{\sqrt{\qhat L}}{E}\,.
\end{align}
The angular scales $\theta_s$ and $\theta_c$ will play crucial roles in the following analysis.

In order to find the relevant approximation to the quantum master equation Eq.\,\eqref{eq:Wigner-ho-final} in this regime, it is useful to express it in terms of dimensionless variables. We begin by rescaling time $t$ using the reduced time $\bar t = t/L$. Since the classical opening angle of the dipole, $\theta_{q\bar q}\sim |\pt|/E$, can be comparable to $\theta_c$, it is natural to make the $\pt$ scale dimensionless by introducing $\ptb = \sqrt{\qhat L^3} \pt/E$, with then $\ptb\sim \theta_{q\bar q}/\theta_c$. Similarly, we rescale $\qt$ by defining $\qtb = \qt/\sqrt{\qhat L}$. These two choices determine how $\rt$ and $\bt$ must be rescaled so that the kinetic terms are of order 1 and take the simple forms $\ptb\cdot\nabla_{\rtb}$ and $\qtb\cdot\nabla_{\btb}$. In the end, we perform in Eq.\,\eqref{eq:Wigner-ho-final} the following change of variables
\begin{align}
    \bar t=t/L\,,\quad \rtb = \sqrt{\qhat L}\rt\,,\quad \ptb=\frac{\sqrt{\qhat L^3}}{E}\pt\,,\quad \qtb=\frac{\qt}{\sqrt{\qhat L}}\,,\quad \btb=\frac{q^+}{\sqrt{\qhat L^3}}\bt\,,\label{eq:dimensionless-variables}
\end{align}
such that Eq.\,\eqref{eq:Wigner-ho-final} becomes
\begin{align}
        &\frac{\partial }{\partial \bar t}\begin{pmatrix}
        \rho_s\\
        \rho_o
    \end{pmatrix}+\left(\ptb\cdot\nabla_{\rtb}+\qtb\cdot\nabla_{\btb}\right)\begin{pmatrix}
        \rho_s\\
       \rho_o
    \end{pmatrix}\nonumber\\
    &=\frac{1}{4C_F}\left\{\begin{pmatrix}
        0 & 0 \\
        0 & N_c
    \end{pmatrix}\nabla^2_{\qtb}+\begin{pmatrix}
        -2C_F & 2C_F\\
        \frac{1}{N_c} & -\frac{1}{N_c}
    \end{pmatrix}\rtb^2+\kappa^2\begin{pmatrix}
        \frac{C_F}{2} &  \frac{C_F}{2}\\
        \frac{1}{4N_c} &  \frac{C_F}{2}-\frac{1}{2N_c}
    \end{pmatrix}\nabla^2_{\ptb}\right\}\begin{pmatrix}
        \rho_s\\
        \rho_o
    \end{pmatrix}\,,\label{eq:Wigner-ho-fullNc-final-dimensionless}
\end{align}
with the dimensionless ratio $\kappa$ defined as
\begin{align}
    \kappa&\equiv\frac{2\theta_s}{\theta_c}=\frac{Q_s}{p_\perp}\times \frac{2\theta_{q\bar q}}{\theta_c}\,.\label{eq:kappa-def}
\end{align}
The second equality in Eq.\,\eqref{eq:kappa-def} illustrates that the coefficient $\kappa$ is formally a power correction in $Q_s/p_\perp$ when $\theta_{q\bar q}\sim \theta_c$.
Typically, one has $\theta_s\ll \theta_c$ in the high energy limit such that $\kappa\ll 1$. For instance, for realistic values of the medium parameters such as $\qhat =1.5$ GeV$^2$/fm and $L=4$ fm, we have $\theta_s\sim 0.02\div 0.002$ for $E$ in the range $100\div 1000$ GeV, while $\theta_c\sim 0.04$ independently of the value of $E$. 

For such small values of $\kappa$, the last term in Eq.\,\eqref{eq:Wigner-ho-fullNc-final-dimensionless} can be neglected. In this case, the only operator capable of driving colour transitions is the $\rt^2$ operator, while the relative-momentum diffusion operator induces octet-to-singlet transitions that are power-suppressed by $\kappa^2$. We will focus on solutions to this simplified equation from now on, and examine the corrections introduced by the $O(\kappa^2)$ term in the final section of this paper. The asymmetry in the way $\qt$ and $\pt$ are rendered dimensionless in Eq.\,\eqref{eq:dimensionless-variables} --- reflecting the strong hierarchy of scales $E \gg p_\perp \gg q_\perp \sim Q_s$ --- is a key feature of our analysis that allows us to neglect the $\ptb$ diffusion term $\propto \nabla^2_{\ptb}$ as a first approximation. 

\subsection{Large $N_c$ limit}

Another convenient approximation to the master equation, albeit not crucial to obtain analytic solutions, is the large $N_c$ limit. In the context of quarkonia dynamics in the QGP, the large $N_c$ limit has been considered in \cite{Escobedo:2020tuc} where it is specified that octet to singlet transitions are suppressed in this limit.
Following up on this work, in order to implement the large $N_c$ limit, one should work in a basis where the singlet and octet component of the density matrix have a similar scaling with $N_c$.  One can indeed observe in the current scaling of Eq.\,\eqref{eq:Wigner-ho-final} that in the large $N_c$ limit the singlet to octet transitions would be suppressed, meaning that if the initial state has only the singlet component, the octet component would never be populated.

To understand how to rescale the density matrix, one can look at Eq.\,\eqref{eq:dm:qqbar:singlet-octet}, where $\rho_o$ is the probability associated to each of the $N_c^2-1$ projectors $\ket{o^c}\bra{o^c}$ on orthonormal spaces. As stated in \cite{Escobedo:2020tuc}, we want the population of the singlet and the octet component to be comparable, therefore to accomplish this we have to rescale
\begin{equation}
            \rho
            =
            \rho_s\ket{s}\bra{s} + (N_c^2-1)\rho_o\sum_c\frac{\ket{o^c}\bra{o^c}}{N_c^2-1}
            \equiv
            \rho_s\ket{s}\bra{s} + \tilde{\rho}_o\sum_c\frac{\ket{o^c}\bra{o^c}}{N_c^2-1},
        \end{equation}
        where the rescaled octet component is defined as
    \begin{equation}
            \tilde{\rho}_o=(N_c^2-1)\rho_o\to\rho_o=\frac{\tilde{\rho}_o}{N_c^2-1}.
        \end{equation}
This rescaling can simply be interpreted as an average over all $N_c^2-1$ colour states gathered in the octet component of the density matrix.
After performing this change of basis in Eq.\,\eqref{eq:Wigner-ho-final} and neglecting the contribution suppressed by $1/N_c^2$, we obtain the following quantum master equation, 
\begin{align}
        \frac{\partial }{\partial \bar t}\begin{pmatrix}
        \rho_s\\
        \tilde \rho_o
    \end{pmatrix}&+\left(\ptb\cdot\nabla_{\rtb}+\qtb\cdot\nabla_{\btb}\right)\begin{pmatrix}
        \rho_s\\
       \tilde \rho_o
    \end{pmatrix}&\nonumber\\
    &=\frac{1}{2}\left\{\begin{pmatrix}
        0 & 0 \\
        0 & 1
    \end{pmatrix}\nabla^2_{\qtb}+\begin{pmatrix}
        -1 & 0\\
        1 & 0
    \end{pmatrix}\rtb^2+\frac{\kappa^2}{4}\begin{pmatrix}
        1 & 0 \\
        1 & 1
    \end{pmatrix}\nabla^2_{\ptb}\right\}\begin{pmatrix}
        \rho_s\\
        \tilde \rho_o
    \end{pmatrix}\,.\label{eq:Wigner-ho-lNc-final}
\end{align}
At this stage, a clarification is in order regarding the precise meaning of the large $N_c$ limit. Recall that the jet quenching parameter $\qhat$ has been defined in the fundamental representation, such that $\qhat \sim \alpha_s C_F m_D^2T\sim \alpha_s^2N_cn$ where $n\sim T^3$ is the colour charge density. For Eq.\,\eqref{eq:Wigner-ho-lNc-final} to remain physically meaningful, we therefore implicitly keep $\qhat$ fixed when taking the large $N_c$ limit. This can be achieved, for instance, by scaling the coupling to the medium through the $\alpha_s^2$ factor or by reducing the colour charge density $n$, so as to compensate for the increase of $N_c$ as $N_c\to\infty$.

A nice feature of this equation is that it is triangular in the single-octet basis:  the singlet component of the density matrix does not receive contributions from the octet state, as for a static quark antiquark pair~\cite{Escobedo:2020tuc}. In the following sections, we shall essentially solve and discuss Eq.\,\eqref{eq:Wigner-ho-lNc-final}  which captures the main quantum aspects of the dynamics of the $q\bar q$ dipole subsystem when propagating inside the environment of a QGP.

\section{Initial condition for the Wigner density}
\label{sec:initial}

Before solving the simplified Lindblad equations, Eq.\,\eqref{eq:Wigner-ho-fullNc-final-dimensionless} or Eq.\,\eqref{eq:Wigner-ho-lNc-final}, we must specify the initial condition for the density matrix, or equivalently for the Wigner distribution. We emphasize that our Lindblad equation does not account for the production mechanism of the $q\bar q$ pair, which is assumed to be there at $t=0$ with a given wave-function on the relative and total transverse momentum space.
To motivate a physically reasonable choice for the initial state, we draw guidance from the cross section for the process $\gamma\to q\bar q$ in the kinematic regime of interest, discussed in section~\ref{sec:kinematics}. As shown in appendix~\ref{app:feynman-TMB}, in the regime $E\gg p_\perp\gg q_\perp$, the probability of producing a $q\bar q$ pair in vacuum with relative transverse momentum $\pt$ and total transverse momentum $\qt$ from the virtual photon is
\begin{align}
\frac{\der P}{\der^2\pt\der^2\qt}&=\frac{\alpha_{\rm em}N_c}{4\pi}\frac{(z_1^2+z_2^2)}{\pt^2}\times\delta(\qt)
\label{eq:proba-vacuum}
\end{align}
to lowest order in the fine structure constant $\alpha_{\rm em}$ and strong coupling $\alpha_s$. Here $z_2=1-z_1$. Strictly speaking, the probability density also depends on $z_1$, but we do not display this dependence explicitly on the left-hand side, as we will subsequently consider the distribution at fixed $z_1=1/2$ for the sake of simplicity. The first factor, which depends on the hard scale $|\pt|$, corresponds to the DGLAP splitting kernel for the process $\gamma \to q\bar q$, while the $1/\pt^2$ behaviour reflects the usual collinear enhancement of the splitting. The second factor encodes the dependence on $\qt$. As discussed in section~\ref{sec:kinematics}, at lowest order and in vacuum, the delta function enforces exact transverse-momentum conservation, implying $\qt=0$. Therefore, Eq.\,\eqref{eq:proba-vacuum} exhibits a simple \textit{factorized} structure, schematically $\der P \sim H(\pt)\times B(\qt)$, where the hard factor $H$ captures the dependence on the scale $p_\perp$, while $B$ encodes the softer scale $q_\perp$. Eq.\,\eqref{eq:proba-vacuum} can be generalised to the decay of a massive vector boson, in which case the $\pt$ distribution is now peaked around the scale $\pt^2=z_1z_2M^2$ where $M$ is the mass of the vector boson~\cite{Motyka:2008ac}.

Motivated by this factorization of the cross section, we choose an initial Wigner distribution that factorizes into a contribution $\rho_R(\pt,\rt;0)$ in the relative phase space and a contribution $\rho_T(\qt,\bt;0)$ in the center-of-mass phase space:
\begin{align}
\rho(\pt,\rt;\qt,\bt;0)&=\rho_R(\pt,\rt;0)\rho_T(\qt,\bt;0)\,.\label{eq:init-factorisable}
\end{align}
In the absence of medium effects, the time evolution of the Wigner distribution resulting from Eq.\,\eqref{eq:Wigner-ho-final} is simply
\begin{align}
\rho(\pt,\rt;\qt,\bt;t)
&=\rho_R\left(\pt,\rt-\frac{\pt t}{E};0\right)
\rho_T\left(\qt,\bt-\frac{\qt t}{q^+};0\right)\,,
\end{align}
so that the probability of finding the pair with relative transverse momentum $\pt$ and imbalance $\qt$, obtained by integrating over $\rt$ and $\bt$, is time independent and fully determined by the initial condition:
\begin{align}
\frac{\der P}{\der^2\pt\der^2\qt}
&=\left(\int\der^2\rt \,\rho_R(\pt,\rt;0)\right)
\left(\int\der^2\bt \,\rho_T(\qt,\bt;0)\right)\,.
\end{align}
In other words, the time evolution of the Wigner distribution preserves the factorized structure of the initial condition.

This simple result provides some guidance in building the $\pt$ and $\qt$ dependence of the initial Wigner density. Regarding the $\qt$ dependence, transverse momentum conservation would impose $\int\der^2\bt \rho_T(\qt,\bt;0)=\delta(\qt)$ in the absence of QCD radiative corrections or non-perturbative effects, as in Eq.\,\eqref{eq:proba-vacuum}. The implementation of radiative corrections in our framework would require adding at least one gluon into the problem ; this is left for future studies. For what concerns non-perturbative effects, we introduce a non-perturbative parameter $\Lambda$ which will kill the density matrix in position space $\rho_T(\Bcalt,\bt,t=0)$ for distances larger than $\sim 1/\Lambda$. A simple ansatz is to consider a Gaussian with width $\Lambda$ of the order of the QCD confinement scale $\Lambda_{\rm QCD}$ such that
\begin{align}
    \rho_T(\qt,\bt;0)&=\frac{1}{\pi^2}e^{-\qt^2/\Lambda^2-\Lambda^2\bt^2}\,.
\end{align}    
This is reminiscent of the modelling of the non-perturbative part of the TMD parton distribution to describe the initial state of hadronic collisions or TMD parton fragmentation function in final state hadro-production~\cite{Boussarie:2023izj}. In the limit $\Lambda\to 0$, one gets $\int\der^2\bt \rho_T(\qt,\bt;0)\to\delta(\qt)$ as expected.

We now turn to the Wigner distribution $\rho_R(\pt,\rt;0)$ on the relative coordinate phase space. 
Based on the result~Eq.\,\eqref{eq:proba-vacuum}, we expect the $\pt$ dependence $\rho_R(\pt,\rt;0)$ to scale like $1/\pt^2$ after marginalising over $\rt$:
\begin{align}
    \int\der^2\rt \rho_R(\pt,\rt;0)\sim \frac{1}{\pt^2}\,,
\end{align}
while the $\rt$ dependence should be given by a function peaked around $\rt=\boldsymbol{0}_\perp$.
However, for such a  dependence, it is too complicated to obtain analytic solution for the Wigner density $\rho_R$ and its time evolution when medium effects are taken into account. Instead, we shall also use a Gaussian model which captures the main features of this dependence: the peak of the $\pt$ distribution is centred around $\Ptz^2\equiv z_1z_2M^2$, 
the peak of the $\rt$ distribution is centred around $\rt=\boldsymbol{0}_\perp$ and the respective widths $\mu_1$, $\mu_2$ are taken as free parameters. We thus use
\begin{align}
    \rho_R(\pt,\rt;0)=\frac{\mu_2^2}{\pi^2\mu_1^2}\exp\left(-\frac{(\pt-\Ptz)^2}{\mu_1^2}\right)\exp(-\mu_2^2\rt^2)\,.\label{eq:Gaussian-wigner}
\end{align}
Last but not least, note that the Heisenberg uncertainty relation principle demands $\mu_1/\mu_2\ge 1/2$ since the distribution cannot be simultaneously localised both in $\pt$ and in $\rt$. We will return to the implications of this constraint in Section~\ref{sec:pt-diffusion}. At this point, Eq.\,\eqref{eq:Gaussian-wigner} should be regarded as a toy model for the initial condition in $\pt$ and $\rt$: although not realistic from a QCD perspective, it is convenient for practical purposes. Most of the conclusions drawn in Sections~\ref{sec:factorisation-violation}-\ref{sec:decoherence} will in fact be independent of this specific choice, as they primarily rely on the analysis of $\rho_T(\qt,\bt;t)$.

\section{Quasi-factorised Wigner density}
\label{sec:factorisation-violation}

We are now in a position to solve the Lindblad equation, Eq.\,\eqref{eq:Wigner-ho-lNc-final}, in the large $N_c$ limit, within the harmonic approximation, and neglecting the $\pt$-diffusion term which is suppressed by $\kappa^2$. We first derive the solution using the standard method of characteristics for partial differential equations, and then use it to obtain and analyze the transverse-momentum broadening distribution in $\qt$ of the dipole. In section \ref{sec:pt-diffusion}, we will also present selected results beyond the large $N_c$ limit, using Eq.\,\eqref{eq:Wigner-ho-fullNc-final-dimensionless}.

\subsection{Solution from the method of characteristics} Equation~\eqref{eq:Wigner-ho-lNc-final} without the $\ptb$ diffusion term $\propto \kappa^2$ can be solved using the method of characteristics. To do so, we go to Fourier conjugate space for the total degrees of freedom, $\qtb\to \Bcaltb, \btb\to \Qcaltb$ such that Eq.\,\eqref{eq:Wigner-ho-lNc-final} becomes
\begin{align}
      \left[\frac{\partial }{\partial \bar t}+\ptb\cdot\nabla_{\rtb}+\Qcaltb\cdot\nabla_{\Bcaltb}\right]\begin{pmatrix}
        \rho_s\\
       \tilde \rho_o
    \end{pmatrix}=\frac{1}{2}&\left\{\begin{pmatrix}
        0 & 0 \\
        0 & 1
    \end{pmatrix}\Bcaltb^2+\begin{pmatrix}
        -1 & 0\\
        1 & 0
    \end{pmatrix}\rtb^2\right\}\begin{pmatrix}
        \rho_s\\
        \tilde \rho_o
    \end{pmatrix}\,.\label{eq:Wigner-ho-lNc-final-approx}
\end{align}
Introducing the function $\varrho$ on the characteristics curves $\Bcaltb(\bar t),\rtb(\bar t)$ of the partial differential equation (to be determined),
\begin{align}
   \varrho(\ptb;\Qcaltb; \bar t)=\rho(\ptb,\rtb(t);\Bcaltb(\bar t),\Qcaltb;\bar t)\,,
\end{align}
the function $\varrho$ satisfies
\begin{align}
    \frac{\der}{\der \bar t}\begin{pmatrix}
        \varrho_s\\
        \tilde \varrho_o
    \end{pmatrix}&=\frac{\qhat}{2}\left\{\begin{pmatrix}
        0 & 0 \\
        0 & -1
    \end{pmatrix}\Bcaltb^2(\bar t)+\begin{pmatrix}
        -1 & 0\\
        1 & 0
    \end{pmatrix}\rtb^2(\bar t)\right\}\begin{pmatrix}
        \varrho_s\\
        \tilde\varrho_o
    \end{pmatrix}\,,\label{eq:varrho-eq}
\end{align}
 provided the characteristic curves coincide with classical equation of motions for the center-of-mass and relative distance of the $q\bar q$ pair; namely the relative transverse coordinates of the $q\bar q$ pair increases with effective velocity $\ptb$ while $\Bcaltb$ increases with effective velocity $\Qcaltb$:
\begin{align}
    \frac{\der \Bcaltb(\bar t)}{\der \bar t}&=\Qcaltb\,,\quad \frac{\der \rtb(\bar t)}{\der \bar t}=\ptb\,.
\end{align}
 Eq.\,\eqref{eq:varrho-eq} for $\varrho$ is now a simple first order linear differential equation.
The solution to this equation, once expressed in terms of the original physical variables $\pt,\rt,\Bcalt,\Qcalt$ finally reads
\begin{align}
   \rho_s(\pt,\rt;\Bcalt,\Qcalt;t)&=\exp\left(-\frac{1}{2}\int_0^t\der s \ \qhat\left[\rt-\frac{\pt}{E}s\right]^2\right)\nonumber\\
   &\times\rho_s\left(\pt,\rt-\frac{\pt}{E}t;\Bcalt-\frac{\Qcalt}{q^+}t,\Qcalt;0\right)\,,\label{eq:rhos-oqs-largeNc}\\
  \tilde\rho_o(\pt,\rt;\Bcalt,\Qcalt;t)&=\Bigg\{\tilde\rho_o\left(\pt,\rt-\frac{\pt}{E}t;\Bcalt-\frac{\Qcalt}{q^+}t,\Qt;0\right)\nonumber\\
    &+\frac{1}{2}\rho_s\left(\pt,\rt-\frac{\pt}{E}t;\Bcalt-\frac{\Qcalt}{q^+}t,\Qcalt;0\right)\times\int_0^t\der s \ \qhat \left[\rt-\frac{\pt}{E}s\right]^2\nonumber\\
    &\times \exp\left[-\int_s^t\der s' \ \frac{\qhat}{2}\left(\left[\rt-\frac{\pt}{E}s'\right]^2-\left[\Bcalt-\frac{\Qcalt}{q^+}s'\right]^2\right)\right]\Bigg\}\nonumber\\
    &\times \exp\left(-\frac{1}{2}\int_0^t\der s \ \qhat\left[\Bcalt-\frac{\Qcalt}{q^+}s\right]^2\right)\,.\label{eq:rhoo-oqs-largeNc}
\end{align}
One should note that this solution remains positive definite with time and thus admits a probabilistic interpretation. In addition, these equations remain valid if $\qhat$ effectively depends on time as in the case of a Bjorken expanding medium: one simply put $\qhat(t)$ inside the time integrals. 

For a colour singlet initial state the total density operator, which is the sum of the colour singlet and octet components, simplify. Indeed, integrating by part in the time integral appearing in the colour octet contribution, we find 
\begin{align}
        \rho(\pt,\rt;\Bcalt,\Qcalt;t)&=\rho_s\left(\pt,\rt-\frac{\pt}{E}t;\Bcalt-\frac{\Qcalt}{q^+}t,\Qcalt;0\right)\left\{e^{-\frac{1}{2}\int_0^t\der s \ \qhat\left[\Bcalt-\frac{\Qcalt}{q^+}s\right]^2}\right.\nonumber\\
    &\hspace{-1.5cm}\left.+\int_0^t\der s \ \frac{\qhat }{2}\left[\Bcalt-\frac{\Qcalt}{q^+}s\right]^2e^{-\int_s^t\der s' \ \frac{\qhat}{2}\left[\rt-\frac{\pt}{E}s'\right]^2}e^{-\frac{1}{2}\int_0^s\der s' \qhat\left[\Bcalt-\frac{\Qcalt}{q^+}s'\right]^2}\right\}\,,
    \label{eq:rhotot-oqs-largeNc-ipp}
\end{align}
where $\rho=\rho_s+\tilde \rho_o$. This expression already displays an interesting feature: even if the initial Wigner density is factorisable, as in Eq.\,\eqref{eq:init-factorisable}, the Lindblad evolution of the density matrix will break this factorisation property because of the dependence on $\rt$ and $\pt$ in the exponential factor of the second line of Eq.\,\eqref{eq:rhotot-oqs-largeNc-ipp}. This dependence comes from the transition of the singlet state to colour octet state as the relative distance between the quark and the antiquark increases.

For a Gaussian initial condition $\rho_{T,s}$ (and for any choice of initial condition for $\rho_{R,s}$), the Wigner density corresponding to Eq.\,\eqref{eq:rhotot-oqs-largeNc-ipp} obtained by Fourier transform over $\Bcalt$ and $\Qcalt$ reads
\begin{align}
    &\rho(\pt,\rt;\qt,\bt;t)=\frac{1}{\pi^2}\rho_{R,s}\left(\pt,\rt-\frac{\pt t}{E};0\right)\left\{e^{-\frac{\qhat }{2}\int_0^t\der s' \left(\rt-\frac{\pt s'}{E}\right)^2}e^{-\frac{\qt^2}{\Lambda^2}-\Lambda^2\left(\bt-\frac{\qt t}{q^+}\right)^2} \right.\nonumber\\
    &\hspace{0.5cm}\left.+\int_0^t\der s \ \frac{3(q^+)^2\qhat L^2}{2d(s)} \left(\rt-\frac{\pt s}{E}\right)^2e^{-\frac{\qhat }{2}\int_s^t\der s' \left(\rt-\frac{\pt s'}{E}\right)^2}e^{-\frac{a(s)\qt^2+b(s)\bt^2+c(s)\qt\cdot\bt}{d(s)}}\right\}\,,\label{eq:rho-tot-gauss-qt}
\end{align}
where the functions $a(s),b(s),c(s)$ and $d(s)$ are defined as 
\begin{align}
    a(s)&=3(q^+)^2+2\Lambda^2\qhat s^3+3\Lambda^4t^2\,,\\
    b(s)&=3\Lambda^2(q^+)^2(\Lambda^2+2\qhat s)\,,\\
    c(s)&=6\Lambda^2 q^+(\qhat s^2+\Lambda^2 t)\,,\\
    d(s)&=6(q^+)^2\qhat s+\Lambda^2(3(q^+)^2+\qhat^2s^4)+2\Lambda^4\qhat s(s^2-3st+3t^2)\,.
\end{align}
In Eq.\,\eqref{eq:rho-tot-gauss-qt}, the first term in the first line comes from the singlet component of the density matrix. 

Note that for the singlet component only, it is straightforward to obtain from Eq.\,\eqref{eq:rhos-oqs-largeNc} the Wigner function for any initial condition (in the large $N_c$ limit):
\begin{align}
 \rho_s(\pt,\rt;\qt,\bt;t)=e^{-\frac{1}{2}\int_0^t\der s \ \qhat\left(\rt-\frac{\pt}{E}s\right)^2}\rho_s\left(\pt,\rt-\frac{\pt}{E}t;\qt,\bt-\frac{\qt}{q^+}t;0\right)\,.
\end{align}
This equation shows that the impact parameter $\bt$ undergoes a classical motion with transverse velocity $\qt/q^+$, in agreement with our discussion in subsection~\ref{sub:scale-analysis}.

\subsection{$\theta_{q\bar q}$ dependence of the transverse momentum broadening $q_\perp$-distribution}
\label{subsect5:2}

It is evident from Eqs.\,\eqref{eq:rhos-oqs-largeNc}-\eqref{eq:rhoo-oqs-largeNc} that, due to the feeding from the singlet component, the Wigner function at time $t>0$ can no longer be expressed as a product of a Wigner function in the relative phase space and a Wigner function in the center-of-mass phase space, even if the initial condition is factorized.
Nevertheless, if the initial Wigner distribution in the relative phase space is sharply peaked around $\rt=0$, one can derive a simpler analytic expression for the probability distribution of observing the dipole with transverse momentum $\qt$. This expression incorporates the effects of factorization breaking, namely its residual dependence on $\pt$, that we shall subsequently interpret.

If one lets $\mu_2\to \infty$ in the initial condition, then the initial Wigner density takes the form
\begin{align}
    \rho_T(\pt,\rt;0)\to H(\pt)\delta(\rt)\,,
\end{align}
with $H(\pt)=\frac{1}{\pi\mu_1^2}e^{-(\pt-\Ptz)^2/\mu_1^2}$ the effective ``hard factor'' associated with our choice of initial condition discussed in section~\ref{sec:initial}. One should note that, in order to satisfy the Heisenberg uncertainty principle, one must have $\mu_1 \ge \mu_2/2$ (as discussed in Section~\ref{sec:initial}), implying that the width of the $\pt$ distribution also diverges in this regime. Nevertheless, our primary goal in this section is to analyse the Wigner density in the $(\bt,\qt)$ phase space. The results we obtain are largely insensitive to the precise form of the initial Wigner density in the $(\rt,\pt)$ phase space, provided that it is initially peaked around $\rt=\boldsymbol{0}_\perp$. Finite $\mu_2$ corrections will also be discussed towards the end of this subsection. 

The delta function, once inserted in Eq.\,\eqref{eq:rho-tot-gauss-qt}, enforces the classical motion of the relative degrees of the freedom, namely $\rt=\pt t/E$. It is then straightforward to integrate Eq.\,\eqref{eq:rho-tot-gauss-qt} over $\rt$ and $\bt$ to obtain the probability density of finding the $q\bar q$ pair with relative transverse momentum $\pt$ and imbalance $\qt$:
\begin{align}
    \frac{\der P}{\der^2\pt\der^2\qt}&=H(\pt)\times\frac{1}{\pi}\left\{e^{-\frac{\qhat \pt^2t^3}{6E^2}}\frac{e^{-\frac{\qt^2}{\Lambda^2}}}{\Lambda^2}+\int_0^t\der s \ \frac{\qhat \pt^2(t-s)^2}{2E^2(\Lambda^2+2\qhat s)}e^{-\frac{\qhat \pt^2(t-s)^3}{6E^2}}e^{-\frac{\qt^2}{\Lambda^2+2\qhat s}}\right\}\,.\label{eq:broadening-qt-init-lambda}
\end{align}
This expression simplifies even further if one takes  the limit $\Lambda\to 0$ and write $\pt^2=\theta_{q\bar q}E^2/2$ such that
\begin{align}
    \frac{\der P}{\der^2\pt\der^2\qt}&=H(\pt)\times\left\{e^{-\frac{\qhat \theta_{q\bar q}^2t^3}{12}}\delta(\qt)+\frac{1}{\pi}\int_0^t\der s \ \frac{\theta_{q\bar q}^2(t-s)^2}{8s}e^{-\frac{\qhat \theta_{q\bar q}^2(t-s)^3}{12}}e^{-\frac{\qt^2}{2\qhat s}}\right\}\,.\label{eq:OQS-factor}
\end{align}
Eqs.\,\eqref{eq:broadening-qt-init-lambda} and \eqref{eq:OQS-factor} admit a simple physical interpretation. The colour-singlet state of the $q\bar q$ pair, which gives rise to the first term in these two equations, does not experience any  transverse momentum broadening (although it is still affected by the medium through the overall exponential suppression; see the discussion just below). Accordingly, its $\qt$ dependence is entirely determined by the initial condition.

However, due to the presence of the $\rt^2$ operator in the Lindblad equation, colour transitions between the singlet and octet states of the $q\bar q$ pair are allowed. After a characteristic time $t_c$, defined as
\begin{align}
t_c\equiv \left(\frac{4}{\qhat \theta_{q\bar q}^2}\right)^{1/3}\,,\label{eq:tc-def-lNc}
\end{align}
the singlet state is no longer significantly populated --- the population being suppressed by the overall factor $e^{-t^3/(3t_c^3)}$ multiplying the $\delta(\qt)$ distribution --- and only the octet state remains. The latter undergoes diffusion in $\qt$, resulting in the broadening of the transverse-momentum imbalance distribution.
In this framework, the time scale $t_c$ can be interpreted as the typical inverse transition rate between singlet and octet states. The critical angle $\theta_c$ is then defined as the value of $\theta_{q\bar q}$ such that $t_c=L$: if $\theta_{q\bar q}\le \theta_c$, the medium is too short for the singlet state to decay, and the dipole does not undergo transverse-momentum broadening. Conversely, if $\theta_{q\bar q}\ge \theta_c$, the octet state is present before the end of the evolution and diffuses in $\qt$ space.

Since the exponential suppression of the $\delta(\qt)$ term in Eq.\,\eqref{eq:OQS-factor} is sufficient to define the time scale $t_c$, one can easily obtain it for a time dependent $\qhat$ coming from the QGP Bjorken-like expansion with $\qhat(t)=\qhat(t_0)(t_0/t)$. In that case, the suppression factor reads
\begin{align}
    \exp\left(-\frac{\theta_{q\bar q}^2}{4}\int_0^t\der s\ s^2\qhat(s)\right)=\exp(-t^2/(2t_c^2))\,,
\end{align}
with $t_c=2/(\qhat(t_0)t_0\theta_{q\bar q}^2)^{1/2}$. The corresponding critical angle, defined by $t_c= L$ is thus $\theta_c=2/\sqrt{\qhat(t_0)t_0L^2}=2/\sqrt{\qhat(L)L^3}$. This expression is essentially identical to that obtained in the static medium case, provided that $\hat{q}$ is evaluated at the final time $t$ of the evolution~\cite{Caucal:2020uic}.

\begin{figure}[t]
    \centering
    \includegraphics[width=0.48\linewidth,page=1]{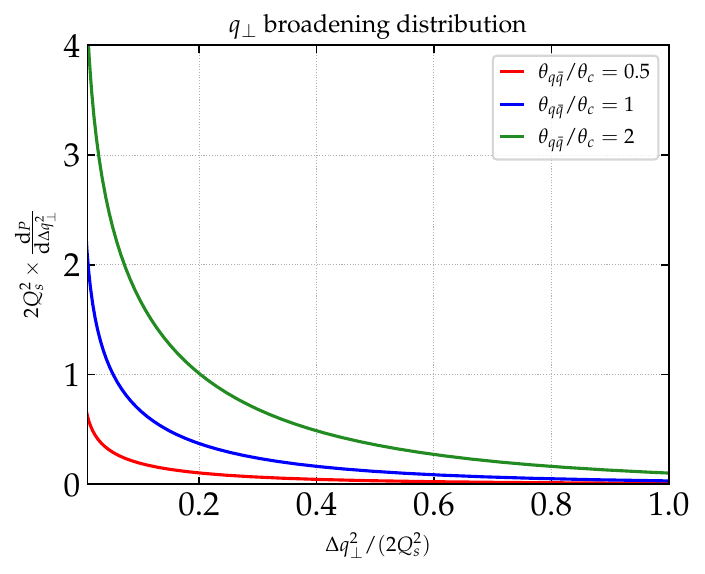}
    \hfill
    \includegraphics[width=0.5\linewidth,page=2]{plot-broad-scaled.pdf}
    \caption{(Left) Probability density for the $q\bar q$ dipole to acquire a transverse momentum imbalance $\Delta q_\perp^2$ after propagating through the medium in the large $N_c$ limit for several values of $\theta_{q\bar q}/\theta_c$ as a function of $\Delta q_\perp^2/(2Q_s^2)$. (Right) Average $\Delta q_\perp^2$ as a function of $\theta_{q\bar q}$ with exact $N_c$ dependence.}
    \label{fig:qt-distrib}
\end{figure}

The behaviour of the $\qt$ probability distribution --- namely Eq.\,\eqref{eq:OQS-factor} without the hard factor $H(\pt)$ --- is shown in Fig.~\ref{fig:qt-distrib} (left) for three values of the ratio $\theta_{q\bar q}/\theta_c$. As this ratio increases, the width of the distribution increases accordingly, reflecting the enhanced broadening of the octet component. Note that in this plot we set $\Lambda=0$. As a consequence, there is a contribution in the $q_\perp=0$ bin that is not visible in the figure but ensures that all three distributions remain properly normalized to unity. One should also note that the shape of the $\qt$-distribution 
differs significantly from a simple Gaussian, due to the 
intricate time convolution appearing in Eq.\,\eqref{eq:OQS-factor}.

In this regard, an interesting quantity to consider is the difference $\langle\Delta \qt^2\rangle$ between the width $\langle \qt^2\rangle $ of the distribution at time $t$ and the width of the distribution at the initial time. Using Eqs.\,\eqref{eq:broadening-qt-init-lambda} or \eqref{eq:OQS-factor}, one finds
\begin{align}
    \langle \Delta \qt^2\rangle_s&=2Q_s^2\left(\frac{\theta_{q\bar q}^2}{\theta_c^2}\right)\int_0^1\der s \  s(1-s)^2\exp\left(-\frac{\theta_{q\bar q}^2}{3\theta_c^2}(1-s)^3\right)\,,\label{eq:meanqt2-oqs}
\end{align}
if the density matrix only contains a singlet component at $t=0$. This result can be generalised to include the exact $N_c$ dependence (using Eq.\,\eqref{eq:Wigner-ho-fullNc-final-dimensionless} without the $\kappa^2$ term) and the case where the density matrix is initially in the colour octet state only. For a colour singlet initial state, the result is 
\begin{align}
    \langle \Delta \qt^2\rangle_s &=2Q_s^2\left\{1+\frac{1}{3}E_{2/3}\left(\frac{N_c^2}{N_c^2-1}\frac{\theta_{q\bar q}^2}{3\theta_c^2}\right)-\Gamma(4/3)\left(\frac{N_c^2-1}{N_c^2}\frac{3\theta_{c}^2}{\theta_{q\bar q}^2}\right)^{1/3}\right\}\,,\label{eq:meanqt-singlet}
\end{align}
and for a colour octet initial state, we find
\begin{align}
    \langle \Delta \qt^2\rangle_o &=2Q_s^2\left\{1-\frac{1}{N_c^2-1}\left[\frac{1}{3}E_{2/3}\left(\frac{N_c^2}{N_c^2-1}\frac{\theta_{q\bar q}^2}{3\theta_c^2}\right)-\Gamma(4/3)\left(\frac{N_c^2-1}{N_c^2}\frac{3\theta_{c}^2}{\theta_{q\bar q}^2}\right)^{1/3}\right]\right\}\,,\label{eq:meanqt-octet}
\end{align}
where we have introduced the generalised exponential integral function defined by
\begin{align}
    E_n(x)\equiv\int_1^{\infty} \der t \ \frac{e^{-xt}}{t^n}\,.
\end{align}
These results are independent of the choice of the initial Wigner function in $(\qt,\bt)$, provided its $\qt$ dependence admits a second moment.

The dependence of $\langle \Delta \qt^2\rangle_s$ and $\langle \Delta \qt^2\rangle_o$ as a function of $\theta_{q\bar q}$ are shown in Fig.~\ref{fig:qt-distrib}. For a colour singlet initial density matrix, when $\theta_{q\bar q}\gg \theta_c$, the contribution expressed in terms of $E_{2/3}$  vanishes exponentially, and the last term in that expression is power suppressed such that one finds $\langle\Delta q_\perp\rangle ^2\approx 2Q_s^2$, where the factor of 2 simply comes from the fact that the two colour charges undergo independent transverse momentum diffusion (hence $\langle \Delta q_\perp^2\rangle=\langle (\Delta k_{1\perp}+\Delta k_{2\perp})^2\rangle=2\langle \Delta k_{1\perp}^2\rangle$). On the other hand, when $\theta_{q\bar q}\ll \theta_c$, using $E_{2/3}(x)=\Gamma(4/3)x^{-1/3}-3+O(x^{1/3})$, one sees that $\langle\Delta q_\perp\rangle ^2\to 0$. Basically, the medium does not resolve the two colour charges and ``sees'' a singlet state which is not affected by the medium. 

On the other hand, for a colour octet initial state, while the limit $\theta_{q\bar q}\gg\theta_c$ yields the same result $\langle \Delta\qt^2\rangle_o$ for the same reason as in the colour singlet initial state case, the limit $\theta_{q\bar q}\ll \theta_c$ is different. Indeed, in that case, there is no transition between colour octet and colour singlet states (even at finite $N_c$) because the size of the dipole is too small. Therefore, the broadening should be that of a gluon and one finds indeed $\langle \Delta \qt^2\rangle_o=C_A Q_s^2/C_F$ (recall that we define $\qhat$ in the fundamental representation, the $C_A/C_F$ factor thus amounts to redefine $\qhat$ in the adjoint representation). This is what is observed in the blue curve of Fig.~\ref{fig:qt-distrib}-right.

To our knowledge, these results, although relatively intuitive, are new in the jet-quenching literature (similar results have been obtained in~\cite{Cougoulic:2017ust} for the broadening of an energetic parton pair traversing a medium, albeit for a fixed transverse separation of the pair, as relevant to the case of nuclear initial-state broadening). They can be used for phenomenological applications, for instance to study azimuthal correlations of hard subjets within jets, as discussed in the conclusion. Moreover, they could be naturally incorporated into a Monte Carlo implementation of a dipole-based parton shower, in order to account for the transverse-momentum broadening of dipoles propagating through the plasma. These results can also be derived using standard Feynman diagram techniques, as shown in appendix~\ref{app:feynman-TMB}. This provides a non-trivial cross-check of the validity of our Lindblad equation and of the subsequent simplifications, and demonstrates the capability of the open-quantum-system formalism to yield phenomenologically relevant predictions.

More generally, the open quantum system approach comes with benefits. First, the initial condition of the quantum master equation naturally provides a modelling of the broadening distribution that incorporates both non-perturbative effects (encoded in the scale $\Lambda$) and medium-induced effects (encoded in the scale $Q_s$), in close analogy with the treatment of initial states in TMD factorization for hadronic reactions.
In addition, the OQS approach allows us to extend the result Eq.\,\eqref{eq:OQS-factor} including the effect of the finite width of the $\rt$ Wigner density. Thus, it enables one to obtain physically motivated analytic expressions for the broadening distribution including non-perturbative corrections encoded in the parameters $\Lambda$ and $\mu_2$. If instead of using Eq.\,\eqref{eq:broadening-qt-init-lambda} to compute the $\qt$ distribution, we keep the Gaussian dependence on $\rt$ as in Eq.\,\eqref{eq:Gaussian-wigner}, we obtain for finite $\Lambda$ and $\mu_2$ the distribution:
\begin{align}
    &\frac{\der P}{\der^2\pt\der^2\qt}=H(\pt)\times\frac{1}{\pi}\left\{\frac{e^{-\frac{\pt^2\qhat t^3}{24E^2}\frac{8\mu_2^2+\qhat t}{2\mu_2^2+\qhat t}}}{1+\frac{\qhat t}{2\mu_2^2}}\frac{e^{-\frac{\qt^2}{\Lambda^2}}}{\Lambda^2}+\int_0^t\der s \  \frac{\qhat }{2(\Lambda^2+2\qhat s)}e^{-\frac{\qt^2}{\Lambda^2+2\qhat s}}\right.\nonumber\\
    &\left.\times  e^{-\frac{\pt^2\qhat (t-s)^3}{24E^2}\frac{8\mu_2^2+\qhat (t-s)}{2\mu_2^2+\qhat (t-s)}}\frac{\mu_2^2[8E^2(2\mu_2^2+\qhat (t-s))+\pt^2(t-s)^2(4\mu_2^2+\qhat (t-s))^2]}{2E^2(2\mu_2^2+\qhat(t-s))^3} \right\}\,.
\end{align}
The $\qt$ distribution in the curly bracket of this expression provides a two parameters model extending Eq.\,\eqref{eq:OQS-factor} which includes non-perturbative physics encoded in the initial condition of the Lindblad equation. In particular, the singlet state survival probability $P_s(t)$ (which is also, by probability conservation, one minus the singlet-to-octet transition probability) can be directly extracted from the coefficient multiplying the $\qt$ distribution originating from the singlet component:
\begin{align}
    P_s(t)&=\frac{1}{1+\frac{\qhat t}{2\mu_2^2}}\exp\left(-\frac{\theta_{q\bar q}^2}{12\theta_c^2}\frac{8\mu_2^2+\qhat t}{2\mu_2^2+\qhat t}\right)\,.\label{eq:Ps-mu2-correction}
\end{align}
where we have used again $\pt^2=\theta_{q\bar q}^2E^2/2$.
This formula shows that the singlet survival probability is not one even for vanishing $\theta_{q\bar q}$ and reads $(1+\qhat t/(2\mu_2^2))^{-1}\le 1$ which can be small for $\mu_2^2\ll \qhat t$. Indeed, in coordinate space, the initial state is made of a quantum superposition of dipole sizes $r_\perp$ with typically $r_\perp\sim 1/\mu_2$ which is then large as compared to $1/Q_s$ in the regime $\mu_2^2\ll \qhat t=Q_s^2$. Therefore, colour transitions are allowed even if the classical trajectory does not lead to an increase of the $q\bar q$ transverse separation (because $\theta_{q\bar q}$ is small). Eq.\,\eqref{eq:Ps-mu2-correction} naturally extends the transition probability obtained from Eq.\,\eqref{eq:OQS-factor} in the case of a localised initial state $\rt$ and given by $P_s(t)=\exp(-\theta_{q\bar q}^2/(3\theta_c^2))$ by incorporating a genuine quantum description of the initial state satisfying the Heisenberg uncertainty principle.

\section{Quantum decoherences}
\label{sec:decoherence}

As stated in the introduction, our original motivation for addressing this problem within the open quantum system formalism --- and in particular for describing the quantum state of the system by means of a density matrix --- is to clarify the role of the time scale $t_c$ defined in the previous section (and the associated critical angle $\theta_c$) in the genuine quantum decoherence of the system. In the results of the previous section concerning the $\qt$ distribution of the $q\bar q$ dipole, $t_c$ was interpreted physically as an \textit{inverse transition rate} between singlet and octet states. In this picture, the $q\bar q$ pair can be viewed as an effective two-level system, composed of a singlet and an octet component. Since the evolution allows transitions between these two species, a non-trivial dynamics emerges, leaving in particular a clear imprint of $\theta_c$ on the final transverse-momentum distribution through the ratio $\theta_{q\bar q}/\theta_c$.

In this section, we aim to elucidate the role of $t_c$ and $\theta_c$, as well as any additional relevant scales, in the decoherence of the density matrix in the momentum-space basis. We will investigate decoherence both in the colour (discrete) degrees of freedom and in the center-of-mass momentum (continuous) degrees of freedom $(\qt,\Qcalt)$.

\subsection{Colour decoherence, colour equilibration and colour transitions}
\label{sub:colour-decoherence}

We recall that, by construction, the colour density matrix contains no coherences in the singlet-octet basis, as required by gauge invariance, cf.~Eq.\,\eqref{eq:dm:qqbar:singlet-octet}. As a consequence, the interaction with the medium can drive colour transitions between the singlet and octet sectors, while no singlet-octet coherences can be generated by the quantum master equation: the colour evolution of the densities (diagonal components of the density matrix in the singlet-octet basis) decouples from the evolution of the possible ``coherences" (off-diagonal elements).

It may therefore appear somewhat artificial to discuss colour decoherence in this setting. In this section, however, we show that colour decoherence occurs in a stronger sense than is usually encountered in the quantum physics of non-relativistic systems. Instead of merely becoming diagonal in a particular preferred basis, the colour density matrix approaches, at late times, a state proportional to the identity matrix and is therefore diagonal in any basis. This behaviour underlies the notion of colour equilibration. We thus demonstrate that colour transitions, colour equilibration, and colour decoherence are simply different facets of the same underlying phenomenon, and that the choice of terminology is largely a matter of convenience dictated by the physical context.

First, we thus demonstrate using 
the present open quantum system approach that the time scale $t_c=(4/(\qhat \theta_{q\bar q}^2))^{1/3}$ represents the characteristic time scale for the suppression of the off-diagonal elements in colour space of the density matrix, in agreement with the terminology used in the field. In analogy with decoherence in e.g.~spin physics, we consider an initial colour state which is a quantum superposition such that the initial colour density matrix is non-diagonal (in a given basis) and we check that for $t\gg t_c$, the density matrix becomes diagonal. 
A natural such initial state to consider is the colour singlet state $\ket{\psi_s}$ for the $q\bar q$ pair, with
\begin{align}
   \ket{\psi_s}= \frac{1}{\sqrt{3}}\left(
                \ket{r\bar r}+\ket{g\bar g}+\ket{b\bar b}
            \right),
\end{align}
in the natural basis $\ket{\alpha\bar\beta}$ built from colour and anti-colour states of the quark and the antiquark. As a quantum superposition of states in this basis, the density matrix $\ket{\psi_s}\bra{\psi_s}$ at $t=0$ is non-diagonal, since 
\begin{align}
    \braket{\alpha\bar\beta|\psi_s}\braket{\psi_s|\gamma\bar\delta}&=\frac{1}{3}\delta_{\alpha\beta}\delta_{\gamma\delta}\,.
\end{align}
(With our notation, the diagonal part of the density matrix would be instead $\delta_{\alpha\gamma}\delta_{\beta\delta}$.)

Let us consider the evolution equation for the Wigner function integrated of the total degrees of freedom ($\bt$ and $\qt$). When doing so, the kinetic term $\qt\cdot\nabla_{\bt}$ and the diffusion term $\nabla^2_{\qt}$ disappear (for suitable boundary condition at infinity). Moreover, we also neglect the $\pt$ diffusion term by working again in the regime $\theta_s\ll\theta_c$ such that $\kappa^2\ll 1$. The equation we have to solve is then
\begin{align}
    \frac{\partial }{\partial t}\begin{pmatrix}
        \rho_s\\
        \rho_o
    \end{pmatrix}&+\frac{1}{E}\pt\cdot\nabla_{\rt}\begin{pmatrix}
        \rho_s\\
        \rho_o
    \end{pmatrix}=\frac{\qhat\rt^2}{4C_F}\begin{pmatrix}
        -2C_F & 2C_F\\
        \frac{1}{N_c} & -\frac{1}{N_c}
    \end{pmatrix}\begin{pmatrix}
        \rho_s\\
        \rho_o
    \end{pmatrix}\,.\label{eq:exactNc-rho(rt,pt)}
\end{align}
For an initial colour singlet state, the solution to this coupled set of partial differential equation is obtained using the method of characteristics as above followed by the diagonalization of the matrix on the right hand side of Eq.\,\eqref{eq:exactNc-rho(rt,pt)}. It reads
\begin{align}
  \rho_s(\pt,\rt,t)&=\left[\frac{1}{N_c^2}+\frac{N_c^2-1}{N_c^2}\exp\left(-\int_0^t\der t' \ \frac{\qhat\rt^2(t')N_c}{4C_F}\right)\right]\rho_s(\pt,\rt-\pt t/E,0)\label{eq:rhoRs-fullNc}\\
    \rho_o(\rt,\pt,t)&=\frac{1}{N_c^2}\left[1-\exp\left(-\int_0^t\der t' \ \frac{\qhat\rt^2(t')N_c}{4C_F}\right)\right]\rho_s(\pt,\rt-\pt t/E,0)\label{eq:rhoRo-fullNc}
\end{align}
with $\rt(t')=\rt-\pt t'/E$.
Integrating over $\rt$ using and initial condition sharply peaked around $\rt=\boldsymbol{0}_\perp$ to simplify the calculation of this integral, one readily obtains that
\begin{align}
    \rho_s(\pt,t)&=\frac{1}{N_c^2}+\frac{N_c^2-1}{N_c^2}\exp\left(-\frac{N_c\qhat\theta_{q\bar q}^2t^3 }{24C_F}\right)\label{eq:rhos-finiteNc}\\
    \rho_o(\pt,t)&=\frac{1}{N_c^2}\left[1-\exp\left(-\frac{N_c\qhat\theta_{q\bar q}^2t^3 }{24C_F}\right)\right]\label{eq:rhoo-finiteNc}
\end{align}
Note that although we use the same notation as in Eqs.\,\eqref{eq:rhoRs-fullNc}-\eqref{eq:rhoRo-fullNc}, $\rho_{s/o}(\pt,t)$ here refers to the marginalized probability distributions over the relative position of the quark-antiquark pair.
We also recall that here, $\theta_{q\bar q}^2=2\pt^2/E^2$ implicitly depends on the Wigner variable $\pt$.
Hence, for $t\gg t_c$ with the characteristic time scale $t_c$ defined like in Eq.\,\eqref{eq:tc-def-lNc} but keeping the exact $N_c$ dependence:
\begin{align}
    t_c&=\left(\frac{8C_F}{N_c\qhat\theta_{q\bar q}^2}\right)^{1/3}\,,\label{eq:tc-def}
\end{align}
we have $\rho_s=\rho_o=1/N_c^2$. This equality demonstrates two interesting physical phenomena. The first, which is the central focus of this paper, is colour decoherence, reflected in the disappearance of the off-diagonal matrix elements in the colour-anticolour basis. Indeed, we have
\begin{align}
\braket{\alpha\bar\beta|\rho|\gamma\bar\delta}&=\frac{1}{N_c}(\rho_s-\rho_o)\delta_{\alpha\beta}\delta_{\gamma\delta}+\rho_o\delta_{\alpha\gamma}\delta_{\beta\delta}
\end{align}
so that $\rho_s=\rho_0$ for $t\gg t_c$ implies that the density matrix becomes diagonal. Moreover, the diagonal coefficients are given by $\rho_o=1/N_c^2$: they are therefore all identical and independent of the medium’s properties. This is the colour equilibration phenomenon, which essentially states that each colour state $\ket{\alpha\bar\beta}$ is equally populated after the characteristic time scale $t_c$. Beyond this time, the colour state of the dipole becomes a statistical mixture of colour-anticolour states with uniform probability $1/N_c^2$. Obviously, the same is true in the singlet-octet bases $\{\ket{s},\ket{o^c}\}$ since the density matrix for $t\gg t_c$ is proportional to the identity matrix in colour space. This colour equilibration (or randomization) phenomenon has also been investigated for an in-medium $g\to gg$ splitting in~\cite{Zakharov:2018hfz}.

As stated in the previous section, we have also interpreted $t_c$ as the typical time scale for colour transitions. The mathematical structure of the quantum master equation resembles that of a multi-species Fokker-Planck equation and as such, it is therefore natural to interpret $\rho_s$ and $\rho_o$ as two independent “species” of $q\bar q$ dipoles. As is clear from Eqs.\,\eqref{eq:rhos-finiteNc}-\eqref{eq:rhoo-finiteNc}, $t_c$ also represents the time scale governing the transitions between the singlet and octet species. In particular, in the large $N_c$ limit, we have
\begin{align}
\rho_s(\pt,t)&=\exp\left(-\frac{t^3}{3t_c^3}\right)=P_s(t)\,,\quad \tilde\rho_o(\pt,t)=1-\exp\left(-\frac{t^3}{3t_c^3}\right)\,,\label{eq:singlet-to-octet-rate}
\end{align}
with $t_c=(4/(\qhat\theta_{q\bar q}^2))^{1/3}$ the \textit{inverse rate} of the singlet to octet transition. Here, the exponential factor can be interpreted as the survival probability for the dipole to remain in a colour singlet state, which is one of the standard interpretations of the time scale $t_c$ in the jet quenching literature.

To summarise, the open quantum system approach allows us to interpret $t_c$ in three distinct yet closely related ways: it is the time scale for colour decoherence in the quantum mechanical sense (the diagonalisation of the density matrix in colour space, which occurs here in any colour basis), the time scale for colour equilibration of the subsystem, and the time scale governing colour transitions between the singlet and octet states of the $q\bar q$ dipole.

\subsection{Decoherence in the transverse momentum imbalance basis}
\label{subsub:decoherence-qt}

So far, we have only discussed the colour space structure of the density matrix. In this section, we will consider its transverse coordinate and momentum structure via the study of the density matrix $\rho_T(\qt,\Qcalt;t)$  as a function of the diagonal variable $\qt$ and its off-diagonal $\Qt$ counter-part and we want to study how the off-diagonal component behaves as a function of time. We focus on the center-of-mass degrees of freedom where the interesting medium effects occur in the regime $p_\perp\gg q_\perp\sim Q_s$. We shall discuss the decoherence in the relative momentum in the next section, as it requires to deal with the $\kappa^2$ corrections which have been neglected so far. To illustrate the physical features of the solution, we work again in the large $N_c$ approximation, but similar results can be obtained at finite $N_c$. We thus consider
\begin{align}
    \rho_{T}(\qt,\Qcalt;t)&=\int\der^2\bt\  e^{i\Qcalt\cdot \bt} \int\der^2\rt\int\der^2\pt \ \rho(\pt,\rt;\qt,\bt;t)\,,\\
    &=\int \der^2\Bcalt e^{i\qt\cdot\Bcalt}\int\der^2\rt\int\der^2\pt \ \rho(\pt,\rt;\Bcalt,\Qcalt;t)\,.
\end{align}
For the following calculation, we also note $\rho_{T,s}(\Bcalt,\Qcalt;0)$ the double Fourier  transform of the Wigner distribution on the $(\qt,\bt)$ phase space:
\begin{align}
    \rho_{T,s}(\Bcalt,\Qcalt;0)&=\int\der^2\bt\int\frac{\der^2\qt}{(2\pi)^2}e^{i\Qcalt\cdot\bt-i\Bcalt\cdot\qt}\frac{1}{\pi^2}e^{-\qt^2/\Lambda^2-\Lambda^2\bt^2}\,,\\
    &=\frac{1}{(2\pi)^2}e^{-\Bcalt^2\Lambda^2/4-\Qcalt^2/(4\Lambda^2)}\,.\label{eq:init-BQ-variable}
\end{align}
Using Eqs.\,\eqref{eq:rhos-oqs-largeNc}-\eqref{eq:rhoo-oqs-largeNc} with the initial condition given by Eq.\,\eqref{eq:init-BQ-variable}, we obtain $\rho_{T}(\Bcalt,\Qcalt,t)$ at any time $t$. After Fourier transform from $\Bcalt$ to $\qt$, we find  the following density matrices in the momentum space basis:
\begin{align}
    \rho_{T,s}(\qt,\Qcalt;t)&=\frac{1}{\pi \Lambda^2}e^{-t^3/(3t_c^3)}e^{i\frac{(\Qcalt\cdot \qt)t}{q^+}}e^{-\qt^2/\Lambda^2}e^{-\Qcalt^2/(4\Lambda^2)}\,,\label{eq:density-matrix-rhos}\\
    \rho_{T,o}(\qt,\Qcalt;t)&=\frac{1}{\pi}\int_0^t\der s \ e^{-\frac{(t-s)^3}{3t_c^3}}\frac{\qhat \theta_0^2(t-s)^2}{4(\Lambda^2+2\qhat s)}e^{-\frac{1}{\Lambda^2+2\qhat s}\left[\qt-\frac{i(\qhat s^2+\Lambda^2t)}{2q^+}\Qcalt\right]^2}\nonumber\\
    &\times\exp\left\{-\Qcalt^2\left[\frac{1}{4\Lambda^2}+\frac{t^2\Lambda^2}{4(q^+)^2}+\frac{\qhat s^3}{6(q^+)^2}\right]\right\}\,.\label{eq:density-matrix-rhoo}
\end{align}
When $t\ll t_c$, the density matrix is dominated by the singlet component. From the first expression above, one sees that the width of the non-vanishing off-diagonal elements remains constant, of order of $\Lambda^2$, and is controlled by the initial form of the density matrix. When $t$ becomes larger than $t_c$, the transition to the octet state has occurred and the $\Qcalt$ dependence is now given by Eq.\,\eqref{eq:density-matrix-rhoo}.
Hence, the typical width $\sigma_{\Qcalt}$ of the non-vanishing off-diagonal elements of the density matrix, clearly identified after $\qt$ integration, is given by
\begin{align}
    \frac{\sigma^{2}_{\Qcalt}}{4\Lambda^2}&\sim\frac{1}{1+\frac{\Lambda^4t^2}{(q^+)^2}+\frac{2\qhat\Lambda^2 t^3 }{3(q^+)^2}}\,,
\end{align}
and therefore, it goes to 0 (meaning, the density matrix becomes diagonal) for $t\gg q^+/\Lambda^2$ or $t\gg [(q^+)^2/(\qhat \Lambda^2)]^{1/3}$. The two previous time scales have already been discussed in~\cite{Barata:2023uoi} and following the notations of this paper, we introduce
\begin{align}
    t_0\equiv \frac{q^+}{\Lambda^2}\,,\quad t_2\equiv\left(\frac{(q^+)^2}{\qhat\Lambda^2}\right)^{1/3}\,.\label{eq:t0-t2-def}
\end{align}
Yet, their interplay with the decoherence time $t_c$ is novel in our analysis. Unlike the case where a single quark propagates through the medium studied in~\cite{Barata:2023uoi}, the kinematic decoherence in transverse momentum imbalance space depends on the relative ordering between $t_c$ and the time scales $t_0$ and $t_2$ (with $t_2$ typically the smallest in the high-energy limit considered here). When $t_c < \min(t_0,t_2)$, the system first undergoes colour decoherence/transition while the density matrix is still in a genuinely quantum regime. The transition to a classical description only occurs later, at times of order $\min(t_0,t_2)\simeq t_2$, after which the subsequent evolution in transverse momentum imbalance is effectively classical. In the opposite regime, $t_c > \min(t_0,t_2)$, the evolution remains quantum (the density matrix retains the off-diagonal elements inherited from the initial condition, cf.~Eq.\,\eqref{eq:density-matrix-rhos}) up to times of order $t_c$, since the dipole has not yet undergone the transition to an octet state. Once $t_c$ is reached, colour decoherence and the transition to a classical regime occur simultaneously, and the subsequent dynamics is directly classical. The condition $t_2 < t_c$ translates into $\theta_{q\bar q} < \Lambda/q^+$, implying that the regime where decoherence and classicalization coincide is restricted to very small dipole angles, close to the hadronization angle $\theta_\Lambda \equiv \Lambda/q^+$.

To summarize, although $t_c$ truly controls the \textit{colour} decoherence of the density matrix as demonstrated in subsection~\ref{sub:colour-decoherence}, the \textit{kinematic} decoherence of the transverse momentum imbalance wave-function of the dipole is typically controlled by the time scale $t_2$, which is a much larger time scale. It is smaller than the medium size for $q^+\le \sqrt{\qhat \Lambda^2L^3}$, a scale of order $20$ GeV for $\qhat = 1.5$ GeV$^2$/fm, $\Lambda=0.5$ GeV and $L=4$ fm. For high energy dipoles, with $q^+\gg \sqrt{\qhat \Lambda^2L^3}$, the dipole is still described by a coherent quantum superposition of $\qt$ wave-functions as it escapes the medium, with the coherence property set by the parameter $\Lambda$ in the initial condition. This is an important outcome of our study, as it underscores the role of QCD radiation and hadronization processes at later times (which are, of course, not included in our analysis) in driving the decoherence of the $q\bar q$ dipole wave function. The interaction with the medium alone is not sufficient to achieve complete decoherence, either in colour space or in kinematic space.

\begin{figure}[t]
    \centering
    \includegraphics[width=0.95\textwidth]{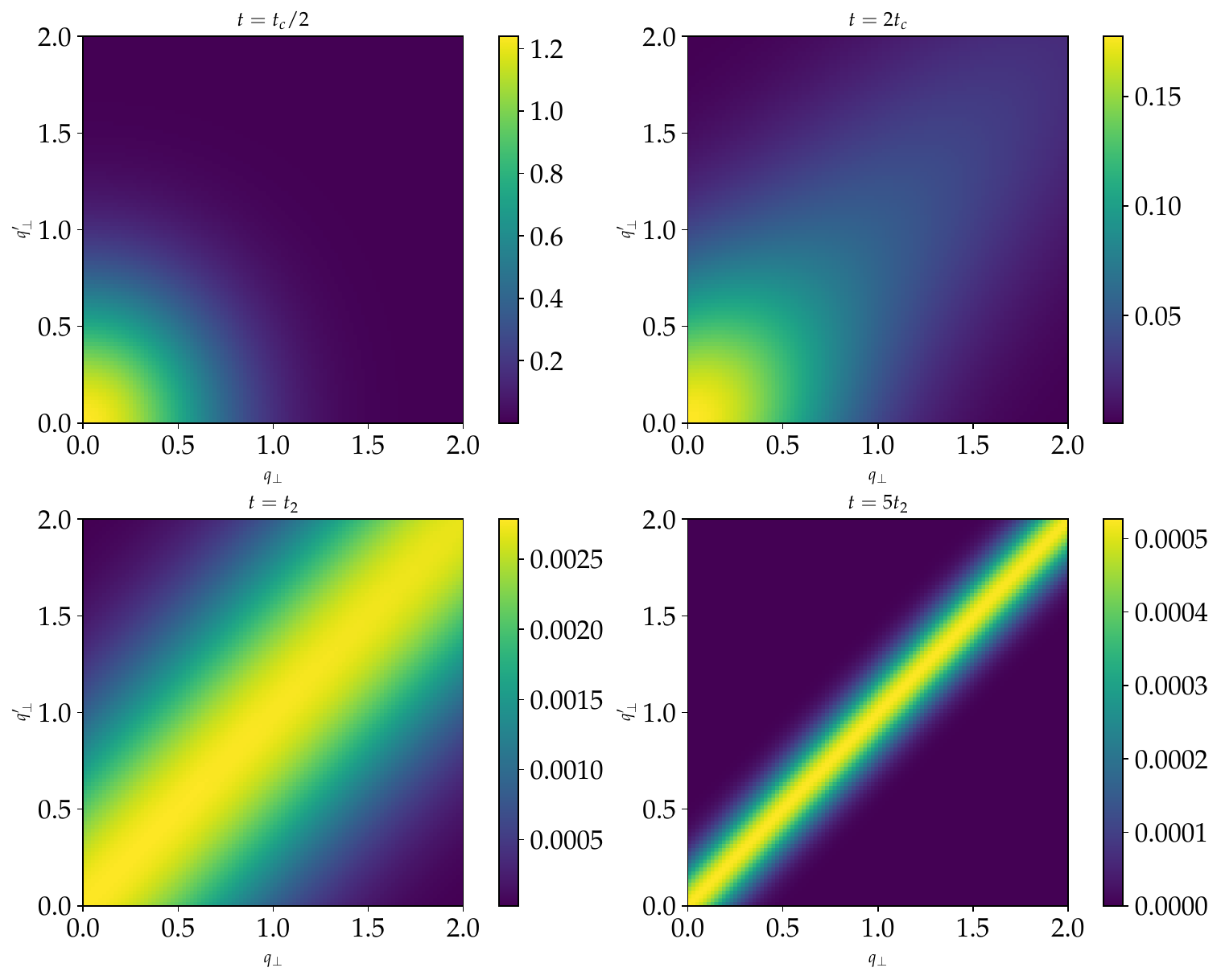}
    \caption{Density matrix in the total transverse momentum basis for at $t=0$ (upper left), $t=2t_c$ (upper right), $t=t_2$ (lower left) and $t=5t_2$ (lower right). Details in text.}
    \label{fig:density-matrix-qt-Qt}
\end{figure}

In Fig.~\ref{fig:density-matrix-qt-Qt}, we illustrate the behaviour of the density matrix in the transverse momentum imbalance basis $\rho_T(\qt,\Qcalt;t)$. More precisely, we show the real part of $\rho_T(\qt,\Qcalt;t)$ with the column variable $q_\perp'=|\qt-\Qcalt/2|$ of the matrix on the vertical axis and its line variable $q_\perp=|\qt+\Qcalt/2|$ on the horizontal axis, assuming that $\qt$ and $\qt'$ are collinear. The four panels correspond to four choices of times $t$: $t=0$, $t=2t_c$, $t=t_2$ and $t=5t_2$. We use $q^+=800$ GeV, $\qhat =1.5$ GeV$^2$/fm, $\theta_{q\bar q}=0.1$ and $\Lambda=0.5$ GeV such that $t_c\simeq 2.2$~fm and $t_2\simeq 40$ fm. The upper-left panel at $t=0$ shows the two-dimensional Gaussian initial condition for the singlet state, with a characteristic width set by the scale $\Lambda$. As is evident, the initial singlet state contains quantum superpositions of dipole configurations in $\qt$ space. After a time of order $t_c$, the singlet state begins to decay into the octet state. The corresponding $\qt$ distribution of the octet becomes broader due to collisions with the medium constituents, which explains why the density matrix in the upper-right panel is distorted relative to the initial condition. From that point on, the qualitative behaviour of the density matrix closely resembles what was observed in~\cite{Barata:2023uoi} for a single quark. In the present case, since the effective degree of freedom in the octet state is most closely analogous to a single gluon, the comparison is particularly natural. Around the time scale $t_2$, the density matrix becomes increasingly diagonal, signaling the onset of kinematical decoherence and the progressive classicalization of the system, as clearly illustrated in the lower-left panel at $t = 5 t_2$. The lower-left panel should be regarded as a formal illustration of the asymptotic regime, since $t_2$ is typically larger than the jet path length through the medium (and therefore, $5t_2$ as well) for high-energy dipoles. Nevertheless, it shows that medium interactions are not sufficient to fully classicalize the dipole in momentum space, which therefore typically remains a quantum object as it emerges from the quark-gluon plasma.

\section{Effect of the $p_\perp$-diffusion term: perturbative approach}
\label{sec:pt-diffusion}

In the two previous sections, we have systematically ignored the effect of the quantum diffusion term in $\pt$ in the master equation satisfied by the Wigner function. This was motivated by the fact that this term scales like $\theta_s^2/(2\theta_c^2)\ll 1$, which is small in the high energy limit. Accordingly, we did not focus on the quantum aspects of the evolution in the $(\pt,\rt)$ phase space and we have essentially considered that the dynamics of the relative degrees of freedom of the $q\bar q $ pair is driven by the geometry of the classical $q\bar q$ trajectory, namely $\rt\simeq \pt t/E$, provided the initial condition is peaked around $\rt=\boldsymbol{0}_\perp$. 
In this section, we investigate the validity of this approximation and we explore the effect of the $\pt$-diffusion proportional to $\kappa^2$ in Eq.\,\eqref{eq:Wigner-ho-fullNc-final-dimensionless} on the singlet to octet transition rate and on the decoherence of the density matrix in the relative transverse momentum basis $(\pt,\Pcalt)$.

\subsection{The Gaussian ansatz}

In the large $N_c$ limit, the master equation for the colour singlet component of the density matrix is a closed equation. Indeed, considering the singlet component of Eq.\,\eqref{eq:Wigner-ho-lNc-final} and integrating over $\bt$ and $\qt$, we find that the Wigner quasi-distribution over the relative degrees of freedom phase space satisfies:
\begin{align}
    \frac{\partial \rho_{R,s}}{\partial \bar t}+ \ptb\cdot\nabla_{\rtb} \rho_{R,s}=-\frac{1}{2}\rtb^2\rho_{R,s}+\frac{\kappa^2}{8}\nabla^2_{\ptb}\rho_{R,s}\,.
    \label{eq:rhos-section-ptdiffusion}
\end{align}
This partial differential equation cannot be solved using the method of characteristics as done in section~\ref{sec:factorisation-violation} because of the combination of the  $\nabla^2_{\pt}$ and $\rt^2$ terms. In the following paragraphs, we develop a general strategy to find approximate solutions for $\rho_s$ which will allow us to discuss the effect of the $\pt$ diffusion term on the single to octet transition probability\footnote{One can see from Eq.\,\eqref{eq:Wigner-ho-lNc-final} that the solution for the octet component can be computed analytically without any perturbation scheme if the initial condition is purely octet, see later on in Eq.\,\eqref{eq:rhooctet-full solution}.}. 
For the Gaussian initial condition Eq.\,\eqref{eq:Gaussian-wigner}, we can try to find the solution to Eq.\,\eqref{eq:rhos-section-ptdiffusion} with the general Gaussian ansatz
\begin{align}
    \rho_{R,s}(\ptb,\rtb;\bar t)=\frac{\alpha(0)\beta(0)}{\pi^2}\exp&\left[-\alpha(\bar t)\rtb^2-\beta(\bar t)\ptb^2-\delta(\bar t)+\gamma(\bar t)\ptb\cdot \rtb\right.\nonumber\\
    &\left.+\boldsymbol{\epsilon}_\perp(\bar t)\cdot \ptb+\boldsymbol{\zeta}_\perp(\bar t)\cdot \rtb\right]\,,
    \label{eq:Gaussian-ansatz}
\end{align}
in which $\langle \ptb\rangle$ and $\langle \rtb\rangle$ are not necessarily collinear. Plugging this ansatz inside Eq.\,\eqref{eq:rhos-section-ptdiffusion} yields the following non-linear system of first order differential equations
\begin{align}
&\dot\alpha=\frac{1}{2}-\frac{1}{8}\kappa^2\gamma^2\,,
\quad \dot{\beta}-\gamma=-\frac{1}{2}\kappa^2 \beta^2\,, 
\quad \dot{\gamma}-2\alpha=-\frac{\kappa^2}{2}\gamma\beta\,,
\quad \dot{\delta}=\frac{1}{2}\kappa^2\beta-\frac{1}{8}\kappa^2|\boldsymbol{\epsilon}_\perp|^2\,,\nonumber\\
&\dot{\boldsymbol{\epsilon}}_\perp=-\boldsymbol{\zeta}_\perp-\frac{1}{2}\kappa^2\beta \boldsymbol{\epsilon}_\perp\,,
\quad \dot{\boldsymbol{\zeta}}_\perp=\frac{1}{4}\kappa^2\gamma\boldsymbol{\epsilon}_\perp,
\label{eq:pt-diffusion:Gaussian-ansatz-params}
\end{align}
where the last two equations act in a two-dimensional transverse space. This system depends on the effective non-linear ``coupling'' $\kappa$ coming from the $\pt$ diffusion term in the quantum master equation.

It is useful to note that such Gaussian ansatz can be also written under the form
\begin{equation}
 \rho_{R,s}(\ptb,\rtb;\bar t) =
 \frac{1}{\pi^2\sigma_{\bar p,0}^2\sigma_{\bar r,0}^2}
 \exp\left[ - \frac{\sigma_{\bar{p}}^2\delta\rtb^2 +\sigma_{\bar{r}}^2\delta\ptb^2 -2 c_{\bar{r},\bar{p}} \delta\rtb\cdot\delta\ptb }{\sigma_{\bar{r}}^2 \sigma_{\bar{p}}^2 - c_{\bar{r},\bar{p}}^2} -\tilde{\delta}\right]\,,
 \label{eq:pt-diffusion:Gaussian-ansatz-params:physical}
\end{equation}
where $\delta\rtb= \rtb -\langle \rtb \rangle$, $\delta\ptb= \ptb -\langle \ptb \rangle$, while $\sigma_{\bar{p}}^2=
\langle \delta\ptb^2\rangle$, $\sigma_{\bar{r}}^2=
\langle \delta\rtb^2\rangle$, and $c_{\bar{r},\bar{p}}=\langle \delta\ptb\cdot \delta\rtb\rangle$. In this expression, $\tilde{\delta}$, $\langle \rtb \rangle$, $\langle \ptb \rangle$, $\sigma_{\bar{p}}^2$, $\sigma_{\bar{r}}^2$ and $c_{\bar{r},\bar{p}}$ are time-dependent quantities equivalent to the parameters introduced in greek letters in the original ansatz ; yet, they admit a transparent physical interpretation in terms of the norm, mean values, widths and correlation of the transverse size and relative transverse momentum of the colour-singlet dipole. 
They can be shown to satisfy the following equations
\begin{equation}
\frac{\der\langle\rtb\rangle}{\der \bar t} = 
\langle \ptb \rangle  - \frac{\sigma_{\bar{r}}^2}{2}\langle\rtb\rangle,\quad
\frac{\der \langle\ptb\rangle}{\der \bar t} = 
- \frac{1}{2}  c_{\bar{r},\bar{p}}\langle \rtb \rangle,
\label{eq:diffusion-pt:first-moms:eq}
\end{equation}
for the first moments, 
 \begin{equation}
\frac{\der \sigma_{\bar{r}}^2}{\der \bar{t}}=  
2 c_{\bar{r},\bar{p}} - \frac{\sigma_{\bar{r}}^4}{2} ,
\quad
\frac{\der \sigma_{\bar{p}}^2}{\der \bar{t}}=  
\frac{\kappa^2}{2} - \frac{c_{\bar{r},\bar{p}}^2}{2}\, \label{eq:diffusion-pt:second-moms:eq}
\end{equation}
for the variances, and
\begin{equation}
\frac{\der c_{\bar{r},\bar{p}}}{\der \bar{t}} =
 \sigma_{\bar{p}}^2- \frac{\sigma_{\bar{r}}^2 \,c_{\bar{r},\bar{p}} }{2}\,,
 \label{eq:diffusion-pt:second-moms:eq2}
 \end{equation}
 for  the correlation of the $\rtb$ and $\ptb$ variables. Notice that the set of equations for the second moments is closed: the evolution of the Gaussian widths and the correlation do not depend on the mean values, while the inverse is not the case.
 The norm, that is the singlet survival probability, evolves 
 according to
 \begin{equation}
 \frac{\der P_s}{\der \bar{t}}=-\frac{\langle \rtb \rangle^2 +\sigma_{\bar{r}}^2}{2} P_s,
  \label{eq:diffusion-pt:zeroth-mom:eq}
 \end{equation}
with $P_s(0)=1$ for a singlet initial state.

We now turn to the perturbative solution of this system of equations and examine the accuracy of the diffusionless approximation employed throughout the paper thus far. In particular, we are interested in scanning the parameter space for the widths of the Gaussian initial state Eq.\,\eqref{eq:Gaussian-wigner}, which we label here as $\mu_1\to\sigma_{\bar{p},0}\equiv\sigma_{\bar{p}}(0)$, $1/\mu_2\to\sigma_{\bar{r},0}\equiv\sigma_{\bar{r}}(0)$ to make a closer contact with the physical variables we are going to compute. The initial Wigner function Eq.\,\eqref{eq:Gaussian-wigner} for the relative degrees of freedom reads then
\begin{equation}
\rho_{R,s}(\ptb,\rtb;0) \propto e^{-\frac{(\ptb-\boldsymbol{\bar P}_{0\perp})^2}{\sigma_{\bar{p},0}^2}-\frac{(\rtb-\boldsymbol{\bar r}_{0\perp})^2}{\sigma_{\bar{r},0}^2}},
\label{eq:pt-diffusion:initial:sigma}
\end{equation}
where we remind the reader that the uncertainty relation imposes $\sigma_{\bar p,0}\sigma_{\bar r,0}\ge \kappa$.
This analysis is important as, even if $\kappa$ is assumed to be small, the diffusion term can bring non-negligible contributions for certain values of $\sigma_{\bar{p},0}$. Indeed, one can see that at early times
\begin{equation}
    \kappa^2\nabla_{\ptb}^2\rho_s(t\sim0)
    \sim
    \frac{\kappa^2}{\sigma_{\bar{p},0}^2}
    \sim
    \frac{\theta_s^2}{\theta_c^2}
    \cdot
    \frac{\theta_c^2E^2}{\sigma_{p,0}^2}
    =
    \frac{\hat qL}{\sigma_{p,0}^2},
\end{equation}
meaning that for a narrow enough initial $\ptb$ distribution, such as $\sigma_{p,0}^2 \sim \hat qL$ for the non-rescaled momentum, one has the diffusion term of the same order of the kinetic term.

We can thus distinguish two extreme cases. (i) An initial wave packet that is narrow in relative transverse coordinate, corresponding to the situation considered in the previous section where the Wigner density was assumed to be sharply peaked around $\boldsymbol{r}_\perp=\boldsymbol{0}_\perp$. In this case, $\sigma_{\bar r,0}$ is as small as $\kappa$ (or even smaller), while $\sigma_{\bar p,0}$ is of order unity (or even larger), so that the diffusion term is genuinely of order $O(\kappa^2)$ at early times. (ii) An initial wave packet that is narrow in relative transverse momentum, with $\sigma_{\bar p,0}\sim \kappa$. In this case, the diffusion term is of order $O(1)$ at the initial time, as just discussed. In the following, we perform a systematic scan between these two limiting cases, varying $\sigma_{\bar r,0}$ from $\kappa/20$ to $\kappa$ with $\kappa=1$. This choice of $\kappa$ allows us to probe the most challenging region of phase space, where the effects of the neglected $\kappa$-dependent corrections are expected to be largest, thereby providing a conservative test of the approximation.

\subsection{Perturbative computation of physical variables}

\begin{figure}[t]
    \centering
    \includegraphics[width=0.32\linewidth]{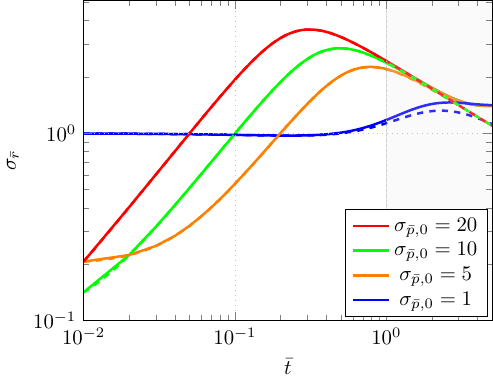}
    \includegraphics[width=0.32\linewidth]{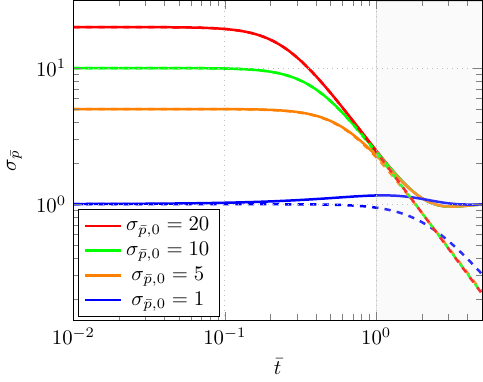}
    \includegraphics[width=0.32\linewidth]{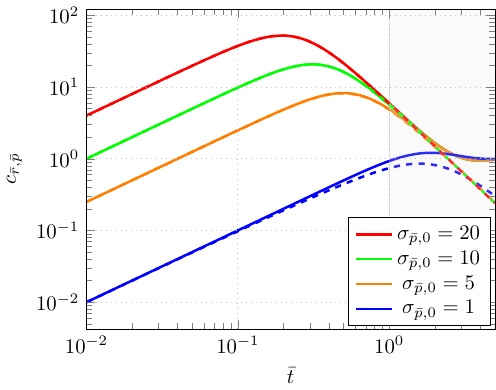}
    \caption{Comparison of the numerical solution (solid curves) and the zeroth order analytic solution (dashed curves) for the time dependence of the widths $\sigma_{\bar p}$ and $\sigma_{\bar r}$, as well as the correlation coefficient $c_{\bar r,\bar p}$ of the Gaussian ansatz satisfying Eq.\,\eqref{eq:rhos-section-ptdiffusion}. The numerical solution includes the diffusion term in $\pt$ with $\kappa=1$, whereas the analytic solution neglects this term. Different colours correspond to different initial values of $\sigma_{\bar p,0}$, with $\sigma_{\bar r,0}=\kappa/\sigma_{\bar p,0}$. The grey area corresponds to the unphysical region with $\bar t\ge 1$, i.e.~$t\ge L$.}
    \label{fig:pt-diff:sigmapt2:numexact-vs-diffusionless}
\end{figure}

We obtain perturbatively the solution of equations Eqs.\,\eqref{eq:diffusion-pt:first-moms:eq}-\eqref{eq:diffusion-pt:zeroth-mom:eq} by expanding each variable $y$ with a series
\begin{equation}
    y(t)
    =
    \sum_{i=0}^{+\infty}
    y^{(i)}(t)\kappa^{2i}.
    \label{eq:diffusion-pt:var-expansion-kappa}
\end{equation}
We plug the expansions inside the equations and we focus on the solutions up to the first order.
To get the zeroth order, that is the diffusionless solutions, the easiest way is to solve the closed set of equations for $\alpha,\beta,\gamma$ --- from Eq.\,\eqref{eq:pt-diffusion:Gaussian-ansatz-params} --- for $\kappa=0$, which gives simple polynomial solutions in $\bar t$.
Then, we know that for a Gaussian ansatz of the form of Eq.~\eqref{eq:Gaussian-ansatz} the second moments read
\begin{equation}
    \sigma_{\bar r}^2
    =
    \frac{4\beta}{4\alpha\beta-\gamma^2},
    \qquad
    \sigma_{\bar p}^2
    =
    \frac{4\alpha}{4\alpha\beta-\gamma^2},
    \qquad
    c_{\bar r,\bar p}
    =
    \frac{2\gamma}{4\alpha\beta-\gamma^2},
\end{equation}
and therefore we get their diffusionless time evolution
\begin{align}
    &(\sigma_{\bar r}^2)^{(0)}
    =8\frac{6 \sigma_{\bar r,0}^2+\sigma_{\bar p,0}^2 \bar t^2 \left(\sigma_{\bar r,0}^2 \bar t+6\right)}{D(\bar t)},
    \label{eq:pt-diff:diffusionless:sol:sigmar2}\\
    &(\sigma_{\bar p}^2)^{(0)}
    =\frac{24 \sigma_{\bar p,0}^2 \left(\sigma_{\bar r,0}^2 \bar t+2\right)}{D(\bar t)},
    \label{eq:pt-diff:diffusionless:sol:sigmap2}\\
    &c_{\bar r,\bar p}^{(0)}
    =\frac{12 \sigma_{\bar p,0}^2 \bar t \left(\sigma_{\bar r,0}^2 \bar t+4\right)}{D(\bar t)},
    \label{eq:pt-diff:diffusionless:sol:corrrp}
\end{align}
with $D(\bar t)\equiv\sigma_{\bar p,0}^2\bar t^3( \sigma_{\bar r,0}^2\bar t+8)+24(\sigma_{\bar r,0}^2 \bar t+2)$.
Consequently, by obtaining the second moments at $O(\kappa^0$), we are able to solve at this order also the equations in Eqs.\,\eqref{eq:diffusion-pt:first-moms:eq} for the first moments. We report the analytical expressions in Appendix~\ref{app:pt-diff:pert-results:first-mom}.
The plot of the diffusionless evolution of the second moments is shown in Figure~\ref{fig:pt-diff:sigmapt2:numexact-vs-diffusionless} together with the numerical solution of the full equations with $\kappa=1$. 
Up to the final time $t=L$ (i.e. $\bar t=1$) we observe a good agreement between the solutions of the equations in the diffusionless regime, used in all the previous sections of this paper, and the full (exact) numerical solutions.
The largest deviations are present at late times for the evolutions with initial width $\sigma_{\bar p,0}=1$ (blue lines).

We can observe in all plots of Fig.\,\ref{fig:pt-diff:sigmapt2:numexact-vs-diffusionless} a characteristic timescale associated with the inversion in the growth of $\sigma_{\bar r}$ and $c_{\bar r,\bar p}$, or the change of slope in the decrease of $\sigma_{\bar p}$. By studying the behaviour of the diffusionless solutions in Eq.\,\eqref{eq:pt-diff:diffusionless:sol:sigmar2}-\eqref{eq:pt-diff:diffusionless:sol:corrrp} (dashed lines), we find that the maxima in $\sigma_{\bar r}^2$ are at $\bar t=(12/\sigma_{\bar p,0}^2)^{1/3}$, which rewritten in terms of physical variables is
\begin{equation}
    t_2'
    =
    \left(
    \frac{12E^2}{\hat q\sigma_{p,0}^2}
    \right)^{\frac{1}{3}}\,,\label{eq:t2prime-def}
\end{equation}
and is similar to the time-scale $t_2$ introduced in Eq.\,\eqref{eq:t0-t2-def} if one replaces $\sigma_{p,0}$ with $\mu$ and recalls that $E=q^+/4$. Note, however, that since we remain agnostic in this section about the value of $\mu$, and hence of $\sigma_{\bar p,0}$, this time scale is not necessarily larger than $t_c$, particularly when $\sigma_{\bar p,0}\gg 1$. The maxima in $c_{\bar r,\bar p}$ is reached at a scale of the same order as $t_2'$ when $\sigma_{\bar p,0}\gg 1$.
The diffusionless solution for $\sigma_{\bar p}$ is monotonically decreasing and changes convexity exactly when $c_{\bar r,\bar p}$ reaches the maximum, again at a typical time set by $t_2'$. Upon introducing the characteristic angle in the initial wave-packet $\theta_\mu^2\equiv 2\sigma_{p,0}^2/E^2$, the time scale $t_2'$ can be expressed as
$t_2'\sim (\qhat \theta_\mu^2)^{-1/3}$.
Its form is therefore identical to that of $t_c$, modulo the replacement $\theta_{q\bar q}\to \theta_\mu$. For a Gaussian wave packet centred at vanishing initial relative transverse momentum with $\langle \ptb \rangle=0$ at $t=0$, $t_2'$ characterizes the time scale associated with the singlet-to-octet colour transition, as we shall observe below when computing the singlet survival probability.

 \begin{figure}[t]
    \centering
    \includegraphics[width=0.32\linewidth]{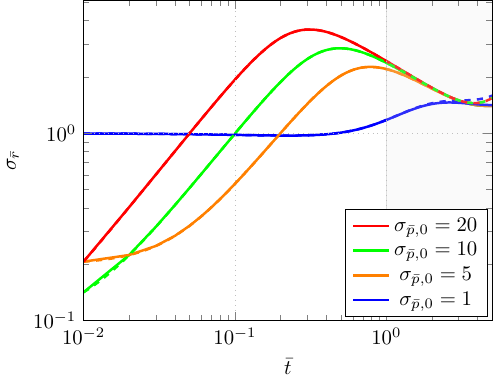}
    \includegraphics[width=0.32\linewidth]{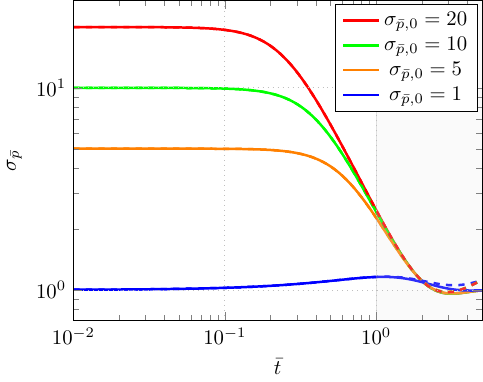}
    \includegraphics[width=0.32\linewidth]{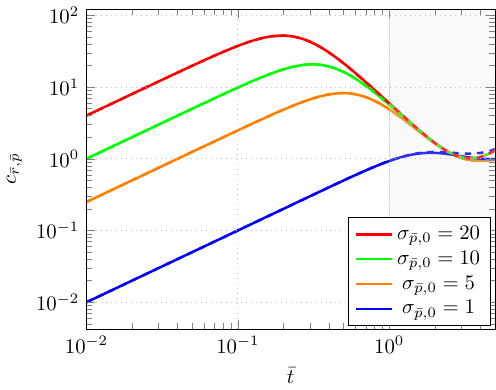}
    \caption{Same as Fig.\,\ref{fig:pt-diff:sigmapt2:numexact-vs-diffusionless}, but with the analytic curves (dashed lines) obtained from the first-order expansion in $\kappa^2$ of $\sigma_{\bar p}$, $\sigma_{\bar r}$, and $c_{\bar r,\bar p}$.}
    \label{fig:pt-diff:sigmapt2:numexact-vs-0+1}
\end{figure}

At the first order, that is $O(\kappa^2)$, the equations Eqs.\,\eqref{eq:diffusion-pt:first-moms:eq}-\eqref{eq:diffusion-pt:second-moms:eq}-\eqref{eq:diffusion-pt:second-moms:eq2}, whose variables have been expanded through the series in Eq.\,\eqref{eq:diffusion-pt:var-expansion-kappa}, read
\begin{align}
&\frac{\der \langle\rtb\rangle^{(1)}}{\der t} = 
\langle \ptb \rangle^{(1)}  - \frac{(\sigma_{\bar{r}}^2)^{(0)}\langle\rtb\rangle^{(1)}+(\sigma_{\bar{r}}^2)^{(1)}\langle\rtb\rangle^{(0)}}{2},\\
&\frac{\der \langle\ptb\rangle^{(1)}}{\der t} = 
- \frac{c_{\bar{r},\bar{p}}^{(0)}\langle \rtb \rangle^{(1)}+c_{\bar{r},\bar{p}}^{(1)}\langle \rtb \rangle^{(0)}}{2}  ,
\label{eq:diffusion-pt:first-moms:eq:1st-order}
\end{align}
and
\begin{align}
&\frac{\der  (\sigma_{\bar{r}}^2)^{(1)}}{\der \bar{t}}=  
2 c_{\bar{r},\bar{p}}^{(1)} - 
(\sigma_{\bar{r}}^2)^{(0)}(\sigma_{\bar{r}}^2)^{(1)} ,
\label{eq:diffusion-pt:second-moms:eq:1st-order:sigmar2}\\
&\frac{\der  (\sigma_{\bar{p}}^2)^{(1)}}{\der \bar{t}}=  
\frac{1}{2} - c_{\bar{r},\bar{p}}^{(0)}c_{\bar{r},\bar{p}}^{(1)},\label{eq:diffusion-pt:second-moms:eq:1st-order:sigmap2}\\
&\frac{\der  c_{\bar{r},\bar{p}}^{(1)}}{\der \bar{t}} =
 (\sigma_{\bar{p}}^2)^{(1)}- \frac{(\sigma_{\bar{r}}^2)^{(0)} c_{\bar{r},\bar{p}}^{(1)}+(\sigma_{\bar{r}}^2)^{(1)} c_{\bar{r},\bar{p}}^{(0)} }{2}.
 \label{eq:diffusion-pt:second-moms:eq:1st-order:corr}
 \end{align}
 Hence, by exploiting the $0$-th order solution obtained above, we can solve analytically this set of coupled --- now linear --- differential equations. The plot is shown in Figure~\ref{fig:pt-diff:sigmapt2:numexact-vs-0+1} and the analytical expressions are summarised in Appendix~\ref{app:pt-diff:pert-results:second-mom}.

\begin{figure}[t]
    \centering
    \includegraphics[width=0.32\linewidth]{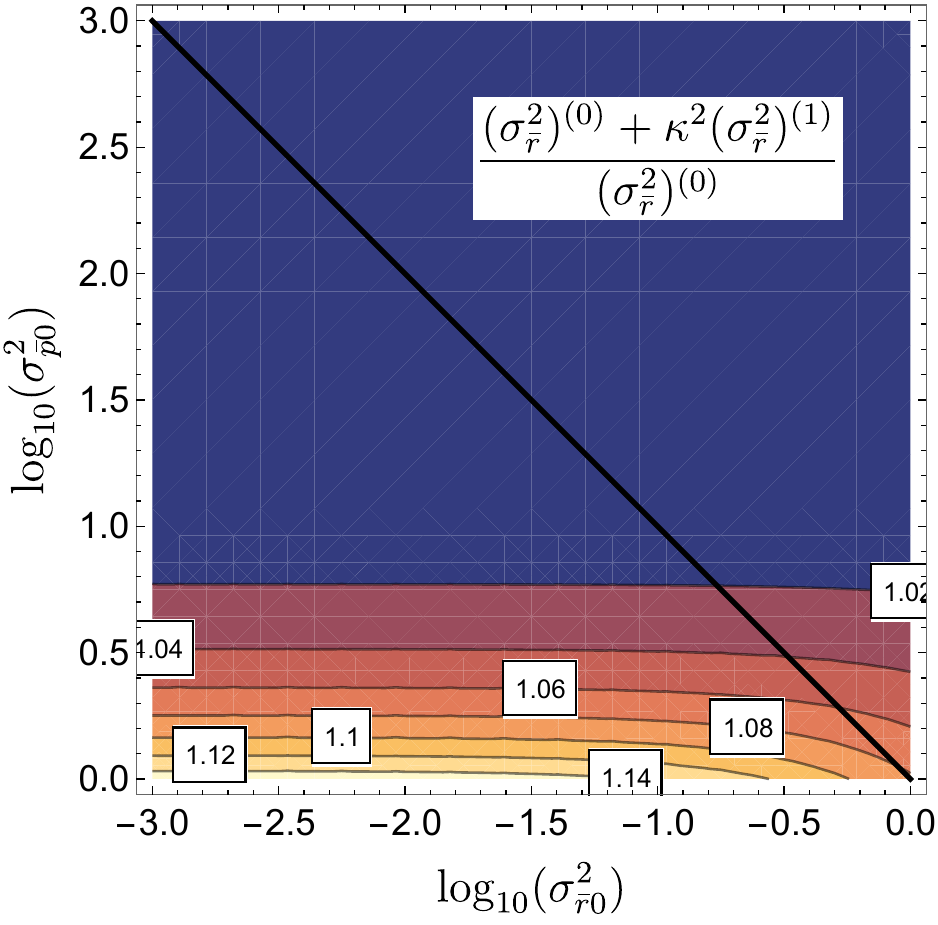}
    \includegraphics[width=0.32\linewidth]{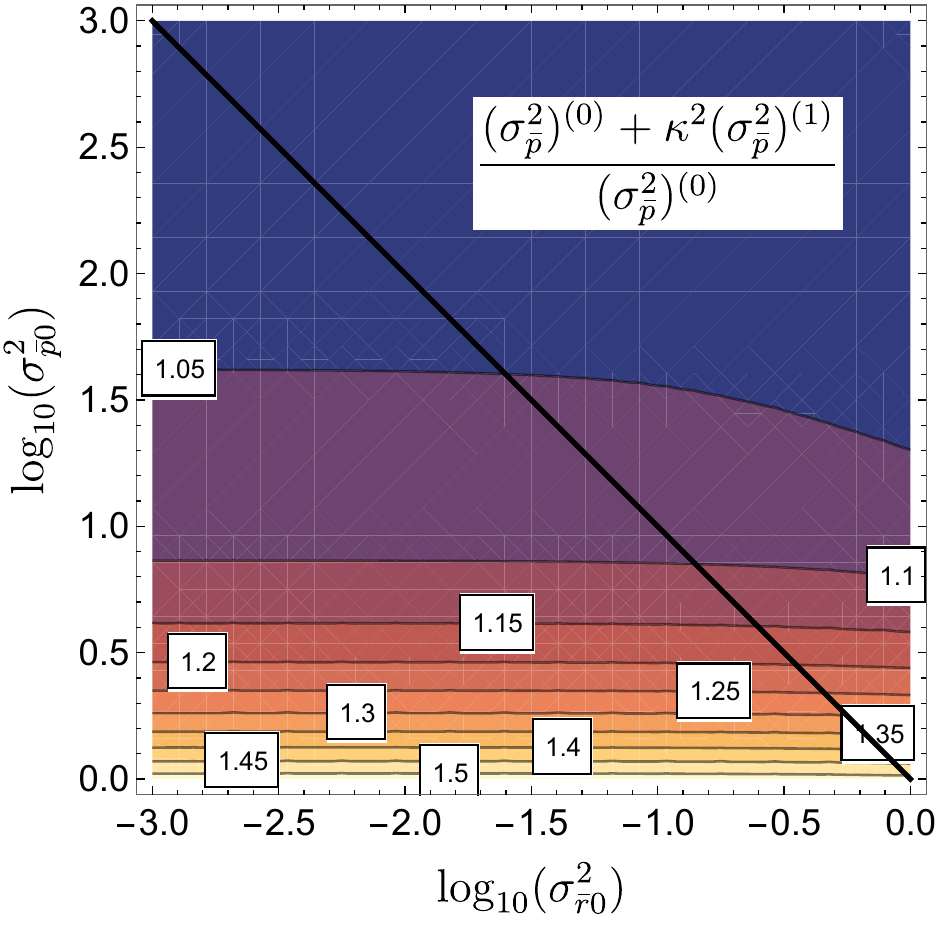}
    \includegraphics[width=0.32\linewidth]{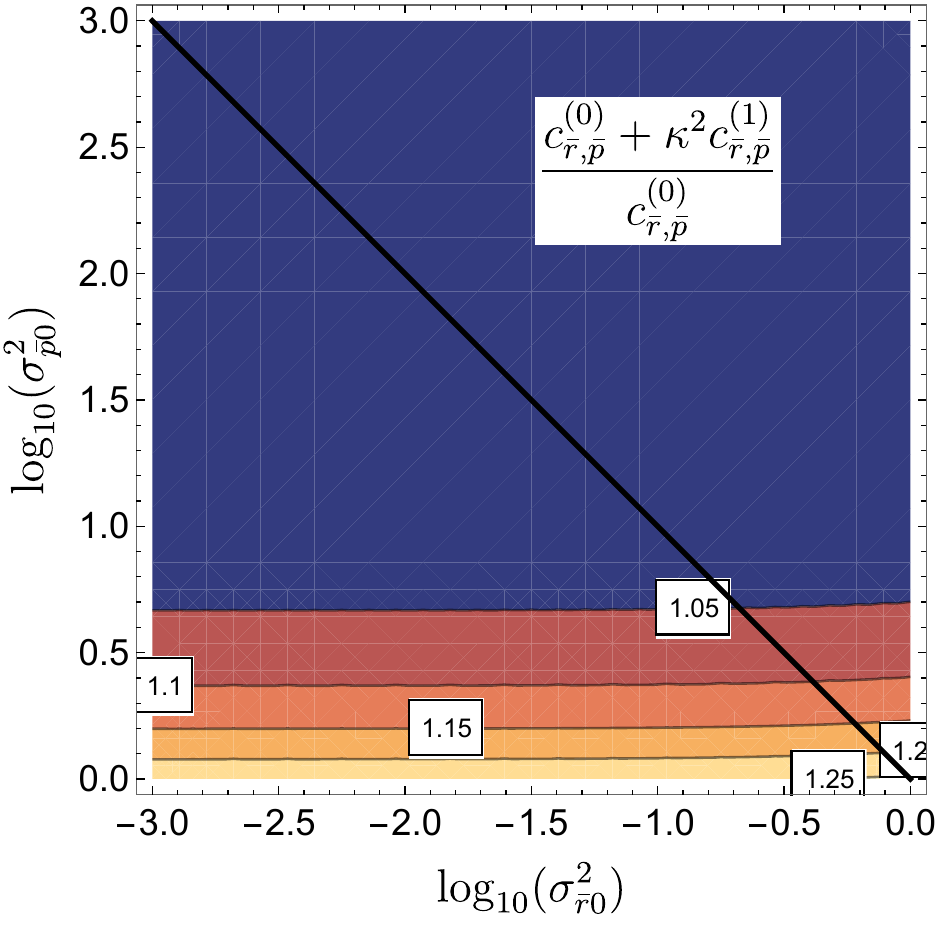}
    \caption{Final-time ($t=L$) values of the ratios between the first-order and zeroth-order perturbative results for the variance in $\rtb$ (left), the variance in $\ptb$ (middle), and the $\rtb$-$\ptb$ correlation coefficient (right), shown as functions of the initial widths of the Gaussian ansatz. The black line corresponds to the Heisenberg's uncertainty constraint $\sigma_{\bar r,0}\times\sigma_{\bar p,0}=\kappa=1$.}
    \label{fig:pt-diff:sigmapt2:first-order:contour}
\end{figure}

By looking at Figure~\ref{fig:pt-diff:sigmapt2:numexact-vs-0+1}, we observe a very good agreement of the perturbative solution with the exact (numerical) one for all the four initial conditions displayed. This shows the fast convergence and stability of the perturbative scheme we are employing to solve Eq.\,\eqref{eq:rhos-section-ptdiffusion}
These conditions have been chosen by varying the width in the rescaled momentum and by consequence varying also the width in the rescaled position of the initial condition, minimizing the uncertainty relation constraint with $\sigma_{\bar r,0}\sigma_{\bar p,0}=\kappa$.
However, it is not trivial to ensure that this dependence on $\kappa$ in the initial condition is not spoiling the perturbative expansion we are using to solve the equations.
Hence, we show in Figure~\ref{fig:pt-diff:sigmapt2:first-order:contour} the ratio $(y^{(0)}+\kappa^2 y^{(1)})/y^{(0)}$ for the second moments, i.e.~$y=\sigma_{\bar r}^2,\sigma_{\bar p}^2, c_{\bar r,\bar p}$, at final time $\bar t=1$ as a function of the variances in $\rtb$ and $\ptb$ of the initial condition. We observe that the values for these ratios vary smoothly and remain close to one for all the choices in a large range of initial widths, in particular in the upper right triangle corresponding to initial conditions which satisfy the Heisenberg uncertainty relation $\sigma_{\bar p}\sigma_{\bar r}\ge \kappa$. This plot demonstrates again the good convergence of the perturbative expansion, independently of the choice of physically reasonable Gaussian initial conditions.

\subsection{Singlet survival probability}

Following the same perturbative scheme of the previous section and applying it to the singlet survival probability Eq.\,\eqref{eq:diffusion-pt:zeroth-mom:eq}, we obtain the analytical solution up to the first order in $\kappa^2$.
At the leading order $O(\kappa^0)$, the equation reads
\begin{equation}
    \frac{\der P_s^{(0)}}{\der \bar{t}}=-\frac{\left(\langle \rtb \rangle^{(0)}\right)^2 +(\sigma_{\bar{r}}^2)^{(0)}}{2} P_s^{(0)}\,,
    \label{eq:pt-diff:Ps:zeroth-order:eq}
\end{equation}
and its solution is
\begin{equation}
    P_s^{(0)}(\bar t)
    =
    \frac{48}{D(\bar t)}
    \exp\left\{-\frac{\bar t \left[\bar{\boldsymbol{P}}_{0\perp}^2 \bar t^2 \left(\sigma_{\bar r,0}^2 \bar t+8\right)+24 \bar{\boldsymbol{P}}_{0\perp}
    \cdot
    \bar{\boldsymbol{r}}_{0\perp} \bar t+\bar{\boldsymbol{r}}_{0\perp}^2 \left(\sigma_{\bar p,0}^2 \bar t^3+24\right)\right]}{D(\bar t)}\right\}\,,
    \label{eq:pt-diff:Ps:sol:0th}
\end{equation}
with $D(\bar t)\equiv\sigma_{\bar p,0}^2\bar t^3( \sigma_{\bar r,0}^2\bar t+8)+24(\sigma_{\bar r,0}^2 \bar t+2)$. Here $\boldsymbol{P}_{0\perp}$ and $\boldsymbol{r}_{0\perp}$ denote the average relative transverse momentum $\langle \pt\rangle$ and relative transverse separation $\langle \rt\rangle $ at initial time, respectively.

For $\bar{\boldsymbol{r}}_{0\perp}=0$, the solution becomes
\begin{equation}
    P_s^{(0)}(\bar t)
    \big|_{\bar{\boldsymbol{r}}_{0\perp}=0}
    =
    \frac{48}{D(\bar t)}
    \exp\left\{-\frac{\bar{\boldsymbol{P}}_{0\perp}^2 \bar t^3 \left(\sigma_{\bar r,0}^2 \bar t+8\right)
    }{D(\bar t)}\right\}\,,
    \label{eq:pt-diff:Ps:sol:0th:r0eq0}
\end{equation}
and shows manifestly the dependence on the ratio $\theta_{q\bar q}/\theta_c$, through the rescaled initial relative transverse momentum $\bar{\boldsymbol{P}}_{0\perp}^2=\hat qL^3\boldsymbol{P}_{0\perp}^2/E^2=2\theta_{q\bar q}^2/\theta_c^2$. It also reproduces Eq.\,\eqref{eq:singlet-to-octet-rate} when
$\sigma_{\bar p,0}$, $\sigma_{\bar r,0}\to0$, provided one identifies the Wigner variable $\pt$ in that equation with the mean value $\boldsymbol{P}_{0,\perp}$.
As expected, in this ``classical" limit having both a vanishing separation $\bar{\boldsymbol{r}}_{0\perp}$ and vanishing relative transverse momentum $\bar{\boldsymbol{P}}_{0\perp}$ implies $P_s^{(0)}(\bar t)=1$, meaning that the dipole cannot be resolved and stays in the singlet representation.

\begin{figure}[t]
    \centering
    \begin{subfigure}{0.49\textwidth}
        \includegraphics[width=\linewidth]
        {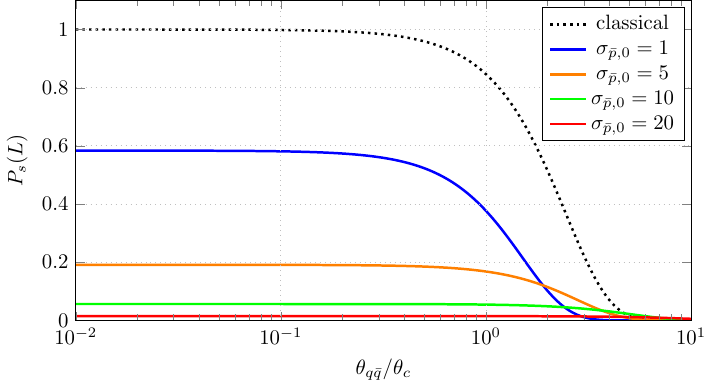}
    \end{subfigure}
    \hfill
    \begin{subfigure}{0.49\textwidth}
        \includegraphics[width=\linewidth]{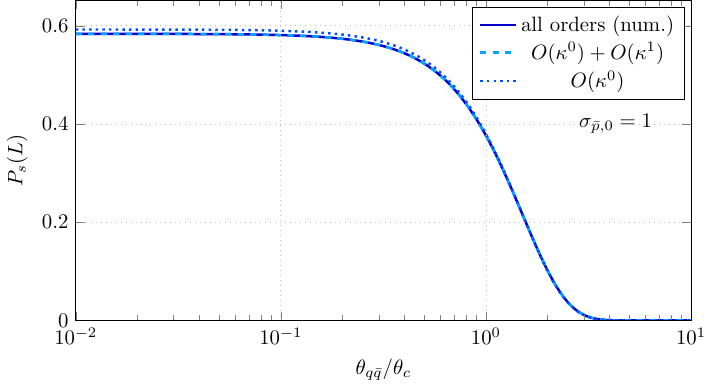}
    \end{subfigure}
    \caption{
    Singlet survival probability as function of $\theta_{q\bar q}/\theta_c$ at final time $t=L$. In both figures, we use a vanishing initial separation $\bar{\boldsymbol{r}}_{\perp,0}=0$ in the Gaussian ansatz. The numerical solution of Eq.\,\eqref{eq:diffusion-pt:zeroth-mom:eq} at all orders in $\kappa^2$ (left) shows a damped probability for different values of the initial width $\sigma_{\bar p,0}$ (coloured lines) with respect to the \textit{classical} value (black dotted line), obtained using $\sigma_{\bar p,0}=\sigma_{\bar r,0}=0$ and $\kappa=0$. Indeed, the momentum uncertainty allows singlet-to-octet transitions even for narrow dipoles ($\theta_{q\bar q}\ll\theta_c$).
    The perturbative scheme for the analytical solutions converges quickly for the reference value of $\sigma_{\bar p,0}=1$ (right).
    The dotted line is the analytical solution for the diffusionless equation ($\kappa=0$ in Eq.\,\eqref{eq:rhos-section-ptdiffusion}).
    The dashed line is the analytical solution to first order in $\kappa^2$.
    The plain line shows the same numerical non-perturbative solution (all orders in $\kappa^2$) shown in the left figure.
    }
    \label{fig:pt-diff:Ps:0-1-cl}
\end{figure}

\begin{figure}[t]
    \centering
    \begin{subfigure}{0.49\textwidth}
        \includegraphics[width=\linewidth]{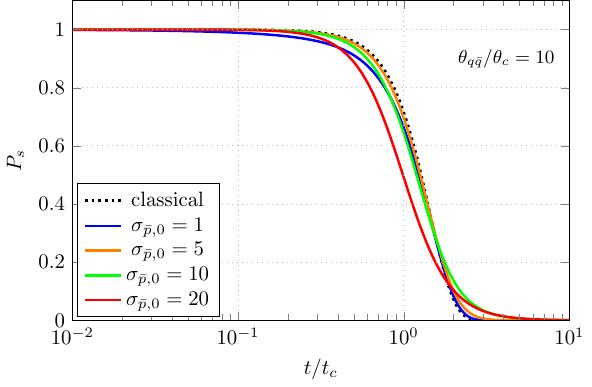}
    \end{subfigure}
    \hfill
    \begin{subfigure}{0.49\textwidth}
        \includegraphics[width=\linewidth]{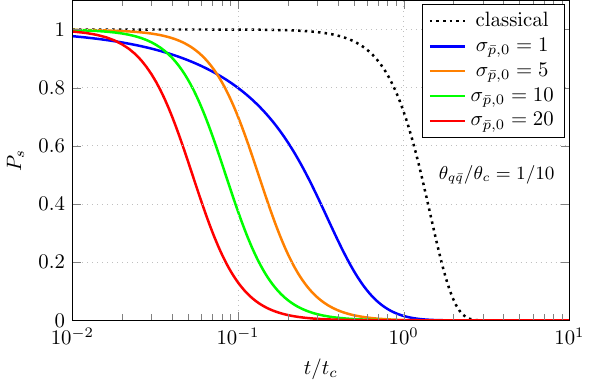}
    \end{subfigure}
    \caption{
   Singlet survival probability as a function of $t/t_c$ for $\theta_{q\bar q}/\theta_c=10$ (left) and $\theta_{q\bar q}/\theta_c=1/10$ (right).
     We use a vanishing initial separation $\bar{\boldsymbol{r}}_{\perp,0}=0$ in the Gaussian ansatz. As in Fig.\,\ref{fig:pt-diff:Ps:0-1-cl}, the black dotted line shows the \textit{classical} value, obtained using $\sigma_{\bar p,0}=\sigma_{\bar r,0}=0$ and $\kappa=0$.
    The coloured lines show the numerical non-perturbative solutions (all orders in $\kappa^2$) of Eqs.\,\eqref{eq:diffusion-pt:first-moms:eq}-\eqref{eq:diffusion-pt:second-moms:eq} for different values of the initial width $\sigma_{\bar p,0}$.
    The characteristic time scale for the wide opening angle case (left) is set by $t_c$, which depends on the mean value of the relative transverse momentum distribution, while for the narrow opening angle case (left) the timescale is set by the widths of the distribution.
    }
    \label{fig:pt-diff:Ps:0-1-cl-ttc}
\end{figure}

The first order corrections, at $O(\kappa^2)$, are obtained from the equations
\begin{equation}
    \frac{\der P_s^{(1)}}{\der \bar{t}}=-\frac{\left(\langle \rtb \rangle^{(1)}\right)^2 +(\sigma_{\bar{r}}^2)^{(1)}}{2} P_s^{(0)}
    -\frac{\left(\langle \rtb \rangle^{(0)}\right)^2 +(\sigma_{\bar{r}}^2)^{(0)}}{2} P_s^{(1)}.
    \label{eq:pt-diff:Ps:first-order:eq}
\end{equation}
We report the analytical solution to the differential equation Eq.\,\eqref{eq:pt-diff:Ps:first-order:eq} in Appendix~\ref{app:pt-diff:pert-results:Ps} and show the result in Figures~\ref{fig:pt-diff:Ps:0-1-cl} and~\ref{fig:pt-diff:Ps:0-1-cl-ttc} as a function of $\theta_{q\bar q}/\theta_c$ and $t/t_c$ respectively in order to compare the classical result corresponding to both $\sigma_{\bar p,0}=\sigma_{\bar r,0}=0$ in the initial condition and the full (exact) numerical solution --- with different initial widths, and confirm the convergent behaviour of the perturbative series. As expected from the smallness of the $O(\kappa^2)$ corrections to the physical parameters of the Gaussian ansatz, the agreement between the numerical result to all orders in $\kappa^2$ and the zeroth or first order truncation of the perturbative series remains very good, even for the choice $\kappa=1$, as shown in Fig.\,\ref{fig:pt-diff:Ps:0-1-cl} (right).

In Fig.\,\ref{fig:pt-diff:Ps:0-1-cl-ttc}, we show the singlet survival probability as a function of $t/t_c$ for two values of the dipole opening angle, $\theta_{q\bar q}/\theta_c=10$ and $1/10$ respectively. The dotted curve, corresponding to Eq.\,\eqref{eq:singlet-to-octet-rate}, neglects the initial spread of the wave packet in relative transverse momentum. In contrast, the coloured curves account for this effect in the specific case of a Gaussian wave packet. For $\theta_{q\bar q}/\theta_c=10\sim \bar P_{0\perp}$, such that $t_c\lesssim t_2'$ for all values of $\sigma_{p,0}$ since $t_c/t_2'\sim (\sigma_{\bar p,0}/\bar P_{0\perp})$, the characteristic time scale for the singlet-to-octet transition remains of order $t_c$. Nevertheless, accounting for the quantum nature of the initial condition softens the transition, while approximatively preserving this $t/t_c$ scaling. By contrast, for $\theta_{q\bar q}/\theta_c=1/10$, such that $t_2'\lesssim t_c$ for all values of $\sigma_{p,0}$, the characteristic time scale governing colour decoherence and the singlet-to-octet transition is instead set by $t_2'$: if plotted in terms of $t/t_2'$, we would observe a scaling of all coloured curves in term of this new variable. This behaviour originates from the large initial width of the dipole in relative momentum space  and, equivalently, in opening angle, so that even if the initial \textit{mean} opening angle is smaller than $\theta_c$, there is still the possibility for the dipole to be in a kinematic configuration where its opening angle is larger than $\theta_c$ thanks to the large width of the distribution.

Finally, we comment on the perturbative treatment of the full quantum master equation in the large $N_c$ limit, including the octet component. 
Thanks to the triangularity of the master equation in the large $N_c$ limit, Eq.\,\eqref{eq:Wigner-ho-lNc-final}, we have been able to solve --- perturbatively --- the closed equation for the singlet component. 
On the contrary, the master equation for the octet component\footnote{Here we have integrated over the center-of-mass degrees of freedom, but the following discussion remains valid also in the case of the full equation.}
\begin{align}
    \frac{\partial \tilde\rho_{R,o}}{\partial \bar t}+
        \ptb\cdot\nabla_{\rtb}
        \tilde\rho_{R,o}
    =
    \frac{\kappa^2}{8}\nabla^2_{\ptb}
    \tilde\rho_{R,o}
    +
    \frac{1}{2}\left(
        \rtb^2+\frac{\kappa^2}{4}\nabla^2_{\ptb}
    \right)\rho_{R,s}\,,
    \label{eq:pt-diff:meq:octet:red}
\end{align}
contains an inhomogeneous term sourced by the singlet component. This inhomogeneous term accounts for octet-to-singlet transitions induced by the relative-momentum diffusion operator, which are formally suppressed by $\kappa^2$ and are briefly discussed at the end of subsection~\ref{sub:ho-approx}. This mechanism now competes with the dominant octet-to-singlet transition mechanism associated with the increasing classical size of the colour dipole, represented by the $\rt^2$ term inside the parenthesis.

Hence, the solution of Eq.\,\eqref{eq:pt-diff:meq:octet:red} can be written as
\begin{equation}
    \tilde\rho_{R,o}(\ptb,\rtb,\bar t)
    =
    \int
        \der^2\ptb'\der^2\rtb'
        \der \bar t'
        \mathcal{G}(\ptb-\ptb',\rtb-\rtb';\bar t-\bar t')
        j(\ptb',\rtb';\bar t')\, ,
\end{equation}
where the Green function is the usual Klein-Kramers propagator, which coincides with the exact solution of the homogenous part of the equation with a classical initial state $\delta^{(2)}(\pt-\boldsymbol{p}_{\perp0})\delta^{(2)}(\rt-\boldsymbol{r}_{\perp0})$. The source $j$ can be determined from the perturbative results obtained previously, by computing
\begin{equation}
    j(\ptb,\rtb;\bar t)
    =
    \frac{1}{2}\left(
        \rtb^2+\frac{\kappa^2}{4}\nabla^2_{\ptb}
    \right)\rho_{R,s}(\ptb,\rtb;\bar t)\, ,
\end{equation}
with $\rho_{R,s}(\ptb,\rtb;t)$ given by the Gaussian ansatz with the physical parameters obtained up to $O(\kappa^2)$.
This means that we have obtained a rigorous perturbative solution to the full propagation equation for the large $N_c$ case.

\subsection{Decoherence in the relative transverse momentum basis}
\label{sub:decoherence-pt}

We finally turn to the discussion of the quantum decoherence of the density matrix in the relative transverse momentum basis $\rho_R(\boldsymbol{p}_\perp,\boldsymbol{\mathcal{P}}_\perp)$, i.e.~as a function of the relative transverse momentum variables $\pt$ and $\Pcalt$, similar to what we have done in the paragraph~\ref{subsub:decoherence-qt} for the transverse momentum imbalance variables. The density matrix is obtained from the Wigner distribution by a 2-dimensional Fourier transform of the Gaussian ansatz written in Eq.\,\eqref{eq:Gaussian-ansatz}:
\begin{align}
    \rho_{R,s}(\boldsymbol{p}_\perp,\boldsymbol{\mathcal{P}}_\perp; \bar t)
    &=\int\der^2\rt e^{i\Pcalt\cdot\rt} \rho_R(\pt,\rt;\bar t)\,,\\
    &=\frac{1}{\pi\alpha Q_s^2}e^{-\delta-\beta\ptb^2+\boldsymbol{\epsilon}_{\perp}\cdot\ptb }e^{i\frac{\Pcalt}{Q_s}\cdot\left(\frac{\gamma\ptb+\boldsymbol{\zeta}_{\perp}}{2\alpha}\right)}e^{\frac{(\gamma\ptb+\boldsymbol{\zeta}_{\perp})^2}{4\alpha}}\exp\left(-\frac{\Pcalt^2}{4\alpha Q_s^2}\right)\,.
\end{align}
Hence, the width of the $\Pcalt$ distribution in the singlet component of the density matrix scales like
\begin{align}
    4\alpha(\bar t) Q_s^2\approx 4\mu^2+2\qhat t\,.
\end{align}
and thus grows with time! The coherence in the relative transverse momentum basis for the singlet component of the density matrix increases over time ; yet one should keep in mind that the overall singlet component is decaying into the octet state because of the overall $e^{-\delta}$ suppression factor. After a time of order $t_c$, the dipole is in a colour octet state anyway. We thus need to estimate the typical time scales governing the coherence of the octet component of the density matrix.

In fact, for a colour-octet initial state, the solution can be obtained to all orders in $\kappa^2$ in the large $N_c$ limit. Indeed, if no colour-singlet component is present at the initial time, none will be generated during the evolution, since transitions from octet to singlet are forbidden in the large $N_c$ limit. As a result, the equation for $\tilde \rho_{R,o}$ becomes closed and can be solved analytically for arbitrary initial conditions in the quantum-variable space $(\Rcalt,\Pcalt)$:
\begin{align}
    \tilde\rho_{R,o}(\Rcaltb,\Pcaltb;\bar t)&=\exp\left(-\frac{\kappa^2}{8}\int_0^{\bar t}\der s \left[\Rcaltb-\Pcaltb s\right]^2\right)\times\tilde \rho_{R,o}\left(\Rcaltb-\Pcaltb \bar t,\Pcaltb;0\right)\,,\label{eq:rhooctet-full solution}
\end{align}
Note that the lowest order in $\kappa^2$, which amounts to remove the exponential factor, reproduces the result of Eq.\,\eqref{eq:rhoo-oqs-largeNc} for the Wigner function in the $(\rt,\pt)$ phase space when there is no singlet state at $t=0$. In terms of the dimension-full momentum variables, we have
\begin{align}
        \tilde\rho_{R,o}(\Rcalt,\Pcalt;t)&=\exp\left(-\frac{\qhat}{8}\int_0^t\der s \ \left[\Rcalt-\frac{\Pcalt}{E}s\right]^2\right)\tilde\rho_{R,o}\left(\Rcalt-\frac{\Pcalt}{E}t,\Pcalt;0\right)\,.
\end{align}
Taking the Fourier transform of the above expression from $\Rcalt$ to $\pt$ we have for the density matrix in the relative transverse momentum basis:
\begin{align}
    \rho_{R,o}(\pt,\Pcalt;t)&=\frac{2}{\pi(\qhat t+8\mu^2)}e^{i\frac{t(16\mu^2+\qhat t)}{2E(\qhat t+8\mu^2)}\pt\cdot\Pcalt}e^{-\frac{2\pt^2}{\qhat t +8\mu^2}}\nonumber\\
    &\times \exp\left(-\Pcalt^2\left[\frac{1}{4\mu^2}+\frac{\qhat t^3(\qhat t+32\mu^2)}{96E^2(\qhat t+8\mu^2)}\right]\right)\,,
\end{align}
which displays the same characteristic time scales as $t_2'$ defined in Eq.\,\eqref{eq:t2prime-def}. This is also the same time scale (modulo the replacement $\Lambda\to \mu$) as the ones appearing in the transverse momentum imbalance part of the density matrix discussed in Section~\ref{subsub:decoherence-qt} and given by Eq.\,\eqref{eq:t0-t2-def}. The $\pt$-diffusion term suppresses coherence in $\Pcalt$ of the density matrix on a time scale of order $t_2'$. This is also the characteristic time scale for the singlet-to-octet colour transition when $\theta_{q\bar q}\lesssim \theta_{\mu}\sim \sigma_{p,0}/E$ (see Fig.~\ref{fig:pt-diff:Ps:0-1-cl-ttc}-right). In this regime, kinematic decoherence in $\Pcalt$ and colour decoherence occur on comparable time scales. Conversely, when $t_c\lesssim t_2'$, the colour transition takes place before $\Pcalt$ decoherence sets in, as discussed in Section~\ref{subsub:decoherence-qt} for the case of the total transverse momentum.

\section{Summary and outlook}

In this paper, we have revisited the phenomenon of decoherence in jet-quenching physics using modern open quantum system techniques to describe the propagation of a colour dipole (the subsystem) through a dense quark-gluon plasma (the environment). Although the coherence time and the associated coherence angle have long been identified and understood through cross-section calculations using Feynman-diagram techniques, their interpretation lacked a firm theoretical foundation within the language of non-relativistic quantum mechanics. While Ref.~\cite{Barata:2023uoi} provided an important first step in this direction, it focused on the propagation of a single colour charge, whereas the notions of decoherence time and coherence angle are intrinsically tied to a colour dipole subsystem (or any subsystem containing at least two colour charges propagating through a dense quark-gluon plasma).

Our first main result has therefore been the derivation of the Lindblad equation for a colour dipole propagating through a quark-gluon plasma in thermal equilibrium and transversely homogeneous, following the seminal work of~\cite{Blaizot:2017ypk} in the context of quarkonia. Compared to the quarkonium case, the operator in the Lindblad equation proportional to the square of the relative separation of the $q\bar q$ pair plays a major role in driving colour decoherence and transitions. This operator couples the two possible colour states of the subsystem (singlet and octet), rendering the equation analytically challenging, as it belongs to the class of multispecies Fokker-Planck equations. To make analytical progress, we specialized to a phenomenologically relevant kinematic regime in which the quark-antiquark pair with relative and total transverse momentum $\pt$ and $\qt$ is highly energetic and strongly boosted, $E\gg p_\perp$, forms a small opening angle, and satisfies the hierarchy $p_\perp\gg q_\perp\sim Q_s$.

Within this regime, we obtained explicit analytic results for the Wigner distribution in the center-of-mass phase space and for the density matrix in the transverse-momentum imbalance basis. Despite the strong scale hierarchy $E\gg p_\perp\gg q_\perp\sim Q_s$, we find that the $\qt$-distribution retains a dependence on the hard scale $p_\perp$, through the ratio $\theta_{q\bar q}/\theta_c$ where $\theta_c=2(\qhat L^3)^{-1/2}$ is the critical angle of the medium. Borrowing the terminology of transverse-momentum-dependent factorization, this behaviour can be interpreted as a medium-induced violation of factorization between two widely separated scales $\qt$ and $\pt$.
The expressions for the $\qt$-distribution of the $q\bar q$ pair and its second moment derived in section~\ref{sec:factorisation-violation} are, to our knowledge, new. They provide simple quantitative input for phenomenological studies and for Monte Carlo implementations of in-medium parton showers formulated in terms of dipole cascades. It can also be used to consistently account for colour coherence effects in the evolution equation for medium-induced minijets, not only in the radiative emission kernel as in~\cite{Barata:2021byj} but also in the transverse broadening kernel. 

Beyond these phenomenological applications, the open quantum system framework offers important conceptual insights. In particular, we find that the $\theta_{q\bar q}/\theta_c$ 
dependence of the $\qt$-distribution --- although involving the 
so-called ``coherence angle'' --- admits a more natural physical 
interpretation in terms of colour transitions between 
singlet and octet states, rather than in terms of quantum 
colour decoherence.
In this context, the coherence time $t_c$ corresponds to the characteristic decay time of the singlet component into the octet level. At the same time, by explicitly analysing the density matrix in colour space, we have demonstrated that the same time scale $t_c$ also governs the loss of coherence in the colour sector as the density matrix becomes diagonal in colour space in any basis after a time of order of $t_c$. By contrast, decoherence in the transverse-momentum basis --- whether for the relative transverse momentum or for the transverse-momentum imbalance --- is controlled by a different time scale, $t_2\sim (E^2/(\qhat \Lambda^2))^{1/3}$, as also observed in~\cite{Barata:2023uoi}. This scale depends on the properties of the initial condition, in particular on its momentum-space dispersion $\Lambda$. The fact that the same time scale governs kinematic decoherence for both a single quark and a $q\bar q$ subsystem highlights that kinematic decoherence is fundamentally distinct from colour decoherence, the latter requiring the notion of compact versus non-compact colour objects.

In the last part of this work, we have systematically investigated power corrections in the ratio $Q_s/p_\perp$ to the Wigner function, focusing in particular on the phase-space integral of the singlet component, which determines the singlet-to-octet transition probability. For a Gaussian initial condition --- analytically convenient, though not necessarily of direct phenomenological relevance --- we found these corrections to be small in the high-energy regime. This suggests that the $\pt$-diffusion Lindblad operator, responsible for $Q_s/p_\perp$ corrections, may have a limited impact on phenomenological applications to high-energy jets. While the linearity of the quantum master equation suggests that this conclusion may extend beyond Gaussian initial wave packets to more general initial conditions that can be expressed as superpositions of Gaussians with widths within the range explored in this work, establishing its robustness for more general initial wave packets remains a subject for further investigation.

This work can be improved in several respects. In particular, some of the approximations adopted here could be relaxed, most notably the harmonic approximation for the scattering potential. One possible direction would be to follow the improved opacity expansion and construct a perturbative scheme around the harmonic oscillator potential considered in this paper. Alternatively, one could employ a more realistic scattering potential --- incorporating the most recent theoretical developments in its determination (see e.g.~\cite{Caron-Huot:2008zna,Moore:2019lgw,Panero:2013pla,Schlichting:2021idr,Moore:2021jwe,Ghiglieri:2022gyv}) as well as a dependence upon the dipole's impact parameter in order to account for medium transverse inhomogeneities --- and solve the resulting quantum master equation numerically. There is also scope for improving the modelling of the initial density operator, in particular regarding the choice of parameters entering the initial condition of the master equation. Its functional form and these  parameters should be grounded in QCD and more consistently connected to the properties of the physical process under consideration. Finally, it would be interesting to investigate, possibly numerically, the solution of the general quantum master equation without restricting to the longitudinally symmetric $q\bar q$ configuration with $z_1=z_2$.

A natural next step is to include gluon radiation. This extension would require working in Fock space and describing a quantum system with a non-conserved particle number. We leave such developments for future work. Nevertheless, the present study constitutes an initial milestone in this direction and should provide useful guidance for setting up the calculation and identifying suitable approximations.

Finally, another promising avenue is to apply the formulas derived here to phenomenology at the LHC (and possibly at RHIC as well, although our analysis has focused on high-energy jets). We plan to carry out a dedicated phenomenological study of azimuthal decorrelations within electroweak jets or flavour-tagged QCD jets, using modern jet-substructure techniques~\cite{Marzani:2019hun,Andrews:2018jcm} or intrajet multiplicity observables as in~\cite{Arleo:2009yu}. Related ideas have recently been explored in the context of boosted top decays, where azimuthal correlations are used to probe the polarization state of the $W$ boson produced in the decay~\cite{Yu:2021zmw} or to probe the formation time dependence of jet quenching~\cite{Apolinario:2017sob}. In our case, the objective would be to access the critical angle $\theta_c$ by measuring the azimuthal decorrelation between the two hardest subjets of a jet~\cite{Chien:2016led}, as a function of their relative opening angle. In addition to the decay of a colour-singlet object, one may also consider azimuthal decorrelation in high-energy quark and gluon decay inside hadronic jets, which are more abundantly produced in heavy-ion collisions, albeit more challenging to control in terms of flavour. As discussed in Section~\ref{sec:lindblad}, addressing this problem would require extending the present analysis to include antennas in higher-dimensional colour representations (following for instance~\cite{Cougoulic:2017ust}). This is also left for future work.

Overall, this work provides a further proof of principle of the usefulness of standard non-relativistic quantum-mechanical tools in jet-quenching physics. It should therefore serve as an analytic benchmark for future numerical studies based on the non-relativistic quantum formalism, whether relying on conventional approaches~\cite{Andres:2020vxs,Li:2021zaw,Li:2023jeh} or on quantum simulations~\cite{DeJong:2020riy,Barata:2022wim,Barata:2023clv,Qian:2024gph,Barata:2026icn} of high-energy partons propagating through a quark-gluon plasma.

\begin{acknowledgments}
 We are grateful to Jo\~ao Barata, Jean-Paul Blaizot, Miguel Escobedo, Jacopo Ghiglieri, Marco Leit\~ao, Aoumeur Daddi Hammou, Stéphane Peigné and Yacine Mehtar-Tani for inspiring discussions. We acknowledge support from the IN2P3 master project QHAT.
 \end{acknowledgments}

\appendix

\section{Feynman graph calculation of the timelike $\gamma^*\to q\bar q$ splitting}
\label{app:feynman-TMB}

In this appendix, we re-derive the expressions presented in the main text for the $q_\perp$-distribution of the $q\bar q$ pair in the boosted and back-to-back limits, using the conventional Feynman diagram approach to compute the parton-level cross section at high energy. Our objectives are twofold: (i) to cross-check the results obtained within the open-quantum-system formalism, thereby confirming the validity of the quantum master equation studied in this paper, and (ii) to examine the production mechanism of the colour dipole, as we will consider the decay of a highly virtual photon. We recall that the production mechanism of the $q\bar q$ pair is not incorporated into the Lindblad equation we solve and is therefore only modelled through our choice of initial condition.

\subsection{Back-to-back limit of the boosted $\gamma^*\to q\bar q$ amplitude}

In this appendix, we compute the amplitude for an in-medium $\gamma^* \to q\bar q$ splitting, focusing on the kinematic regime discussed in section~\ref{sec:kinematics}.
The calculation is similar to the one performed in~\cite{Dominguez:2019ges}, subsequently improved in~\cite{Isaksen:2023nlr} where the eikonal approximation for the produced partons is relaxed (see also~\cite{Blaizot:2012fh,Andres:2026qrt} for the extension to other channels). Indeed, since both $k_1^+$ and $k_2^+$ are assumed to be large, one can take the semi-classical limit for the propagation of the outgoing $q\bar q$ pair: the in-medium quark propagator can be replaced by Wilson lines along the classical path of the particles. However, our calculation differs from the aforementioned references since, in these papers, the transverse momentum imbalance of the $q\bar q$ pair is integrated out, which is precisely what we want to compute here.

We note $q^\mu$ the four-momentum of the incoming photon 
with $q^\mu=(q^+,0,\boldsymbol{0}_\perp)$ and created at light-cone time $t=0$. For the sake of simplicity, we work in a frame where the transverse momentum of the incoming photon vanishes. We note that the incoming photon state is not an asymptotic one, since the photon is produced at $t=0$, such that it can produce the $q\bar q$ pair even in the absence of medium interaction. Applying the effective QCD Feynman rules in the presence of a dense QCD medium represented by a classical gauge field $\mathcal{A}^-$, the amplitude for the $\gamma^* \to q\bar q$ process inside the medium reads
\begin{align}
    \mathcal{M}_{\gamma^* \to q\bar q }&=\frac{1}{q^+}\int_0^L\der t \int\frac{\der^2\boldsymbol{p}_{1\perp}}{(2\pi)^2}\frac{\der^2\boldsymbol{p}_{2\perp}}{(2\pi)^2}\ \mathcal{G}_q(\ktone,L;\boldsymbol{p}_{1\perp},t|k_1^+)\mathcal{G}_{\bar q}(\kttwo,L;\boldsymbol{p}_{2\perp},t|k_2^+) \nonumber\\
    &\times e \gamma^{s\bar s}_\lambda(z_1,z_2)(z_2\boldsymbol{p}_{1\perp}-z\boldsymbol{p}_{1\perp})\cdot\et^{\lambda} \ \mathcal{G}_\gamma(\boldsymbol{p}_{1\perp}+\boldsymbol{p}_{2\perp},t; \boldsymbol{0}_\perp,0|q^+)\,\label{eq:Mgqq-1}
\end{align}
where $ \mathcal{G}_q(\ktone,L;\boldsymbol{p}_{1\perp},t|k_1^+)$ refers to the momentum space propagator of the quark with conserved light cone energy $k_1^+$, initial and final transverse momenta $\boldsymbol{p}_{1\perp}$ et $\ktone$ between light cone time $t$ and $L$ (and likewise for the antiquark $\mathcal{G}^\dagger_{\bar q}$ and the photon $\mathcal{G}_\gamma$). In the semi-classical limit (see e.g.~\cite{Dominguez:2019ges}), we have
\begin{align}
    \mathcal{G}_q(\ktone,L;\boldsymbol{p}_{1\perp},t|k_1^+)\approx e^{-i\frac{\ktone^2}{2k_1^+}(L-t)}\int\der^2\xt e^{-i(\ktone-\boldsymbol{p}_{1\perp})\cdot\xt} V_q(L,t; [\boldsymbol{r}_{q,\rm cl}(s)])\,,
\end{align}
with $V_q$ a light like Wilson line along the classical trajectory $\boldsymbol{r}_{q,\rm cl}$ between $t$ and $L$ of the eikonal quark
\begin{align}
    V_q(L,t; [\boldsymbol{r}_{q,\rm cl}(s)])&=\mathcal{P} \exp\left[ig\int_t^L \der s \ \mathcal{A}^-(s,\boldsymbol{r}_{q,\rm cl}(s))\right]\,,\\\boldsymbol{r}_{q,\rm cl}(s)&=\boldsymbol{x}_\perp +(s-t)\frac{\ktone}{k_1^+}\,.
\end{align}
For the antiquark, $V$ is replaced by $V^\dagger$ and of course, for the propagator of the virtual photon, with have $V(t,0)=1$ since the photon does not couple to the gauge field. In Eq.\,\eqref{eq:Mgqq-1}, the quantity $\gamma_\lambda^{s\bar s}(z_1,z_2)$ encodes the spin-helicity structure of the $\gamma^* \to q\bar q$ vertex and reads
\begin{align}
    \gamma_\lambda^{s\bar s}(z_1,z_2)&=\frac{i}{\sqrt{z(1-z)}}\delta^{s,-\bar s}\left[z_1\delta^{\lambda s}-z_2\delta^{\lambda,- s}\right]\,.
\end{align}
Replacing the various factors in Eq.\,\eqref{eq:Mgqq-1} and using $k^+=k_1^+ + k_2^+=q^+$, we get
\begin{align}
    \mathcal{M}_{\gamma^* \to q\bar q }&=\frac{e}{q^+}\int_0^L\der t\int\frac{\der^2\boldsymbol{p}_{1\perp}\der^2\boldsymbol{p}_{2\perp}}{(2\pi)^2}\int\der^2\xt\der^2\xt' e^{-i(\ktone-\boldsymbol{p}_{1\perp})\cdot\xt-i(\kttwo-\boldsymbol{p}_{2\perp})\cdot\xt'}\nonumber\\
  &  \times e^{-i\frac{\ktone^2}{2z_1q^+}(L-t)-i\frac{\kttwo^2}{2z_2q^+}(L-t)} (V_q(L,t; [\boldsymbol{r}_{q,\rm cl}(s)]) V_{\bar q}^\dagger(L,t; [\boldsymbol{r}_{\bar q,\rm cl}(s)]))_{\beta\gamma}\gamma_\lambda^{s\bar s}(z,1-z)\nonumber\\
  &\times(z_2\boldsymbol{p}_{1\perp}-z_1\boldsymbol{p}_{2\perp})\cdot\et^{\lambda} \delta(\boldsymbol{p}_{1\perp}+\boldsymbol{p}_{2\perp})
\end{align}
We then introduce the change of variables from $\ktone,\kttwo$ to $\pt=z_2\ktone-z_1\kttwo$ and $\qt=\ktone+\kttwo$, where the latter variables are Fourier conjugate to $\rt=\xt-\xt'$ and $\bt=z_1\xt+z_2\xt'$. These manipulations yield
\begin{align}
    &\mathcal{M}_{\gamma^* \to q\bar q }=\frac{e}{(2\pi)^2 q^+}\gamma_\lambda^{s\bar s}(z_1,z_2)\int_0^L\der t \int\der^2\rt\der^2\bt\der^2\boldsymbol{p}_{1\perp} e^{-i\qt\cdot\bt+i(\pt+\boldsymbol{p}_{1\perp})\cdot\rt}\nonumber\\
    &\times e^{-i\frac{L-t}{2z_1z_2q^+}\left(\pt^2+z_1z_2\qt^2\right)}   (V_q(L,t; [\boldsymbol{r}_{q,\rm cl}(s)]) V_{\bar q}^\dagger(L,t; [\boldsymbol{r}_{\bar q,\rm cl}(s)]))_{\beta\gamma}(z_2\boldsymbol{p}_{1\perp}-z_1\boldsymbol{p}_{1\perp})\cdot \et^{\lambda}\,.
\end{align}
In the limit $p_\perp\gg q_\perp$, one has $r_\perp\ll b_\perp$ from the phases, such that one can consider that the dipole is created at the same transverse coordinate point $\bt$. Since the momentum transferred by the medium is small as compared to $p_\perp$ one can also approximate $\boldsymbol{p}_{1\perp}\approx \pt$, such that the integral over $\boldsymbol{p}_{1\perp}$ yields a delta function enforcing $\rt=\boldsymbol{0}_\perp$ in agreement with the scaling $r_\perp\ll b_\perp$. Finally, the $\qt$ dependence in the phase of the exponential can be neglected. In the end, we obtain the following simplified result
\begin{align}
    \mathcal{M}_{\gamma^* \to q\bar q }&\approx \frac{e}{q^+}\gamma_{\lambda} ^{s\bar s}(z_1,z_2)(\pt\cdot \et^{\lambda})\int_0^L\der t \ e^{-i\frac{\pt^2}{2z_1z_2q^+}(L-t)}\nonumber\\
    &\times\int\der^2\bt e^{-i\qt\cdot\bt}  (V_q(L,t; [\boldsymbol{r}_{q,\rm cl}(s)]) V_{\bar q}^\dagger(L,t; [\boldsymbol{r}_{\bar q,\rm cl}(s)]))_{\beta\gamma}\,,\label{eq:amplitude-finite-tf}
\end{align}
where implicitly, the classical trajectories of the quark and antiquark in the Wilson lines are given by $\boldsymbol{r}_{q,\rm cl}(s)=\boldsymbol{b}_\perp +(s-t)\frac{\pt}{z_1q^+}$ and $\boldsymbol{r}_{\bar q,\rm cl}(s)=\boldsymbol{b}_\perp -(s-t)\frac{\pt}{z_2q^+}$.

For $q\bar q$ splitting occurring early inside the medium, such that $t_f\sim 2z_1z_2q^+/P_\perp^2\ll L$, one can extend the time integral up to infinity
\begin{align}
    \lim\limits_{\epsilon\to 0^+}\int_0^\infty \der t \ e^{i\frac{\pt^2}{2z_1z_2q^+}t-\epsilon t}=\frac{2iz_1z_2q^+}{p_\perp^2}\,,
\end{align}
and approximate $V(L,t; [\boldsymbol{r}_{q,\rm cl}(s)])\approx V(L,0; [\boldsymbol{r}_{q,\rm cl}(s)])$ up to power corrections in $t_f/L$ which are suppressed by an additional power of $p_\perp^2$ with respect to the leading term of interest. Our final result for the amplitude is thus, up to a pure phase factor,
\begin{align}
     \mathcal{M}_{\gamma^* \to q\bar q }&\approx e\gamma_{\lambda} ^{s\bar s}(z_1,z_2)\frac{2iz_1z_2(\pt\cdot \et^{\lambda})}{p_\perp^2}\nonumber\\
     &\times\int\der^2\bt e^{-i\qt\cdot\bt}  (V_q(L,0; [\boldsymbol{r}_{q,\rm cl}(s)]) V_{\bar q}^\dagger(L,0; [\boldsymbol{r}_{\bar q,\rm cl}(s)]))_{\beta\gamma}
\end{align}
This expression almost displays factorisation between the hard scale dependence on $p_\perp$ and the soft scale $q_\perp$ ; though, the factorisation is broken by the implicit $\pt$ dependence of the Wilson lines through the eikonal trajectories of the quark-antiquark pair.

\subsection{Cross-section}

The differential cross-section for $q\bar q$ pair production in the regime $p_\perp\gg q_\perp$ is obtained by squaring the amplitude%
\footnote{Note that we average over the transverse polarisation of the photon.} and yields
\begin{align}
    \frac{\der\sigma^{\gamma^*\to q\bar q+X}}{\der z_1\der z_2\der^2\pt\der^2\qt}
 &=\frac{\alpha_{\rm em}N_c}{4\pi}\frac{(z_1^2+z_2^2)}{p_\perp^2}\times\mathcal{P}_s(\qt,\theta_{q\bar q},L)\label{eq:quasi-TMD-factorization}
\end{align}
with
\begin{align}
    \mathcal{P}_s(\qt,\theta_{q\bar q},L)\equiv \int\frac{\der^2\bt\der^2\bt'}{(2\pi)^3}e^{-i\qt\cdot(\bt-\bt')}\frac{1}{N_c}\left\langle \Tr V_q V_{\bar q}^\dagger V_{\bar q}' V_q'^{\dagger}\right\rangle\,,\label{eq:Ps(Qt)-def}
\end{align}
the transverse momentum imbalance distribution. From now on, we use the shorthand notation $V_{q/\bar q} =V_{q/\bar q}(L,0; [\boldsymbol{r}_{q/\bar q,\rm cl}(s)])$ for the Wilson lines, and likewise for $V'_{q/\bar q}$ with $\bt$ replaced by $\bt'$ for the initial transverse position of the classical trajectory in the complex conjugate amplitude. The brackets $\langle ... \rangle$ denotes the medium average over the background gauge field configurations.

The physical meaning of this cross-section is quite transparent: the first factor accounting for the $p_\perp$ dependence encodes the perturbative $\gamma^*\to q\bar q$ QED splitting as given by the corresponding $P_{q\gamma}(z)=z^2+(1-z)^2$ DGLAP splitting function, while the second factor encodes the effect of the medium on the transverse momentum imbalance of the quark-antiquark dipole. As mentioned, the $\qt$ distribution also depends on $p_\perp$ via the opening angle $\theta_{q\bar q}$ of the dipole. 

In order to evaluate the $\qt$ distribution, one needs to compute the medium average of four Wilson lines, i.e.~the quadrupole correlator. For a background field with Gaussian statistics and a two point function which is local in time, this is a standard calculation (see e.g.~\cite{Dominguez:2011wm,Isaksen:2023nlr}). In the large $N_c$ limit and using the harmonic approximation Eq.\,\eqref{eq:harmonic-approx-def}, we have
\begin{align}
   \frac{1}{N_c} \left\langle V_q V_{\bar q}^\dagger V_{\bar q}' V_q'^{\dagger}\right\rangle&=\exp\left(-\frac{1}{4}\int_0^L \der t \  \hat q\left[(\boldsymbol{r}_{q,\rm cl}(t)-\boldsymbol{r}'_{q,\rm cl}(t))^2+(\boldsymbol{r}_{\bar q,\rm cl}(t)-\boldsymbol{r}'_{\bar q,\rm cl}(t))^2\right]\right)\nonumber\\
   &-\frac{1}{2}\int_0^L \der t'\exp\left(-\frac{1}{4}\int_{t'}^L \der t  \ \hat q\left[(\boldsymbol{r}_{q,\rm cl}(t)-\boldsymbol{r}'_{q,\rm cl}(t))^2+(\boldsymbol{r}_{\bar q,\rm cl}(t)-\boldsymbol{r}'_{\bar q,\rm cl}(t))^2\right]\right)\nonumber\\
    &\times \qhat (\boldsymbol{r}_{q,\rm cl}(t')-\boldsymbol{r}'_{q,\rm cl}(t'))\cdot(\boldsymbol{r}_{\bar q,\rm cl}(t')-\boldsymbol{r}'_{\bar q,\rm cl}(t'))\nonumber\\
    &\times \exp\left(-\frac{1}{4}\int_0^{t'} \der t \ \hat q\left[(\boldsymbol{r}_{q,\rm cl}(t)-\boldsymbol{r}_{\bar q,\rm cl}(t))^2+(\boldsymbol{r}'_{\bar q,\rm cl}(t)-\boldsymbol{r}'_{q,\rm cl}(t))^2\right]\right)\,,\\
    &=\exp\left(-\frac{1}{2}\int_0^L \der t \ \hat q(\bt-\bt')^2\right)-\frac{1}{2}\int_0^L \der t' \ \qhat (\bt-\bt')^2\nonumber\\
    &\times \exp\left(-\frac{1}{2}\int_0^{t'} \der s \ \hat q s^2\frac{p_\perp^2}{E^2}\right)\exp\left(-\frac{1}{2}\int_{t'}^L \der s \ \hat q(\bt-\bt')^2\right)\,.
\end{align}
The first term in this expression, dubbed the ``factorisable piece" in the literature~\cite{Isaksen:2023nlr}, represents the independent transverse momentum broadening of the quark and antiquark. The second, non-factorisable, term arises from the geometry of the $q\bar q$ splitting as it depends on the opening angle $\theta_{q\bar q}^2=2P_\perp^2/E^2$ of the dipole. Using this expression, the broadening distribution is obtained by Fourier transform from $\bt-\bt'$ to $\qt$:
\begin{align}
         \mathcal{P}_s(\qt,\theta_{q\bar q},L)&=\frac{S_\perp}{(2\pi)^2 Q_s^2}\exp\left(-\frac{q_\perp^2}{2Q_s^2}\right)\nonumber\\
         &\hspace{-0.5cm}\times \left\{1-\frac{1}{2}\int_0^1\der s\frac{q_\perp^2-2Q_s^2(1-s)}{Q_s^2(1-s)^3}\exp\left[-\frac{q_\perp^2s}{2Q_s^2(1-s)}\right]\exp\left[-\frac{\theta_{q\bar q}^2s^3}{3\theta_c^2}\right]\right\}\,,\label{eq:Pqt-feynman}
\end{align}
with $S_\perp$ the transverse area of the medium and $Q_s^2=\qhat L$. Although not obvious at first sight, this expression is actually equivalent to the $\qt$-distribution obtained in the main text, see Eq.\,\eqref{eq:OQS-factor}. More precisely, once multiplied by $2\pi /S_\perp$, Eq.\,\eqref{eq:Pqt-feynman} is identical to the second term in the curly bracket of Eq.\,\eqref{eq:OQS-factor} which described the $\qt$-broadening of the colour octet component of the density matrix. To see this, one must first identify $t=L$ in Eq.\,\eqref{eq:OQS-factor} and rescale the variable $s\to s/L$. Then, in Eq.\,\eqref{eq:Pqt-feynman},  one must perform the change of variable $s\to 1-s$ in the integral over $s$ and do an integration by part in the first term of the integral over $s$ proportional to $q_\perp^2$. After these manipulations, it is straightforward to see that the two distributions are equal. This provides a non-trivial cross-check of Eq.\,\eqref{eq:OQS-factor} which is a direct consequence of the quantum master equation established in this paper.

\subsection{Generalisation to a colour octet initial state}

The previous calculation of $\mathcal{P}_s(\qt,\theta_{q\bar q},L)$ can easily be generalised for a colour octet initial state, as produced for instance in a gluon to $q\bar q$ hard splitting. The projector on the colour state $c$ is given by $\sqrt{2}t^c_{ii'}$ where $i$ and $i'$ are the colour indices of the quark and antiquark in the initial state. The colour structure that would appear inside the Fourier transform in Eq.\,\eqref{eq:Ps(Qt)-def} reads then
\begin{align}
    2\textrm{Tr}\left[V_q t^cV_{\bar q}^\dagger V_{\bar q}' t^c V_q'^{\dagger}\right]\,.
\end{align}
Averaging over the number of colour states and using the Fierz identity, we find that the relevant operator is actually
\begin{align}
  \frac{1}{N_c^2-1}\left\{\textrm{Tr}\left[ V_{\bar q}'V_{\bar q}^\dagger\right] \textrm{Tr}\left[V_qV_q'^{\dagger}\right]-\frac{1}{N_c} \textrm{Tr}\left[V_qV_{\bar q}^\dagger V_{\bar q}'V_q'^{\dagger}\right]\right\}\,.\label{eq:octet-fierz}
\end{align}
In the large $N_c$ limit, the average of the product of the two traces in the first term --- the second term can be neglected at large $N_c$ anyway --- is equal to the product of the averages of the two traces, hence the product of two dipole correlators. 
We have then, for a static medium,
\begin{align}
  \frac{1}{N_c^2}  \left\langle \textrm{Tr}\left[ V_{\bar q}'V_{\bar q}^\dagger\right]\right\rangle  \left\langle\textrm{Tr}\left[V_qV_q'^{\dagger}\right]\right\rangle
  &=\exp\left(-\frac{1}{2}\qhat L (\bt-\bt')^2\right)
\end{align}
 In the end, the $\qt$ distribution for a colour octet initial state is a simple Gaussian distribution with a width twice as a large as for a single parton,
\begin{align}
        \mathcal{P}_o(\qt,\theta_{q\bar q},L)&=\frac{S_\perp}{(2\pi)^2 Q_s^2}\exp\left(-\frac{q_\perp^2}{2Q_s^2}\right)\,,\label{eq:octet-case-largeNc}
\end{align}
such that $\langle q_\perp^2\rangle_o=2Q_s^2$, in agreement with Eq.\,\eqref{eq:meanqt-octet} after taking the large $N_c$ limit.

\section{Further details on the master equation derivation}
\label{app:master-equation}

In this appendix, we outline the calculation of the quantum master equation satisfied by the density matrix of the subsystem made of the high-energy quark-antiquark pair propagating through a weakly coupled quark-gluon plasma. As discussed in the main text, the derivation closely follows~\cite{Blaizot:2017ypk} obtained in the context of quarkonia.

For the derivation of the master equation is useful to use temporarily the interaction picture of the equation of motion \eqref{eq:dens_mat:eqmot:schr-pic}, which reads
\begin{equation}
    \frac{\der\rho^I(t)}{\der t} = -i[H_\text{int}^I(t),\rho^I(t)],
    \label{eq:dens_mat:eqmot:int-pic}
\end{equation}
with $H_\text{int}^I(t)=U_0^\dagger(t)H_\text{int}U_0(t)$ the free evolution of $H_\text{int}$ defined in Eq.\,\eqref{eq:ham:interaction}.
To set up a perturbative expansion in the small coupling constant $g$, we can first get the formal solution by integration 
\begin{equation}
    \rho^I(t)=\rho(t_0)-i\int^t_{t_0}\der t'[H_\text{int}^I(t'),\rho^I(t')],
\end{equation}
where we consider the total initial density matrix factorized as $\rho(t_0)=\rho_S(t_0)\otimes_{S,E}\rho_{\text{pl}}$.
Then, we insert it back in the Von Neumann's equation Eq.\,\eqref{eq:dens_mat:eqmot:int-pic} and we perform the partial trace $\Tr_\text{pl}(\cdot)$ over the environment degrees of freedom, yielding
\begin{equation}
    \frac{\der\rho^I_S(t)}{\der t}
    =
    -\int^t_{t_0}\der t'\,\Tr_{\text{pl}}\left\{[H_\text{int}^I(t),[H_\text{int}^I(t'),\rho^I(t')]]\right\}.
\end{equation}
Notice, by also using the factorized expression of the interaction Hamiltonian Eq.\,\eqref{eq:ham:interaction}, that the first order term $\Tr_{\text{pl}}\{[H_\text{int}^I(t),\rho^I(t_0)]\}$ vanishes as the partial trace gives an average $\langle A\rangle_\text{pl}=0$, which vanishes when the QGP is assumed to be colour neutral~\cite{Blaizot:2017ypk}.
Now, we perform two useful approximations which are important in usual derivations of Lindblad equations.
First, we substitute $\rho^I(t')\to\rho^I(t)$, since it is a second order correction in the coupling constant. The master equation is now \textit{local} in time as the density matrix $\rho^I(t)$ can be obtained without knowing its past values (\textit{Markov approximation}).
Second, we assume that the factorization $\rho^I(t)=\rho_S^I(t)\otimes \rho_E$ is valid at all times. This approximation (\textit{Born approximation}) is motivated by the weak coupling between the system and the environment, and the fact that the environment is ``larger'' than the system, meaning that it modifies the system without feeling any modification itself (as a thermal bath).
The result is
\begin{multline}
    \frac{\der\rho_S^I(t)}{\der t}=-g^2\int_{t_0}^t \der t' \int_{\xt \xt'}\biggl\{\left[\hat n^a(t, \xt), \hat n^b\left(t', \xt'\right) \rho_S^I\left(t\right)\right] \left\langle A^-_a(t,\xt)A^-_b(t',\xt')\right\rangle_\text{pl}+ \\
    +\left[\rho_S^I\left(t\right) \hat n^a\left(t', \xt'\right), \hat n^b(t, \xt)\right] \left\langle A^-_a(t',\xt')A^-_b(t,\xt)\right\rangle_\text{pl}
    \biggr\}.
    \label{eq:dm:evolution:derivative}
\end{multline}
Following the derivation in \cite{Blaizot:2017ypk}, we can adopt the definition
\begin{equation}
    \left\langle T_\mathcal{C} A^-_a(t,\xt)A^-_b(t',\xt')\right\rangle_\text{pl}
    =
    \delta^{ab}\Delta^{>}(t-t',\xt-\xt'),
\end{equation}
where $T_\mathcal{C}$ orders on the left the operators with the largest time along the Schwinger-Keldysh contour~\cite{Keldysh:1964ud}.
Since in Eq.\,\eqref{eq:dm:evolution:derivative}, this ordering is already implemented, as $t\geq t'\geq t_0$, we can write
\begin{multline}
    \frac{\der\rho_S^I(t)}{\der t}=-g^2\int_{t_0}^t \der t' \int_{\xt \xt'}\biggl\{\left[\hat n^a(t, \xt), \hat n^a\left(t', \xt'\right) \rho_S^I\left(t\right)\right] \Delta^{>}\left(t-t', \xt-\xt'\right)+ \\
    +\left[\rho_S^I\left(t\right) \hat n^a\left(t', \xt'\right), \hat n^a(t, \xt)\right] \Delta^{<}\left(t-t', \xt-\xt'\right)\biggr\},
\end{multline}
with $\Delta^>$ and $\Delta^<$ ordered respectively in the upper and lower branch of the contour $\mathcal{C}$.
Now, as the propagator $\Delta(t-t',\rt)$ vanishes for large time intervals $\tau\equiv t-t'\gtrsim m_D^{-1}$, with $m_D\sim gT$ the Debye mass, the relevant values of the integrand will be for values of $\tau$ which are
\begin{equation}
    \tau = t-t'\lesssim m_D^{-1}.
    \label{eq:tau:debye-mass}
\end{equation}
This allows us to extend the domain of integration from $t-t_0$ to $+\infty$ and establish the \textit{Markovianity} of the equation.
This approximation is often used in derivations of the Lindblad equation and it is related to the hierarchy between the \textit{environment correlation time} $\tau_E\sim 1/(gT)$ and the \textit{system relaxation time} $\tau_R$ of the order of the mean-free path $1/(g^2T)$.
Then, by using the Schr\"odinger picture (and dropping the ``subsystem'' $S$ subscript) in order to make explicit the time evolution operators for future approximations, we get
\begin{multline}
    \frac{\der \rho(t)}{\der t}+i[H_0,\rho(t)]
    =\\
    =-g^2\int_{\xt \xt'}
        \int^{+\infty}_0 \der\tau 
        \biggl\{
            \left[\hat n^a(\xt), U_0(\tau)\hat n^a(\xt')U_0^\dagger(\tau) \rho(t)\right] \Delta^{>}\left(\tau, \xt-\xt'\right)+ \\
            +\left[\rho(t)U_0(\tau)\hat n^a(\xt')U_0^\dagger(\tau),\hat n^a(\xt)\right] \Delta^{<}\left(\tau, \xt-\xt'\right)
    \biggr\}.
    \label{eq:app:meq:schr}
\end{multline}
This equation can be additionally approximated since, as stated above, only small values of $\tau$ will be significant for the integral. 
In particular, we can consider the projection of the master equation onto the (continuous spectrum of) eigenstates of the free Hamiltonian operator: $H_0\ket{n}=\epsilon_n\ket{n}$, where $\ket n=\ket{n_1}\!\otimes\!\ket{n_2}$ belongs to the quark-antiquark Hilbert space. For each of the four terms on the right-hand side of Eq.\,\eqref{eq:app:meq:schr} one can verify that the relative-time dependence comes from phases with a frequency given by energy differences. We show it for the first term
\begin{multline}
    \sum_{r,s}\bra{n}\hat n^a(\xt) \ket{r}\!\bra{r}U_0(\tau)\hat n^a(\xt')U_0^\dagger(\tau)\ket{s}\!\bra{s}\rho(t)\ket{m}\Delta^{>}\left(\tau, \xt-\xt'\right)
    =\\=
    \sum_{r,s}e^{-i(\epsilon_r-\epsilon_s)\tau}
    [\hat n^a(\xt)]_{nr}
    [\hat n^a(\xt')]_{rs}[\rho(t)]_{sm}
    \Delta^{>}\left(\tau, \xt-\xt'\right),
    \label{eq:app:energy-gaps:coh}
\end{multline}
and it holds also the other ones.
This means that if
\begin{equation}
    \tau\sim\tau_E\ll\frac{1}{|\epsilon-\epsilon'|},
    \label{eq:timescale:brownian}
\end{equation}
then we can expand to the first order the operators
\begin{equation}
    U_0(\tau)\hat n^a(\xt)U_0^\dagger(\tau)
    =
    \hat n^a(\xt)
    -i\tau[H_0,\hat n^a(\xt)]
    +O(\tau^2).
    \label{eq:app:tau-expansion}
\end{equation}
In order to estimate the size of the energy gaps one needs to look at the other matrix elements involved. Considering Eq.\,\eqref{eq:app:energy-gaps:coh} and knowing the expression of the number density operator from Eq.\,\eqref{eq:numberdensity:dipole}, we evaluate the matrix elements $[\hat n^a(\xt)]_{nr}[\hat n^a(\xt')]_{rs}$.
We know that
\begin{align}
    [\hat n^a(\xt)]_{nr}
    &\propto
    t^a\bra n\left(\ket \xt\!\bra \xt\otimes \mathbbm{1}_{\bar q}\right)\ket r
    -
    \tilde t^a\bra n\left(\mathbbm{1}_{q}\otimes \ket \xt\!\bra\xt\right)\ket r\nonumber\\
    &=
    t^a\braket{n_1| \xt}\!\braket{\xt|r_1}\!\braket{n_2|r_2}
    -
    \tilde t^a\braket{n_1|r_1}\!\braket{n_2| \xt}\!\braket{\xt|r_2}\nonumber\\
    &=t^a
    e^{i(\boldsymbol{k}_{\perp1,r}-\boldsymbol{k}_{\perp1,n})\cdot\xt}
    \delta^{(2)}(\boldsymbol{k}_{\perp2,n}-\boldsymbol{k}_{\perp2,r})
    -
    \tilde t^a
    \delta^{(2)}(\boldsymbol{k}_{\perp1,r}-\boldsymbol{k}_{\perp1,n})
    e^{i(\boldsymbol{k}_{\perp2,r}-\boldsymbol{k}_{\perp2,n})\cdot\xt},
\end{align}
where the eigenstates $\ket n=\ket{n_1}\otimes\ket{n_2}$ of the free Hamiltonian are eigenstates also of quark-antiquark transverse momentum. We have used the convention $\braket{\xt|\boldsymbol{k}_{\perp,i}}=e^{i\boldsymbol{k}_{\perp,i}\cdot \xt}$. Hence, the product reads
\begin{align}
    &[\hat n^a(\xt)]_{nr}
     [\hat n^a(\xt')]_{rs}
    \propto
    \nonumber\\
    &\Big[
     (t^a\otimes\mathbbm1)
     e^{-2i(\boldsymbol p_{\perp,n}
           -\boldsymbol p_{\perp,r})\cdot\xt}
     \delta^{(2)}\!\left(
       \boldsymbol k_{2\perp,n}
       -\boldsymbol k_{2\perp,r}
     \right)
    -
     (\mathbbm1\otimes\tilde t^a)
     e^{2i(\boldsymbol p_{\perp,n}
           -\boldsymbol p_{\perp,r})\cdot\xt}
     \delta^{(2)}\!\left(
       \boldsymbol k_{1\perp,n}
       -\boldsymbol k_{1\perp,r}
     \right)
    \Big]
    \nonumber\\
    &\Big[
     (t^a\otimes\mathbbm1)
     e^{-2i(\boldsymbol p_{\perp,r}
           -\boldsymbol p_{\perp,s})\cdot\xt'}
     \delta^{(2)}\!\left(
       \boldsymbol k_{2\perp,r}
       -\boldsymbol k_{2\perp,s}
     \right)
    -
     (\mathbbm1\otimes\tilde t^a)
     e^{2i(\boldsymbol p_{\perp,r}
           -\boldsymbol p_{\perp,s})\cdot\xt'}
     \delta^{(2)}\!\left(
       \boldsymbol k_{1\perp,r}
       -\boldsymbol k_{1\perp,s}
     \right)
    \Big],
    \label{eq:app:density-product}
\end{align}
and the Dirac deltas are constraining $\boldsymbol{p}_{\perp,i}-\boldsymbol{p}_{\perp,j}=\pm(\boldsymbol{q}_{\perp,i}-\boldsymbol{q}_{\perp,j})/2$. The definitions of $\boldsymbol{p}_{\perp,i}$ and $\boldsymbol{q}_{\perp,i}$ are the ones introduced in Eq.\,\eqref{eq:mom-def:rel}-\eqref{eq:mom-def:tot} (with $z_1=z_2=1/2$), used now for any $i$th eigenstate.
For example, one of the four phase contributions to the integral over $\xt$ and $\xt'$, coming from Eq.\,\eqref{eq:app:meq:schr}, yields
\begin{align}
    &
    \int_{\xt\xt'}
    e^{2i(\boldsymbol{p}_{\perp,n}-\boldsymbol{p}_{\perp,r})\cdot\xt}
    e^{-2i(\boldsymbol{p}_{\perp,r}-\boldsymbol{p}_{\perp,s})\cdot\xt'}
    \Delta^{>}\left(\tau, \xt-\xt'\right)\propto\nonumber\\
    &\propto
    \int_{\kt}
    \delta^{(2)}(\kt+2(\boldsymbol{p}_{\perp,n}-\boldsymbol{p}_{\perp,r}))
    \delta^{(2)}(\kt+2(\boldsymbol{p}_{\perp,r}-\boldsymbol{p}_{\perp,s}))
    \Delta^{>}\left(\tau, \kt\right),
\end{align}
and three other terms whose delta functions equivalently constrain the size of the momentum differences $|p^2_\perp-p'^2_\perp|$, as we know that $\Delta^{>}\left(\tau, \kt\right)$ is significantly different from zero for soft momentum exchanges of the order $|\kt|\sim m_D$.
Hence, by using the definition of $\tau_E$ in \eqref{eq:tau:debye-mass} and knowing the kinetic energy from Eq.\,\eqref{eq:ham:kinetic}, we can make an estimation of the condition \eqref{eq:timescale:brownian}, which is satisfied by
\begin{align}
    \frac{1}{m_D}
    \ll
    \frac{2E}{|p_\perp^2-p_\perp'^2|}
    \sim
    \frac{1}{\theta_{q\bar q} m_D}
    \quad
    \implies
    \quad
    \theta_{q\bar q}\ll1,
\end{align}
where we have used $|p_\perp-p'_\perp|\sim m_D$, as motivated above, and $|p_\perp+p'_\perp|\sim E\theta_{q\bar q}$. In computing the energy difference $|\epsilon-\epsilon'|$ we have also neglected the contribution from the center of mass $(\qt^2-\qt'^2)/2q^+$, much smaller than the one of the relative degree of freedom due to the hierarchy $|\qt|\ll|\pt|$.

This condition is what usually in open quantum systems literature is referred to as the \textit{quantum Brownian regime}, where the environment correlation time $\tau_E$ is much smaller than the system-intrinsic time scale $\tau_S$, and where this gradient expansion in orders of $\tau$ is permitted.
The first perturbative order in the expansion \eqref{eq:app:tau-expansion} is the one responsible for friction, which is neglected here. This term is however relevant in the case of quarkonia master equations where $E$ is typically replaced by the mass of the bound state which is not as large as $E$ in jet quenching.

Hence, by taking Eq.\,\eqref{eq:app:tau-expansion} at the leading order, it gives
\begin{align}
    \frac{\der \rho(t)}{\der t}+i[H_0,\rho(t)]
    \simeq
    &-g^2\int_{\xt \xt'} 
            \left[\hat n^a(\xt),\hat n^a(\xt') \rho(t)\right]
        \int^{+\infty}_0 \der\tau\,\Delta^{>}\left(\tau, \xt-\xt'\right)
    \label{eq:dm:evolution:derivative:approx:1:term1}\\
    &
    -g^2\int_{\xt \xt'} 
            \left[\rho(t)\hat n^a(\xt'),\hat n^a(\xt)\right]
        \int^{+\infty}_0 \der\tau\,\Delta^{<}\left(\tau, \xt-\xt'\right).
    \label{eq:dm:evolution:derivative:approx:1:term2}
\end{align}
The integrals over $\tau$ of the correlators can get a transparent physical meaning after few manipulations.
First is important to recall the definition of the time-ordered propagator
\begin{equation}
    \left\langle T A^-_a(t,\xt)A^-_b(t',\xt')\right\rangle_\text{pl} = -i\delta_{ab}\Delta(t-t',\xt-\xt'),
\end{equation}
and transform the time component in Fourier space
\begin{equation}
    \Delta(\omega) 
    = 
    \int^{+\infty}_{-\infty}\der t\,e^{i\omega t}
        \Delta(t),
    \label{eq:prop:Tord:omega}
\end{equation}
where the transverse position dependence are kept implicit for the moment.
The time-ordered propagator in Eq.\,\eqref{eq:prop:Tord:omega} can be decomposed in two parts by exploiting the operational relation
\begin{align}
    T[A(x)A(y)]
    =
    \theta(x^0-y^0)[A(x),A(y)]+A(y)A(x),
\end{align}
Inserted into Eq.\,\eqref{eq:prop:Tord:omega} gives
\begin{align}
    \Delta(\omega)
    &=
    i\int^{+\infty}_{-\infty}\der t\,e^{i\omega t}\theta(t)\left\langle[A(t),A(0)]\right\rangle_{\text{pl}}
    +i\int^{+\infty}_{-\infty}\der t\,e^{i\omega t}\left\langle A(0),A(t)\right\rangle_{\text{pl}}\\
    &= \Delta^{\text{ret}}(\omega)+i\Delta^<(\omega),
\end{align}
where we used the definition of the retarded propagator and of the lower branch propagator in the Schwinger-Keldysh contour.
In order to obtain the integrals in Eq.\,\eqref{eq:dm:evolution:derivative:approx:1:term1} we compute (ignoring the diagonal colour indices in our notations)
\begin{align}
    \int^{+\infty}_{-\infty}\der\tau\,
        \Delta(\tau)
    &=
    i\int^{+\infty}_{-\infty}\der\tau\,
        \Big[
            \theta(\tau)
            \underbrace{\left\langle A(\tau)A(0)\right\rangle_{\text{pl}}}_{\Delta^>(\tau)}
            +
            \theta(-\tau)
            \underbrace{\left\langle A(0)A(\tau)\right\rangle_{\text{pl}}}_{\Delta^<(\tau)}
        \Big]\\
    &=
    i\int^{+\infty}_0\der\tau\,\Delta^>(\tau)+
    i\int^{+\infty}_0\der\tau\,\Delta^<(-\tau)\\
    &=
    2i\int^{+\infty}_0\der\tau\,\Delta^>(\tau),
\end{align}
where in the last step we exploited the relation $\Delta^<(-\tau) = \Delta^>(\tau)$, which comes from their definition and the spacetime translational invariance of the correlators. Hence, we can compute the two time integrals in Eq.\,\eqref{eq:dm:evolution:derivative:approx:1:term2} as a Fourier transform 
\begin{align}
    \int^{+\infty}_0\der\tau\,\Delta^>(\tau)
    &=
    -\frac{i}{2}\int^{+\infty}_{-\infty}\der\tau\,\Delta(\tau)
    = -\frac{i}{2}\Delta^{\text{ret}}(\omega=0)+\frac{1}{2}\Delta^<(\omega=0)\,,\\
     \int^{+\infty}_0\der\tau\,\Delta^<(\tau)
    &=
    \int^{+\infty}_{-\infty}\der\tau\,\Delta^<(\tau)
    -\int^{0}_{-\infty}\der\tau\,\Delta^<(\tau)
=\frac{i}{2}\Delta^{\text{ret}}(\omega=0)+\frac{1}{2}\Delta^<(\omega=0).
\end{align}
Re-introducing the position dependence we define the \textit{real} and \textit{imaginary} potentials respectively as \cite{Blaizot:2017ypk}
\begin{equation}
   \Delta^{\text{ret}}(\omega=0,\xt)
   \equiv
   -V(\xt),
   \qquad
   \Delta^<(\omega=0,\xt)
   \equiv
   -W(\xt).
   \label{eq:app:potentials}
\end{equation}
In the context of high-energy parton propagation, we remind the reader that the ``time" refers to the plus light-cone component in these equations. Thus, the space-time interval appearing in the definition of the retarded propagator is always space-like, such that this retarded propagator and $V$ are actually zero by causality. Contrarily to the heavy quarks case, in the high-energy limit of the propagation of two light quarks there is no relevant contribution related to the \textit{screening} of the binding potential. The resulting equation is then Eq.\,\eqref{eq:dm:EoM}.

\section{Analytical results of relative diffusion perturbation scheme}
\label{app:pt-diff:pert-results}

In this last appendix, we provide the first correction in the $\kappa^2=\qhat^2L^4/E^2=2\theta_s/\theta_c$ expansion of the physical parameters in the Gaussian ansatz solving the partial differential equation Eq.\,\eqref{eq:rhos-section-ptdiffusion} satisfied by the singlet Wigner density. Higher orders corrections in $\kappa^2$ can be obtained from these formulas by iteration.

\subsection{Second moments}
\label{app:pt-diff:pert-results:second-mom}

    The solutions of Eqs.\,\eqref{eq:diffusion-pt:second-moms:eq:1st-order:sigmar2}, \eqref{eq:diffusion-pt:second-moms:eq:1st-order:sigmap2} and \eqref{eq:diffusion-pt:second-moms:eq:1st-order:corr} are
    \begin{align}
        (\sigma_{\bar r}^2)^{(1)}
        &=\frac{t^3}
        {210 D^2(\bar t)}
        \Big[
            2 \sigma_{\bar r,0}^2 \bar t \left(11 \sigma_{\bar p,0}^4 \bar t^6+1008 \sigma_{\bar p,0}^2 \bar t^3+25200\right)+\nonumber\\
            &+32 \left(4 \sigma_{\bar p,0}^4 \bar t^6+147 \sigma_{\bar p,0}^2 \bar t^3+2520\right)+\sigma_{\bar r,0}^4 \bar t^2 \left(\sigma_{\bar p,0}^4 \bar t^6+126 \sigma_{\bar p,0}^2 \bar t^3+8064\right)
        \Big],\\
        (\sigma_{\bar p}^2)^{(1)}
        &=\frac{\bar t}
        {70 D^2(\bar t)}
        \Big[
            18 \sigma_{\bar r,0}^2 \bar t \left(13 \sigma_{\bar p,0}^4 \bar t^6+560 \sigma_{\bar p,0}^2 \bar t^3+4480\right)+\nonumber\\
            &+32 \left(34 \sigma_{\bar p,0}^4 \bar t^6+525 \sigma_{\bar p,0}^2 \bar t^3+2520\right)+\sigma_{\bar r,0}^4 \bar t^2 \left(13 \sigma_{\bar p,0}^4 \bar t^6+840 \sigma_{\bar p,0}^2 \bar t^3+20160\right)
        \Big],\\
        c_{\bar r,\bar p}^{(1)}
        &=\frac{\bar t^3}
        {420 D^2(\bar t)}
        \Big[
            4 \sigma_{\bar r,0}^2 \bar t \left(55 \sigma_{\bar p,0}^4 \bar t^6+3528 \sigma_{\bar p,0}^2 \bar t^3+50400\right)+\nonumber\\
            &+576 \left(2 \sigma_{\bar p,0}^4 \bar t^6+49 \sigma_{\bar p,0}^2 \bar t^3+420\right)+\sigma_{\bar r,0}^4 \bar t^2 \left(11 \sigma_{\bar p,0}^4 \bar t^6+1008 \sigma_{\bar p,0}^2 \bar t^3+40320\right)
        \Big],
    \end{align}
    with $D(\bar t)\equiv\sigma_{\bar p,0}^2\bar t^3( \sigma_{\bar r,0}^2\bar t+8)+24(\sigma_{\bar r,0}^2 \bar t+2)$.

\subsection{First moments}
\label{app:pt-diff:pert-results:first-mom}

The perturbative solutions of the first moments equations in Eq.\,\eqref{eq:diffusion-pt:first-moms:eq} up to $O(\kappa^2)$ read
\begin{align}
    \langle r_{\perp}^i\rangle
    &=
    -\frac{12 \left[r_{\perp,0}^i \sigma_{\bar p,0}^2 \bar t^2-2 p_{\perp,0}^i \left(\sigma_{\bar r,0}^2 \bar t+2\right)\right]}{D(\bar t)}
    +
    \frac{\kappa ^2 \bar t^3}{840 D^2(\bar t)}
    \Big\{\sigma_{\bar r,0}^2 \bar t \big[36 \sigma_{\bar p,0}^2 \bar t^3 (14 r_{\perp,0}^i-27 p_{\perp,0}^i \bar t)\nonumber\\
    &-40320 (p_{\perp,0}^i \bar t+r_{\perp,0}^i)+13 r_{\perp,0}^i \sigma_{\bar p,0}^4 \bar t^6\big]+8 \big[-18 \sigma_{\bar p,0}^2 \bar t^3 (26 p_{\perp,0}^i \bar t+21 r_{\perp,0}^i)\nonumber\\
    &-2520 (3 p_{\perp,0}^i \bar t+4 r_{\perp,0}^i)+19 r_{\perp,0}^i \sigma_{\bar p,0}^4 \bar t^6\big]-18 p_{\perp,0}^i \sigma_{\bar r,0}^4 \bar t^3 \left(3 \sigma_{\bar p,0}^2 \bar t^3+280\right)\Big\},\\
    \langle p_{\perp}^i\rangle
    &=
    \frac{4 \left[3 p_{\perp,0}^i \bar t \left(\sigma_{\bar r,0}^2 \bar t+4\right)+r_{\perp,0}^i \left(12-\sigma_{\bar p,0}^2 \bar t^3\right)\right]}{D(\bar t)}\nonumber\\
    &
    +
    \frac{\kappa ^2 \bar t^4}{840 D^2(\bar t)}
    \Big\{\sigma_{\bar r,0}^2 \bar t \left[4 \sigma_{\bar p,0}^2 \bar t^3 (63 r_{\perp,0}^i-65 p_{\perp,0}^i \bar t)-336 (61 p_{\perp,0}^i \bar t+54 r_{\perp,0}^i)+3 r_{\perp,0}^i \sigma_{\bar p,0}^4 \bar t^6\right]\nonumber\\
    &+8 \left[-6 \sigma_{\bar p,0}^2 \bar t^3 (22 p_{\perp,0}^i \bar t+7 r_{\perp,0}^i)-504 (11 p_{\perp,0}^i \bar t+15 r_{\perp,0}^i)+5 r_{\perp,0}^i \sigma_{\bar p,0}^4 \bar t^6\right]\nonumber\\
    &-13 p_{\perp,0}^i \sigma_{\bar r,0}^4 \bar t^3 \left(\sigma_{\bar p,0}^2 \bar t^3+168\right)\Big\},
\end{align}
with $D(\bar t)\equiv\sigma_{\bar p,0}^2\bar t^3( \sigma_{\bar r,0}^2\bar t+8)+24(\sigma_{\bar r,0}^2 \bar t+2)$.

\subsection{Singlet survival probability}
\label{app:pt-diff:pert-results:Ps}

    We have obtained with the perturbative scheme described in Sec.~\ref{sec:pt-diffusion} the analytical expression for the singlet survival probability $P_s$ up to $O(\kappa^2)$.
    The expression $P_s^{(0)}$ at the leading order $O(\kappa^0)$ is shown in Eq.\,\eqref{eq:pt-diff:Ps:sol:0th}.
    The solution for the first order Eq.\,\eqref{eq:pt-diff:Ps:first-order:eq} reads
    \begin{align}
        P_s^{(1)}(\bar t)
        =
        -\frac{1}{35}
        \frac{P_s^{(0)}(\bar t)}{48}
        \bar t^4
        \left(
        \sum_{i=1}^8c_i \bar t^i
        \right),
        \label{eq:pt-diff:Ps:sol:1st}
    \end{align}
    with
    \begin{align}
        &c_0=80640,\\
        &c_1=-8064 \left(3 \bar{\boldsymbol{r}}_{0\perp}^2-7 \sigma_{\bar r,0}^2\right),\\
        &c_2=2688 \left(3 \sigma_{\bar r,0}^4-13 \bar{\boldsymbol{P}}_{0\perp}\cdot\bar{\boldsymbol{r}}_{0\perp}\right),\\
        &c_3=-384 \left(33 \bar{\boldsymbol{P}}_{0\perp}^2+14 \bar{\boldsymbol{P}}_{0\perp}\cdot\bar{\boldsymbol{r}}_{0\perp} \sigma_{\bar r,0}^2-37 \sigma_{\bar p,0}^2\right),\\
        &c_4=96 \left(\sigma_{\bar r,0}^2 \left(50 \sigma_{\bar p,0}^2-41 \bar{\boldsymbol{P}}_{0\perp}^2\right)+7 \bar{\boldsymbol{r}}_{0\perp}^2 \sigma_{\bar p,0}^2\right),\\
        &c_5=8 \left(-39 \bar{\boldsymbol{P}}_{0\perp}^2 \sigma_{\bar r,0}^4+64 \bar{\boldsymbol{P}}_{0\perp}\cdot\bar{\boldsymbol{r}}_{0\perp} \sigma_{\bar p,0}^2+45 \sigma_{\bar p,0}^2 \sigma_{\bar r,0}^4\right),\\
        &c_6=8\left( 11\bar{\boldsymbol{P}}_{0\perp}\cdot\bar{\boldsymbol{r}}_{0\perp} \sigma_{\bar p,0}^2 \sigma_{\bar r,0}^2+16 \sigma_{\bar p,0}^4\right),\\
        &c_7=-8 \sigma_{\bar p,0}^4 \left(\bar{\boldsymbol{r}}_{0\perp}^2-3 \sigma_{\bar r,0}^2\right),\\
        &c_8=\sigma_{\bar p,0}^4 \sigma_{\bar r,0}^4.
    \end{align}

\bibliographystyle{JHEP} 
\bibliography{ref}

\end{document}